\documentclass{appolb}
\usepackage{graphicx}
\usepackage{amssymb,amsmath,tikz}
\usepackage{pgfplots} 
\usepackage{amsthm}
\usepackage{mathrsfs}
\usepackage{pifont}
 \usepackage{tensor}
\usepackage{bigints}
\usepackage{amsmath,graphicx}
\usetikzlibrary{patterns.meta}
\pgfplotsset{compat=newest}
\usepgfplotslibrary{fillbetween}
\usetikzlibrary{arrows.meta,decorations.markings} 

\tikzset
{
   ->-/.style={decoration={markings,mark=at position 0.5 with {\arrow{Straight Barb}}},
               postaction={decorate}}
}

\newcommand{\Beta}{\mathrm{B}}
\newcommand{\widesim}[2][1.5]{
  \mathrel{\overset{#2}{\scalebox{#1}[1]{$\sim$}}}}

\DeclareMathOperator*{\x}{x}

\DeclareMathAlphabet{\mathpzc}{OT1}{pzc}{m}{it}

\newtheorem{lem}{Lemma}
\newtheorem{rem}{Remark}
\newtheorem*{twr*}{Theorem}

\newmuskip\pFqmuskip

\newcommand*\pFq[6][8]{%
  \begingroup 
  \pFqmuskip=#1mu\relax
  \mathchardef\normalcomma=\mathcode`,
  \mathcode`\,=\string"8000
  \begingroup\lccode`\~=`\,
  \lowercase{\endgroup\let~}\pFqcomma
  {}_{#2}F_{#3}{\left[\genfrac..{0pt}{}{#4}{#5};#6\right]}%
  \endgroup
}
\newcommand{\pFqcomma}{{\normalcomma}\mskip\pFqmuskip}

\begin{document}
\title{REPRESENTATION STRUCTURE OF THE $SL(2, \mathbb{C})$ 
ACTING IN THE HILBERT SPACE OF THE QUANTUM COULOMB FIELD
 %
}


\author{Jaros{\l}aw Wawrzycki
\address{\small{and}}
\\[3mm]
{Tomasz Wawrzycki${}^\dag$ 
\address{\small{${}^\dag$ Faculty of Mathematics and Computer Science,
Jagiellonian University, ul. {\L}ojasiewicza 6, 30-348 Krak\'ow, Poland}}
}
\\[3mm]
}

\maketitle
\begin{abstract}
We give a complete description of the representation of $SL(2,\mathbb{C})$ acting in the Hilbert 
space of the quantum Coulomb field and a constructive consistency proof of the axioms of the quantum theory 
of the Coulomb field.
\end{abstract}

\section{Introduction}\label{intro}

In the paper \cite{wawrzyckiSL2C}  we have announced the following

\vspace{0.5cm}

\begin{twr*}
Let $U|_{{}_{\mathcal{H}_{{}_{n}}}}$ be the restriction of the unitary representation $U$ of $SL(2, \mathbb{C})$
in the Hilbert space of the quantum phase field $S$ to the invariant eigenspace ${\mathcal{H}_{{}_{n}}}$
of the total charge operator $Q$ corresponding to the eigenvalue $n\mathpzc{e}$ for some integer $n$. Then 
for all $n$ such that
\[
|n| >  {\textstyle\sqrt{\frac{\pi}{\mathfrak{e}^2}}}
\]
the representations $U|_{{}_{\mathcal{H}_{{}_{n}}}}$ are unitarily equivalent:
\[
U|_{{}_{\mathcal{H}_{{}_{n}}}} \cong_{{}_{U}}
U|_{{}_{\mathcal{H}_{{}_{n'}}}}
\,\,
\textrm{whenever} 
\,\, 
|n| >  {\textstyle\sqrt{\frac{\pi}{\mathfrak{e}^2}}}, \,\,\,
|n'| >  {\textstyle\sqrt{\frac{\pi}{\mathfrak{e}^2}}}.
\]
 
If the two integers $n,n'$ have different absolute values $|n| \neq |n'|$ and 
\[
|n| <  {\textstyle\sqrt{\frac{\pi}{\mathfrak{e}^2}}}, \,\,\,
|n'| < {\textstyle\sqrt{\frac{\pi}{\mathfrak{e}^2}}},
\]
then the representations $U|_{{}_{\mathcal{H}_{{}_{n}}}}$ and $U|_{{}_{\mathcal{H}_{{}_{n'}}}}$
are inequivalent. Each representation $U|_{{}_{\mathcal{H}_{{}_{n}}}}$ contains a unique discrete 
supplementary component if and only if
\[
|n| <  {\textstyle\sqrt{\frac{\pi}{\mathfrak{e}^2}}}, \,\,\,
\] 
and the supplementary components contained in $U|_{{}_{\mathcal{H}_{{}_{n}}}}$ with different 
values of $|n|$ fulfilling the last inequality are inequivalent.
\end{twr*}

We have based the proof on the false lemma asserting that the representation $U|_{{}_{\mathcal{H}_{{}_{n}}}}$ of 
$G = SL(2,\mathbb{C})$ acting in the eigenspace $\mathcal{H}_{n}$
corresponding to the eigenvalue $n\mathfrak{e}$ of the total charge $Q$, has the tensor product structure, 
$U|_{{}_{\mathcal{H}_{{}_{n}}}} = U|_{{}_{|u\rangle}} \otimes U|_{{}_{\mathcal{H}_{{}_{0}}}}$,
where $U|_{{}_{|u\rangle}}$ is the cyclic subrepresentation with the cyclic spherically symmetric vector $|u\rangle = e^{-inS(u)}|0\rangle$. 
However, this (false) lemma was inferred from calculations containing a computational error,
as was recognized only after publication and announced in \cite{Herdegen}.
In this paper we give a proof of the above theorem going along a different line.  
The general idea of the proof consists in decomposition 
of the cyclic representations with cyclic vectors  
\begin{equation}\label{x}
x= c_{\alpha_1}^+ \ldots c_{\alpha_\mathfrak{q}}^+ e^{-inS(u)}|0\rangle \in \mathcal{H}_{n}, \,\,\, \mathfrak{q} = 1,2, \ldots
\end{equation}
and then by recovering inclusion relations between the cyclic representations, where $c_{\alpha_i}^+$
are the creation operators of the transversal infrared photons, and $S(u)$ is the phase operator of \cite{Staruszkiewicz},
all computed in the reference frame with the time-like unit versor $u$. For the construction of decompositions
of the cyclic representations, we generalize the method of \cite{Staruszkiewicz1992ERRATUM}. Contrary to the cyclic state $|u\rangle$
of the cyclic representation investigated in \cite{Staruszkiewicz1992ERRATUM}, 
which is spherically symmetric, our cyclic states are not spherically symmetric. Therefore, the reproducing kernel Hilbert spaces, 
corresponding to our cyclic representations must necessarily be constructed on the whole group $SL(2, \mathbb{C})$, and not just on the 
Lobachevsky space $SL(2, \mathbb{C})/SU(2,\mathbb{C})$. Except this difference, our method of constructing decomposition 
of the cyclic representation is the same as that of \cite{Staruszkiewicz1992ERRATUM}. The lack of tensor product structure is compensated for
by the analyticity of the Fourier transform of the kernels, induced by the cyclic representations with the cyclic vectors (\ref{x}). 
The basic properties of decomposition may be read of from the orthogonality relations 
of the matrix elements of the unitary irreducible representations
and properties of the poles of the Fourier transforms of the positive definite functions
associated with the cyclic representations. The orthogonality and analyticity we use also 
to investigate the mutual inclusion relations of the cyclic representations.    
We obtain results going further than is asserted by the above theorem.
In particular, we obtain a complete description of the representation $U$, as well as 
the action of the operators $c_{\alpha_i}^+, S(u)$ on the subspaces invariant for $U$. For example, let $(l_0,l_1)$
be the unitary irreducible representations of \cite{Geland-Minlos-Shapiro}, then we show that
\begin{equation}\label{decomposition}
U|_{{}_{\mathcal{H}_{{}_{n}}}} =
\underset{l_0 \in \mathbb{Z}}{\bigoplus} \int\limits_{0}^{+\infty} (l_0,i\rho) \, \nu(\rho,z) d\rho \, \bigoplus \nu(z) \, (l_0=0, 1-z),
\,\,\,\,\, z = {\textstyle\frac{n^2\mathfrak{e}^2}{\pi}},
\end{equation}
where for each $z>0$ the weight $\nu(\rho,z)$ is almost everywhere $>0$ and with a positive weight $\nu(z)$ 
of the supplementary component $(l_0,l_1) = (0,1-z)$, nonzero  if and only if $0<z<1$. From (\ref{decomposition})
the theorem follows. Decomposition (\ref{decomposition}) can be also used to solve the problems concerning $U$ and raised
in \cite{Herdegen}.

In \cite{wawrzyckiKernel} we have given a proof that the invariant kernel 
\[
\langle u|v\rangle  = \langle 0| e^{inS(u)}e^{-inS(v)}|0\rangle
\]
on the Lobachevsky space is positive
definite, independently of the theory of the quantum Coulomb field \cite{Staruszkiewicz}. 
In the paper \cite{wawrzyckiKernel} we have written one sentence (repeated twice) 
that the proof presented there  ``gives us a proof of (relative) consistency 
of the theory \cite{Staruszkiewicz}''. This sentence was based 
on the above-mentioned false lemma. Therefore, in Subsection \ref{Consistency}, 
we are giving a constructive consistency proof of the axioms, finishing a proof 
initiated in the works \cite{Staruszkiewicz},\cite{Staruszkiewicz1992ERRATUM}, 
\cite{Staruszkiewicz1992}, \cite{Staruszkiewicz1995}, by adding one last step: 
we compute explicitly the vector valued function $B$ on $G$, which respect, together with 
the matrix valued function $A$ (both introduced in \cite{Staruszkiewicz1992}), 
the requirements formulated in \cite{Staruszkiewicz1992ERRATUM}, 
\cite{Staruszkiewicz1992}, \cite{Staruszkiewicz1995}. Another, more ganeral, consistency 
proof was given in \cite{Herdegen}, including a proof of existence, indecomposability and positivity of a one-parameter of states 
of the $C^*$-algebra $\mathcal{A}$ of \cite{Herdegen}, with unitarily implementable Lorentz automorphisms, among
them, the state giving the theory \cite{Staruszkiewicz}, which is the subject of our paper.

Section \ref{01} has preparatory character, and gives all ingredients, which are then used in the proof.
In Subsection \ref{Consistency} we define $A,B$.
In Subsection \ref{FormulasForA,B} we are giving explicit formulas for $A,B$. 
In Subsection \ref{explicit(l0,l1)}
we give explicit formulas for the matrix elements of all unitary irreducible representations $(l_0,l_1)$ of $G$, and 
point out the properties of these matrices, which we then use in the proof.
In Subsection \ref{DecompositionGeneralCyclic} we give general formulas for the Fourier transforms of the positive definite functions
canonically associated with the cyclic representations with cyclic vectors (\ref{x}), and construct their analytic continuations. We give there
decomposition of these cyclic representations. In Section \ref{RelationCycRep} we give the method for investigation of inclusion relation
of the cyclic subspaces. This method is then applied in Sections \ref{PolesAsymptotic} and \ref{CyclicDomains},
in which a complete desctription of  $U|_{{}_{\mathcal{H}_{{}_{n}}}}$ is given.

Summing up, in this paper we present a constructive consistency proof of the axioms of \cite{Staruszkiewicz}, then give
a proof of (\ref{decomposition}), and the theorem as a corollary of (\ref{decomposition}).

Contribution of the second-named author is the derivation of the series expansion (\ref{GeneralSeries}) 
of the Fourier transforms, valid for the whole class of the Fourier transforms of positive definite 
functions on $SL(2,\mathbb{C})$, which are defined by the cyclic representations with cyclic vectors (\ref{x}).

\section{Cyclic subrepresentations}\label{01}

In the sequel $\mathcal{H}$ be the Hilbert space of the operators $S,Q,c_\alpha, c_\alpha^+$, $\alpha = (l,m)$, $l=1,2, \ldots$,
$-l \leq m \leq l$, introduced in \cite{Staruszkiewicz}. Let for $n\in \mathbb{Z}, n\neq 0$, 
$\mathcal{H}_{n}$ be the eigenspace of the total charge $Q$ corresponding to 
the eigenvalue $n \mathfrak{e}$. We will use the notation of \cite{Staruszkiewicz1992ERRATUM}, \cite{Staruszkiewicz1992}, \cite{Staruszkiewicz1995},
\cite{Staruszkiewicz2009}.  Let $G = SL(2,\mathbb{C})$, and let $U$ be the unitary representation of $G$ acting in $\mathcal{H}$. 
We consider cyclic subrepresentation of $U$, with the invariant subspace $\mathcal{H}_x \subset \mathcal{H}_n$  
generated by a single cyclic $x\in \mathcal{H}_n$, \emph{i.e.}, 
with $\mathcal{H}_x$ spanned by Lorentz transforms $|g\rangle = U(g) x\in \mathcal{H}_n$, $g \in G$. Below we consider special $x$
of the form (\ref{x})
or $x = e^{-inS(u)}|0\rangle = |u\rangle$, with $\alpha_i = (l_i,m_i)$, $i,l=1,2, \ldots$, $-l_i \leq m_i \leq l_i$.
With this convention $x$ will sometimes be denoted by $x= U(e)x = |e\rangle$. 
In what follows $e$ stands for the basis of natural logarithms or the unit element in $G$, 
and $\mathfrak{e}^2$ in $z = n^2\mathfrak{e}^2/\pi$ stands for the square of the elementary charge, 
experimental value of which is approximately equal to $1/137$ in units in which $\hbar = c = 1$. 

To compute the decomposition of the cyclic subrepresentation with cyclic vector (\ref{x}) 
we consider the corresponding invariant subspace $\mathcal{H}_x$
as the reproducing kernel Hilbert space 
with the corresponding left invariant kernel 
\[
\langle g| h\rangle = \langle U(g)x | U(h)x\rangle,
\]
on $G$. Next we use the Fourier transform (Gelfand-Neumark Plancherel formula on $G$, \cite{NeumarkLorentzBook}) 
and compute the Fourier transform of the corresponding left-invariant kernel. 

Let $U^{{}^{(l_0,l_1)}}$ be the irreducible unitary representation $(l_0,l_1)$ of $G$, \cite{Geland-Minlos-Shapiro},  \cite{NeumarkLorentzBook},
with $l_0$ any half-integer or integer, and $l_1 = i\rho$, $\rho \in \mathbb{R}$ (principal series) 
or $l_0=0, l_1 \in (-1,1)$ (supplementary series).
The representations $(l_0,l_1)$, $(-l_0, -l_1)$ are not only equivalent, but the matrices of these representations, found in  \cite{Geland-Minlos-Shapiro},  \cite{NeumarkLorentzBook}, are identical. Recall that the representation
of the subgroup $SU(2, \mathbb{C})\subset G$ defined by restriction of $U^{{}^{(l_0,l_1)}}$ to $SU(2, \mathbb{C})$, is equal to the dierct sum of the standard irreducible representations with the weight $l=|l_0|, |l_0|+1, \ldots$, each entering with multiplicity one, and with $|l_0|$ being the lowest weight.
We have the following range of the matrix component indices $\alpha = (l,m)$: $l \in \{|l_0|, |l_0| + 1, \ldots \}$,
$\leq -l \leq m \leq l$ of the representations $(l_0,l_1)$.
Thus, the only irreducible unitary representations containing nonzero
$SU(2,\mathbb{C})$-invariant state are those with $l_0=0$, and are called spherical.
Representations of the principal and supplementary series
exhaust all equivalence classes of unitary irreducible representations of $G$, \cite{NeumarkLorentzBook}. 
Representations $(l_0,l_1=i\rho)$, $\rho \geq 0$, exhaust all equivalence classes of the principal series. Equivalently,
we will also use $(l_0,l_1=i\rho)$, with $l_0 > 0$,$\rho \in \mathbb{R}$ or with $l_0=0$, $\rho \geq 0$, 
as the representants of the principal series. We found the following
coordinate system $\left(\theta_{{}_{1}}, \varphi_{{}_{1}}, \vartheta_{{}_{1}}, \vartheta, \varphi, \lambda \right)$ on $G$ useful. Let $g\in G$. Let $g_{{}_{03}}(\lambda) \in G$ be the (matrix representing) hyperbolic rotation along the $03$ 
plane with hyperbolic angle $\lambda$. Let $g_{{}_{ik}}(\theta)$, for $i,k\in\{1,2,3\}$, be the (matrix representing) spatial rotation along the $ik$ plane 
with the rotation angle $\theta$. Then each $g\in G$ can be uniquely decomposed as follows
\begin{multline}\label{g}
g=g_{{}_{12}}(\theta_{{}_{1}})g_{{}_{13}}(\varphi_{{}_{1}})g_{{}_{12}}(\vartheta_{{}_{1}})g_{{}_{12}}(-\vartheta)
g_{{}_{13}}(-\varphi)g_{{}_{03}}(\lambda)g_{{}_{13}}(\varphi)g_{{}_{12}}(\vartheta) 
\\
= a_1(\theta_{{}_{1}},\varphi_{{}_{1}},\vartheta_{{}_{1}})^*a_2(\vartheta,\varphi)^*g_{{}_{03}}(\lambda)a_2(\vartheta,\varphi)
\end{multline}     
with the invariant measure on $G$
\begin{multline}\label{dg}
dg = {\textstyle\frac{1}{8\pi^2} \sin \varphi_{{}_{1}}}d\theta_{{}_{1}}d\varphi_{{}_{1}}d\vartheta_{{}_{1}} \pi^2 \textrm{sinh}^2 \lambda \sin \varphi
d\lambda d\vartheta d\varphi 
\\
= da_{{}_{1}} \pi^2 \textrm{sinh}^2 \lambda \sin  \varphi
d\lambda d\vartheta d\varphi
\end{multline}
normalized as in \cite{NeumarkLorentzBook}, where $da_{{}_{1}} $ is the invariant measure on $SU(2,\mathbb{C})$ normalized to unity.  
Here
\[
0\leq \varphi_{{}_{1}},\varphi \leq \pi, \,\,\, 0\leq \vartheta_{{}_{1}},\vartheta \leq 2\pi, \,\,\, 0 \leq \lambda < \infty.
\]

\subsection{Definition of $A$ and $B$. Consistency}\label{Consistency}

Before we continue, let us introduce the basic quantities $A,B$ used in the proof and 
give a constructive consistency proof of the axioms of \cite{Staruszkiewicz}. 
Consistency, for each $\mathfrak{e}^2\geq 0$, has in principle 
been almost completely shown already in the works \cite{Staruszkiewicz}, 
\cite{Staruszkiewicz1992}, \cite{Staruszkiewicz1995}. We finish the proof 
initiated there. As remarked in
\cite{Staruszkiewicz}, in a concrete Lorentz frame with the time-like unit versor $u = (1,0,0,0)$, 
the operators $S_0= S(u),Q,c_\alpha, c_\alpha^+$, can be constructed on the Hilbert space 
tensor product $\mathcal{H} = L^2(\mathbb{S}^1) \otimes \Gamma(H_{{}_{1,0}})$ 
of the Hilbert space $L^2(\mathbb{S}^1, d\phi)$
of square summable functions on the unit circle with the standard invariant measure $d\phi$,
with the bosonic Fock space $\Gamma(H_{{}_{1,0}})$ 
over the single particle Hilbert space $H_{{}_{1,0}}$ of infrared electric-type transversal
states, putting, respectively, $S_0= S'_0 \otimes \boldsymbol{1}$,$Q = Q'\otimes \boldsymbol{1}$, $c_\alpha  = \boldsymbol{1} \otimes c'_\alpha$, 
$c_\alpha^+ = \boldsymbol{1} \otimes {c'}_\alpha^+$. Here $H_{{}_{1,0}}$ is the Hilbert space of the unitary irreducible
representation $(l_0, l_1) = (1,0)$ of $SL(2,\mathbb{C})$. $Q'$ is the self-adjoint extension 
of the operator $i\mathfrak{e}\tfrac{d}{d\phi}$ on $L^2(\mathbb{S}^1)$.
$S'_0$ is the operator of multiplication by the (periodic) angle $\phi$ acting on $L^2(\mathbb{S}^1)$. $c'_\alpha, {c'}_\alpha^+$, 
are the standard annihilation-creation operators in the Fock space $\Gamma(H_{{}_{1,0}})$, of the single paricle states $\xi_{{}_{\alpha}}$, $\alpha
= (l,m)$, which are the normalized eigenstates of the operators $\Delta = M_{{}_{23}}^2 +  M_{{}_{31}}^2 +  M_{{}_{12}}^2$, $M_{{}_{12}}$: 
$\Delta \xi_{{}_{l,m}} = l(l+1) \xi_{{}_{l,m}}$, $ M_{{}_{12}}\xi_{{}_{l,m}} = m\xi_{{}_{l,m}}$, $l \in \{l_0 = 1,l_0+1 = 2, \ldots$, 
$-l \leq m \leq l$. $|0\rangle = v_{{}_{0}} \otimes \tfrac{1}{\sqrt{2\pi}} 1_{{}_{\mathbb{S}^1}}$,
where $v_{{}_{0}}$ is the vacuum in $\Gamma(H_{{}_{1,0}})$ and $\tfrac{1}{\sqrt{2\pi}} 1_{{}_{\mathbb{S}^1}} \in L^2(\mathbb{S}^1)$ 
is the normalized constant function equal $\tfrac{1}{\sqrt{2\pi}}$ for each angle in $\mathbb{S}^1$. For a rigorous 
construction of the Fock space with the creation-annihilation operators, compare e.g. \cite{obata-book}. 
In the construction of \cite{obata-book} we are using the single particle Gelfand triple $E \subset H_{{}_{1,0}} \subset E^*$, 
determined by the standard operator $\Delta$ on $H_{{}_{1,0}}$, and its Fock lifting 
$(E) \subset \Gamma(H_{{}_{1,0}}) \subset (E)^*$, determined by the standard operator
$\Gamma(\Delta)$ on $\Gamma(H_{{}_{1,0}})$. For details we refer to \cite{obata-book}. 
For the dense nuclear domain $(\mathcal{E})$ of the operators $Q,c_\alpha, c_\alpha^+$,
we are using the (unique) tensor product of the Hida test space $(E) \subset \Gamma(H_{{}_{1,0}})$, with the dense nuclear subspace 
$\mathscr{C}^\infty(\mathbb{S}^1) = \underset{n \in \mathbb{N}}{\cap} \textrm{Dom} \, Q'^n \subset L^2(\mathbb{S}^1)$. 
$S(u)$ being bounded has the whole Hilbert space $\mathcal{H}$ as its domain.
By construction $|0\rangle \in (\mathcal{E})$. The operators $e^{-iS(u)},Q,c_\alpha, c_\alpha^+$, transform continuously 
$(\mathcal{E})$ into itself with respect to the nuclear countably normed
topology, making $(\mathcal{E})$ a standard countably Hilbert nuclear space \cite{obata-book}. Thus, the existence of the operators  $S_0= S(u),Q,c_\alpha, c_\alpha^+$, fulfilling the commutation rules, with a cyclic 
vacuum in their Hilbert space, is clear. Let us mention that irreduciblity of the algebra generated by 
$S_0= S(u),Q,c_\alpha, c_\alpha^+$, is an almost immmediate corollary of the irreducibilty theorem, \emph{i.e.} corollary 4.7 of \cite{luo},
and the construction of $S_0= S(u),Q,c_\alpha, c_\alpha^+$ given here.

As shown in \cite{Staruszkiewicz1992}, any representation $U$ of $SL(2,\mathbb{C})$, giving transformation rule 
$S(u)'= US(u)U^{-1}, Q' = UQU^{-1} = Q, \ldots, {c'}^{+}_{\alpha}=Uc_{\alpha}U^{-1}$, of the operators 
$S(u),Q,c_\alpha, c_\alpha^+$, preserving the commutation rules 
\begin{align}
[Q,S(u)] = i \mathfrak{e}, \,\, \left[Q,c_\alpha\right] = \left[S(u),c_\alpha\right] =0, &
\left[c_\alpha, c_\beta^+ \right] = 4\pi \mathfrak{e}^2 \delta_{{}_{\alpha \,\,\, \beta}},
\label{CommutationRules}
\\
c_\alpha |0\rangle = \langle 0|c_\alpha^+ = Q|0\rangle = \langle 0| Q =0, &
\label{Vacuum}
\end{align}
of the theory \cite{Staruszkiewicz}, necessary has the general form 
\begin{equation}\label{c'S'}
c'_{\alpha} = 
\sum\limits_{\beta} c_{{}_{\beta}}A_{{}_{\beta \,\, \alpha}} + B_{{}_{\alpha}}Q, \,\,\,\,\,
S(u)' = S(u) -
{\textstyle\frac{1}{4\pi i\mathfrak{e}}}
\sum\limits_{\alpha,\beta}[c_{{}_{\alpha}} A_{{}_{\alpha \,\, \beta}} \overline{B_{{}_{\beta}}}
- c^+_{{}_{\alpha}} \overline{A_{{}_{\alpha \,\, \beta}}} B_{{}_{\beta}}],
\end{equation}
where  
$A$ is a unitary matrix-valued function, and $B$ vector-valued function on $G$ preserving the conditions
\begin{enumerate}
\item[(I)] 
$A(gh)=A(g)A(h)$, 
\item[(II)]
$B(gh) = B(g)A(h)+B(h)$, $g,h \in G$, 
\item[(III)] 
$B(g) = 0$, $g\in SU(2,\mathbb{C}) \subset G$, 
\end{enumerate}
with the products understood as the ordinary matrix products of a matrix $A$ with a vector $B$. Conditions (I)-(II) immediately
follow from the assumed representation (or homomorphism) property of $U$, for (III), compare \cite{Staruszkiewicz1995}.
(Here we are using the convention with right multiplication by $A$ in (II), 
in order to keep ordinary representation property of $A$. In the convention used in \cite{Staruszkiewicz1992}, 
$A$ is a representatation of the group opposite to $G$). 
The operators (\ref{c'S'}) are well-defined with dense domains $(\mathcal{E})$, 
and with $S(u)'$ essentially self-adjoint iff $||B(g|| < \infty$ for $g \in G$.
Next, we observe that preservation of the commutation rules by $U$ implies
preservation of the orthogonality of the complete system of vectors (\ref{x}) and of their norms, and thus implies unitarity of $U$.
It was shown in \cite{Staruszkiewicz1995} that $A=U^{{}^{(l_0=1,l_1=0)}}$ is the matrix of the irreducible unitary 
representation $(l_0=1, l_1=0)$ of \cite{Geland-Minlos-Shapiro}. 
It remains to determine $B$. Consistency will be proved if we construct explicitly  
$B$ on $G$ with $||B(g)|| < \infty$, $g\in G$, which together with $A$ preserves the conditions (I)-(III). 
The equation
\begin{equation}\label{infinitesimalB}
{\textstyle\frac{d}{d\lambda}}B_{{}_{l,m}}(\lambda=0) = \mathfrak{e} {\textstyle\sqrt{\frac{8}{3}}} \delta_{{}_{l \, 1}} \delta_{{}_{m \, 0}}
\end{equation}
for $B(\lambda)=B(g_{{}_{03}}(\lambda))$, was found in \cite{Staruszkiewicz1995}, 
with the initial conditions
\begin{equation}\label{B(0)}
B_{{}_{l,m}}(\lambda=0)=0, \,\,\, l=1,2, \ldots, \,\,\, -l \leq m \leq l,
\end{equation}
and the square 
\begin{equation}\label{||B||}
\|B(g)\|^2 = \sum_\alpha |B(g)_{{}_{\alpha}}|^2 = 8 \mathfrak{e}^2(\lambda \textrm{coth}\lambda - 1)
\end{equation}
of the norm of $B$, computed in \cite{Staruszkiewicz}, for hyperbolic rotation $g$ with hyperbolic angle $\lambda$. 
To continue our proof, and for the further part of the paper, we need the formulas ($l,l' \in\{ l_0 =1, l_0+1, \ldots\}$, $-l \leq m \leq l$,
$-l' \leq m' \leq l'$)
\begin{multline}\label{A(g)}
A_{{}_{lm \,\, l'm'}}\left(g_{{}_{03}}(\lambda)\right)= A_{{}_{lm \,\, l'm'}}(\lambda) 
= U_{{}_{lm \,\, l'm'}}^{{}^{(l_0=1,l_1=0)}}\left(g_{{}_{03}}(\lambda)\right) 
\\
= \delta_{{}_{m,m'}}
\sqrt{\textstyle{\frac{l(l+1)(2l+1)(l-m)!(2l'+1)(l'-m')!}{l'(l'+1)2(l+m)!2(l'+m')!}}}
\,
\int\limits_{-1}^{1} P_{{}_{l,m}}(y) P_{{}_{l',m'}}\big(\textstyle{\frac{\textrm{tanh}(\lambda)+y}{1+\textrm{tanh}(\lambda)y}}\big)
\, dy
\end{multline}
and 
\begin{equation}\label{SU2Cl0l1}
U^{{}^{(l_0,l_1)}}_{{}_{lm \,\, l'm'}}\left(a\right) = \delta_{{}_{l \, l'}} \, T^{{}^{l}}_{{}_{m \,\,m'}}(a) 
\,\,\, \textrm{for} \,\,\, a \in SU(2, \mathbb{C}).
\end{equation}
Here
\begin{multline*} 
T^{{}^{l}}_{{}_{m \,\, m'}}(a) = (-1)^{2l-m-m'}{\textstyle\sqrt{\frac{(l-m)!(l+m)!}{(l-m')!(l+m')!}}}
\sum\limits_{\alpha = \textrm{max}\{0,-m-m'\}}^{\textrm{min}\{l-m,l-m'\}} \Big[
\\
{\textstyle\binom{l-m'}{\alpha}\binom{l+m'}{l-m-\alpha}
(a_{11})^{\alpha}(a_{12})^{l-m-\alpha}(a_{21})^{l-m'-\alpha}(a_{22})^{m+m'+\alpha}} \Big],
\end{multline*} 
for each fixed $l$, are the standard matrices of irreducible representations of weight $l$ of $SU(2, \mathbb{C})$,
\cite{Geland-Minlos-Shapiro}, \cite{NeumarkLorentzBook}. The formula (\ref{A(g)}) uses the realization 
$\alpha \mapsto f(\Lambda(\alpha)^{-1}p)$ of the representation
$(l_0=1,l_1=0)$ in the space of classes modulo constant, homogeneous of degree zero functions $f$, 
on the cone $p \cdot p =0, p_0>0$, thus living effectively on the unit $\mathbb{S}^2$ sphere in the cone, 
with the standard homomorphism $\alpha \mapsto \Lambda(\alpha)$ into the Lorentz group acting in $\mathbb{R}^4$, with the orthonormal
basis given by the representants $\tfrac{1}{\sqrt{l(l+1)}}Y_{{}_{lm}}$, $l=1,2, \ldots$, $-l \leq m \leq l$, 
and with the Lorentz invariant inner product
\[
(f,g) = - \int\limits_{{}_{\mathbb{S}^2}} \overline{f} \Delta_{{}_{\mathbb{S}^2}} g \,\,\, d\mu_{{}_{\mathbb{S}^2}}
\]
with ordinary Laplace operator and $SU(2,\mathbb{C})$-invariant measure on $\mathbb{S}^2$. 
$A$ coincides with the matrix of $(l_0=1,l_1=0)$, which can be checked by comparing with the matrix
of  $(l_0=1,l_1=0)$, given in \cite{NeumarkLorentzBook}, \cite{Geland-Minlos-Shapiro}.
In the sequel, we are using the orthogonality properties of the matrix elements
$T^{{}^{l}}_{{}_{m \,\, m'}}$ regarded as functions on $SU(2,\mathbb{C})$ (Peter-Weyl theorem). 
Integration of (II) along the one parameter subgroup $g_{{}_{03}}(\lambda)$ of hyperbolic rotations, 
using (\ref{infinitesimalB}) and
the initial condition (\ref{B(0)}), gives $B$ along this subgroup. \emph{I.e.}, writing $B\big(g_{{}_{03}}(\lambda)\big) = B(\lambda)$:
\[
B_{{}_{l,m}}(\lambda'+\lambda) = \sum\limits_{l',m'}B_{{}_{l',m'}}(\lambda')A_{{}_{l',m' \,\,\, l,m}}(\lambda) + B_{{}_{l,m}}(\lambda).
\]
Differentiating $\partial_{\lambda'}$ at $\lambda'=0$ and using (\ref{infinitesimalB}) we obtain
\[
{\textstyle\frac{dB_{{}_{l,m}}(\lambda)}{d\lambda}} = i\mathfrak{e}{\textstyle\sqrt{\frac{8}{3}}} A_{{}_{1,0 \,\,\, l,m}}(\lambda),
\]
which, together with the initial conditions (\ref{B(0)}), determines $B(\lambda)$ uniquely.
Similarly, using (III), we have trivial 
zero for $B$ along any one parameter subgroup of unitary elements of $G$.
Using decomposition (\ref{g}), we immediately see that conditions (I)-(III) imply the value of $B$ at a general element (\ref{g}) to be equal
\begin{equation}\label{B(g)1}
B_{{}_{l,m}}(g) = 
B_{{}_{l,0}}(\lambda)A_{{}_{l,0 \,\,\, l,m}}\Big(g_{{}_{13}}(\varphi)g_{{}_{12}}(\vartheta) \Big) = \left(B(\lambda)A\left(a_2\right)\right)_{{}_{lm}},
\end{equation}
\begin{multline}\label{B(g)2}
B_{{}_{l,m}}\left(g^{-1}\right) = 
(-1)^l B_{{}_{l,0}}(\lambda) A_{{}_{l0 \,\,\, l,m}}\Big(g_{{}_{13}}(\varphi)g_{{}_{12}}(\vartheta)g_{{}_{12}}(-\vartheta_1)g_{{}_{13}}(-\varphi_1)g_{{}_{12}}(-\theta_1) \Big)
\\
 = \left(B(-\lambda)A\left(a_2a_{1}^{*}\right)\right)_{{}_{lm}},
\end{multline}
\begin{equation}
 B_{{}_{l,m}}\left(g_{{}_{03}}(\lambda)\right) = B_{{}_{l,m}}(\lambda) = (-1)^l B_{{}_{l,m}}(-\lambda) 
= i \mathfrak{e} \sqrt{\textstyle{\frac{8}{3}}} \, \int\limits_{0}^{\lambda} A_{{}_{1,0 \,\, l,m}}(\lambda') d\lambda'.
\label{B(g)} 
\end{equation}
Note that $B_{{}_{l,m}}(\lambda) = 0$ for $m\neq 0$, by the property 
\[
A_{{}_{l',0 \,\, l,m}}(\lambda) = \delta_{{}_{m \, 0}} \, A_{{}_{l',0 \,\, l,m}}(\lambda)
\]
of the matrix $A(\lambda)$. 
Here $P_{{}_{l,m}}$ are the associated Legendre ``polynomials''
\[
P_{{}_{l,m}}(y)= (-1)^m 2^l (1-y^2)^{m/2} 
\displaystyle\sum_{k=m}^{l} \textstyle{\frac{k!}{(k-m)!}}\binom{l}{k}\binom{\frac{l+k-1}{2}}{l}y^{k-m}
\] 
with the generalized binomial symbol
\[
{\textstyle\binom{w}{k} = \frac{w(w-1)(w-2)\ldots (w-k+1)}{k(k-1)\ldots 1}}, \,\,\,\, w\in \mathbb{R}, k\in \mathbb{N}.
\]
To finish the proof we have to show that $A,B$ given by the formulas (\ref{A(g)}), (\ref{B(g)1}), (\ref{B(g)}), 
respect conditions (I)-(III). (I) and (III) are trivially fulfilled. In order to show (II) we note the parity property 
$A_{{}_{l,0 \,\,\, l',0}}(-\lambda) = (-1)^{l+l'}A_{{}_{l,0 \,\,\, l',0}}(\lambda)$
of the matrix $A(\lambda)$ for the hyperbolic rotations parallel to the $03$ hyperplane. This parity property 
can be easily seen by the exponentiation of the generator $M_{{}_{03}}$ of the representation
$(l_0=1,l_1=0)$ given explicitly in \cite{Geland-Minlos-Shapiro},  \cite{NeumarkLorentzBook}, and which immediately gives the 
series expansion at $\lambda=0$:
\begin{equation}\label{Aexpansion}
A_{{}_{l',0 \,\,\, l,0}}(\lambda) = 
a_{{}_{l', \, l, \, 0}} \lambda^{|l-l'|}+a_{{}_{l', \, l, \, 2}} \lambda^{|l-l'|+2} 
+a_{{}_{l', \, l, \, 4}} \lambda^{|l-l'|+4}
+ \ldots,
\end{equation}
\begin{equation}\label{Aexpansion'}
\\
a_{{}_{l', \, l, \, 0}} = {\textstyle\frac{(-1)^{l-l'}}{(l-l')!}}
\prod\limits_{j=0}^{l-l'-1}\Bigg[(l-j) \sqrt{{\textstyle\frac{(l-j)^2-1}{4(l-j)^2-1}}}\Bigg],
\,\,\, l'<l,
\end{equation}
\[
a_{{}_{l, \, l', \, r}} = (-1)^{l+l'}a_{{}_{l, \, l', \, r}}.  
\]   
The explicit value of the coefficients (\ref{Aexpansion'}), 
we will use only in the further part of the paper. 
Using the parity property of $A(\lambda)$, and representation property of $A$, 
we can easily see that for any $g\in G$ of the form (\ref{g})
and for $B(g)$ given by (\ref{B(g)1}) and (\ref{B(g)})
\begin{equation}\label{*}
B(g)A\left(g^{-1}\right) = - B(-\lambda)A\left(a_2a_{1}^{*}\right) = - B(g^{-1}).
\end{equation}
From (\ref{*}) it follows that the condition (II) is fulfilled with $h$ of the form $g^{-1}k$, for
any $g,k \in G$, and for $B(g)$ defined by (\ref{B(g)1}), (\ref{B(g)}), 
because the right-hand side of (II) for $h=g^{-1}k$ is equal
\[
B(g) A\left(g^{-1}k\right) + B\left(g^{-1}\right)A(k) + B(k) = B(k) 
\]
by (\ref{*}), and thus equal to the left-hand-side of (II). Putting $k=gh$, we get (II) for all $g,h \in G$, 
which proves consistency of the theory. For another consistency proof compare \cite{Herdegen}, \cite{Herdegen2005}.

From the general formulas (\ref{A(g)}), (\ref{B(g)}), it follows the formula (\ref{||B||}) and
\begin{equation}\label{<u,ugh-1>}
\langle u|u'\rangle = \langle u| gu\rangle  = \langle 0| e^{inS(u)}e^{-inS(gu)}|0\rangle = e^{-z(\lambda \textrm{coth} \lambda -1)},
\end{equation}    
for general $g$ of the form (\ref{g}). 
 Of course, consistency of the theory, implying unitarity of $U$,
implies also positive definiteness of the kernel $\langle g|h\rangle$ associated to the cyclic representation
associated with any cyclic vector, in particular with the cyclic vector of the form (\ref{x}).

\subsection{Explicit formulas for $A,B$}\label{FormulasForA,B}

For further purposes, we give explicit forms of $B_{{}_{l,0}}(\lambda)$ and $A_{{}_{l,m \,\, l',m}}(\lambda)$. 
In what follows we are using the convention that the sum $\Sigma \ldots$ is zero whenever the upper summation limit is less than the lower
summation limit, and analogously for the product $\Pi (\ldots)$, which by definition is equal $1$, whenever the upper product limit is less
than the lower product limit.
The integral (\ref{A(g)}) representing
$A_{{}_{l,0 \,\,\, 1,0}}(\lambda)$ can be explicitly computed in terms of the functions $Q_{{}_{l}}(-1/t)$, $t=\textrm{tanh}\lambda$, 
closely related to the Legendre functions of the second kind.
Namely, we use the identity
\begin{equation}\label{Al,01,0}
\int\limits_{-1}^{1}{\textstyle\frac{1}{1+tx}}P_{l}(x) \, dx = -{\textstyle\frac{2}{t}}Q_l\big(-{\textstyle\frac{1}{t}}\big), \,\,\,\,\,\,
-1 < t < 1,
\end{equation} 
where
\[
Q_l(x) = P_l(x){\textstyle\frac{1}{2}} \textrm{log}{\textstyle\frac{x+1}{x-1}}
- \sum\limits_{k=1}^{l} {\textstyle\frac{1}{k}}P_{k-1}(x)P_{l-k}(x)
= P_l(x){\textstyle\frac{1}{2}} \textrm{log}{\textstyle\frac{x+1}{x-1}}- W_{k-1}(x),
\]
which can be proved using Bonnet's recursion formula for the Legendre
polynomials $P_l = P_{{}_{l,0}}$. $Q_l$ are closely related to the Legendre functions 
of the second kind\footnote{We should note that there are 
two versions of the Legendre functions of the second kind which are in use. In fact our $Q_{{}_{l}}$ constitute one of these versions,
and are called \emph{Legendre functions of degree $l$ of the second kind}, e.g., in \cite{WhittakerWatson}, p. 316, 
and are also named Legendre functions of the second kind in \cite{ET}, cited below by us, and denoted there by $Q_{{}_{l}}$. 
Unfortunately, these versions are sometimes mixed, e.g. the formula 8.825 of \cite{GradshteynRyzhik} is in agreement
with our $Q_{{}_{l}}$, and with that used in \cite{ET}, \cite{WhittakerWatson}, but the formula 8.827 of \cite{GradshteynRyzhik}
gives the other version of the Legendre functions of the second kind, with the other form of the argument in $\textrm{log}$.}  
which can be obtained from
$Q_l$ by replacing the argument of $\textrm{log}$ with $(x+1)/(1-x)$ in $Q_l$, \cite{GradshteynRyzhik}, 8.831. 
From Bonnet's recursion formula 
\[
(2l+1)xP_l(x) -(l+1)P_{l+1}(x) - lP_{l-1}(x) = 0  
\] 
for $P_l$ it follows that $Q_l$ and $W_{l-1}$ respect the same Bonnet's recurrence \cite{WhittakerWatson}
\begin{align*}
(2l+1)xQ_l(x) -(l+1)Q_{l+1}(x) - lQ_{l-1}(x) = 0, 
\\
(2l+1)xW_{l-1}(x) -(l+1)W_{l}(x) - lW_{l-2}(x) = 0.
\end{align*}

Let us compute $B_{{}_{l,0}}(\lambda)$ explicitly.
Applying (\ref{Al,01,0}) and Bonnet's recursion for $P_{l}$ to (\ref{A(g)}), 
and then Bonnet's recursion for $Q_{l}$, we get
\begin{multline}\label{Al0,10}
A_{{}_{l,0 \,\,\, 1,0}}(\lambda) 
\\
= c[l] 2 \left[-Q_{{}_{l}}\big(-{\textstyle\frac{1}{t}}\big) 
+ {\textstyle\frac{(l+1)}{2l+1}} \left(-{\textstyle\frac{1}{t}}\right)  Q_{{}_{l+1}}\big(-{\textstyle\frac{1}{t}}\big) 
+ {\textstyle\frac{l}{2l+1}} \left(-{\textstyle\frac{1}{t}}\right)  Q_{{}_{l-1}}\big(-{\textstyle\frac{1}{t}}\big)\right]
\\
= -2c[l]\big(1-{\textstyle\frac{1}{t^2}}\big)Q_{{}_{l}}\big(-{\textstyle\frac{1}{t}}\big) 
= -(-1)^{l+1}2c[l]\big(1-{\textstyle\frac{1}{t^2}}\big)Q_{{}_{l}}({\textstyle\frac{1}{t}}), 
\end{multline}  
where $c[l] = \sqrt{\tfrac{3}{8}l(l+1)(2l+1)}$. Performing the integration (\ref{B(g)}) with  $A_{{}_{1,0 \,\,\, l,0}}(\lambda)$ 
$= (-1)^{l+1}A_{{}_{l,0 \,\,\, 1,0}}(\lambda)$, we obtain
\begin{multline}\label{BuptoConstant}
B_{{}_{l,0}}(\lambda) = i \mathfrak{e} \sqrt{{\textstyle\frac{8}{3}}} (-1)^{l+1} \int\limits_{0}^\lambda 
A_{{}_{l,0 \,\,\, 1,0}}(\lambda') \, d\lambda' = I(\lambda) + b[l],
\\
I(\lambda) = i \mathfrak{e} \sqrt{{\textstyle\frac{8}{3}}} (-1)^{l+1} \int
A_{{}_{l,0 \,\,\, 1,0}}(\lambda) \, d\lambda.
\end{multline}
Comparing (\ref{Al0,10}) and (\ref{BuptoConstant})  we see that computation of the  indefinite integral $I(\lambda)$
is reduced to the compuation of the indefinite integrals $I[k,\lambda] = \int t^{-k}d\lambda$ 
and $J[k,\lambda] = \int t^{-k}\lambda d\lambda$.
Then we compute the limit of $B$ at infinity, which determines the integration constant $b[l]$.  
We use the variable $u= e^{-2\lambda}$ in all the indefinite integrals. The last two indefinite integrals can be rewritten as
$I[k,\lambda] = -\tfrac{1}{2}\mathbb{F}_{{}_{k}}(u)$, $J[k,\lambda] = \tfrac{1}{4} \mathbb{G}_{{}_{k}}(u)$, where $u = e^{-2\lambda}$ and
\begin{equation*}
\mathbb{F}_{{}_{k}}(u)= \int {\textstyle\frac{(1+u)^k}{(1-u)^ku}}du,
\,\,\,
\mathbb{G}_{{}_{k}}(u) 
=  \int {\textstyle\frac{(1+u)^k}{(1-u)^ku}} \log u du.
\end{equation*}
Using the following identities
\[
{\textstyle\frac{u^j}{(1-u)^k}} = \sum\limits_{s=0}^{j} {\textstyle\frac{\binom{j}{s}(-1)^s}{(1-u)^{k-s}}}, \,\, j < k,
\,\,\,\,\,\,\,\,\,\,
{\textstyle\frac{1}{(1-u)^ku}} = \sum\limits_{s=1}^{k} {\textstyle\frac{1}{(1-u)^{s}}} + {\textstyle\frac{1}{u}},
\]
and integration by parts in case $\mathbb{G}_{{}_{k}}(u)$, we compute the indefinite integrals
\begin{multline*}
\mathbb{F}_{{}_{k}}(u) = \int {\textstyle\frac{(1+u)^k}{(1-u)^ku}}du =
\sum\limits_{s=2}^{k} {\textstyle\frac{1}{(s-1)}}{\textstyle\frac{1}{(1-u)^{s-1}}}
+ \sum\limits_{j=1}^{k-1} \sum\limits_{s=0}^{j-1} {\textstyle\binom{k}{j}\binom{j-1}{s}\frac{(-1)^s}{(k-s-1)}\frac{1}{(1-u)^{k-s-1}}}
\\
+ \sum\limits_{s=0}^{k-2} {\textstyle\binom{k-1}{s}\frac{(-1)^s}{(k-s-1)}\frac{1}{(1-u)^{k-s-1}}}
+ \big((-1) - (-1)^{k-1} \big) \log(1-u) + \log u
\\
= F[k,u] + \big((-1) - (-1)^{k-1} \big) \log(1-u) + \log u,
\end{multline*}
\begin{multline*}
\mathbb{G}_{{}_{k}}(u) = \int {\textstyle\frac{(1+u)^k}{(1-u)^ku}}\log u du =
F[k,u] \log u
\\
-\sum\limits_{s=2}^{k} \sum\limits_{r=2}^{s-1} {\textstyle\frac{1}{(s-1)(r-1)}\frac{1}{(1-u)^{r-1}}}
-\sum\limits_{j=1}^{k-1} \sum\limits_{s=0}^{j-1} {\textstyle\binom{k}{j}\binom{j-1}{s}\frac{(-1)^s}{(k-s-1)}}
\sum\limits_{r=2}^{k-s-1}{\textstyle\frac{1}{(r-1)}\frac{1}{(1-u)^{r-1}}}
\\
- \sum\limits_{s=0}^{k-2} {\textstyle\binom{k-1}{s}\frac{(-1)^s}{(k-s-1)}}
\sum\limits_{r=2}^{k-s-1}
{\textstyle\frac{1}{(r-1)}}{\textstyle\frac{1}{(1-u)^{r-1}}} 
\end{multline*}
\begin{multline*}
+ \big((-1) - (-1)^{k-1} \big) \log(1-u)\log u + {\textstyle\frac{1}{2}}\log^2 u 
\\
- F[k]\big( \log(1-u) - \log u \big)
+ \big((-1) - (-1)^{k-1} \big) \textrm{Li}_{{}_{2}} u 
\end{multline*}
\begin{multline*}
 = G[k,u] + \big((-1) - (-1)^{k-1} \big) \log(1-u)\log u + {\textstyle\frac{1}{2}}\log^2 u 
\\
- F[k]\big( \log(1-u) - \log u \big)
+ \big((-1) - (-1)^{k-1} \big) \textrm{Li}_{{}_{2}} u,
\end{multline*}
where $F[k] = - F[k,u=0]$.

Let us introduce the coefficients $p_{{}_{l,k}}$ and $w_{{}_{l-1,k}}$ of the polynomials
\[
P_l(x) = P_{{}_{l,0}}(x) = \sum\limits_{k=0}^{l} p_{{}_{l,k}} x^k,
\,\,\,
W_{l-1}(x)
= \sum\limits_{k=0}^{l-1} w_{{}_{l-1,k}} x^k.
\] 

We have the following identity
\begin{equation}\label{pF}
\sum\limits_{k=0}^{l} p_{{}_{l,k}} \left(F[k] - F[k+2] \right)
= 0.
\end{equation}
To prove (\ref{pF}) we notice that $F[k]$, defined as above, can be simplified
to the following form
\begin{equation}\label{simpleF[k]}
F[k] = -2(1+(-1)^k) \sum\limits_{j=1}^{\lfloor k/2 \rfloor} {\textstyle\frac{1}{2j-1}}.
\end{equation}
To prove (\ref{simpleF[k]}) we apply to $F[k]$ the following standard identities. First, the standard identity
\[
\sum\limits_{s=1}^{k-1}{\textstyle\binom{k-1}{s}\frac{(-1)^s}{s}} = - \sum\limits_{s=1}^{k-1}{\textstyle\frac{1}{s}}
\]
coming from the Stern series for the digamma function and the relation of digamma to the harmonic numbers, we apply
to the last sum in $F[k]$, converting it into the following form 
\[
\sum\limits_{s=0}^{k-2} {\textstyle\binom{k-1}{s}\frac{(-1)^s}{(k-s-1)}}
= (-1)^k \sum\limits_{s=1}^{k-1}{\textstyle\frac{1}{s}}.
\]

Next, the standard identity
\[
\sum\limits_{s=0}^{j-1}{\textstyle\binom{j-1}{s}\frac{(-1)^s}{k-s-1}} = {\textstyle\frac{(-1)^{j+1}}{j\binom{k-1}{j}}},
\]
coming from the fact that the l.h.s is equal to value $-\Beta(j,-k+1)$ of the Euler beta function $\Beta$, 
we apply  to the first summation of the double summation contribution to $F[k]$, converting it into
the following form
\[
\sum\limits_{j=1}^{k-1}\sum\limits_{s=0}^{j-1}{\textstyle\binom{k}{j}\binom{j-1}{s}\frac{(-1)^s}{k-s-1}}
= \sum\limits_{j=1}^{k-1}{\textstyle\frac{k(-1)^{j+1}}{(k-j)j}} = - (1+(-1)^k)\sum\limits_{j=1}^{k-1}{\textstyle\frac{(-1)^j}{j}}.
\]

The identity (\ref{pF}) and the fact that the poynomials $(1-x^2)P_{{}_{l}}(x), (1-x^2)W_{{}_{l-1}}(x)$ are zero
at $x = \pm 1$, imply that the total 
contribution to the linear combination $I(\lambda)$
of the indefinite integrals, $I[k,\lambda] = \mathbb{F}_{_{k}}(u)$ and
$J[k,\lambda] = \mathbb{G}_{_{k}}(u)$, 
comes only from the parts $F[k,u]$ and $G[k,u]$ of the indefinite itegrals  
$\mathbb{F}_{_{k}}(u)$ and $\mathbb{G}_{_{k}}(u)$, and the remaining terms
of the indefinitne intergrals $\mathbb{F}_{_{k}}(u), \mathbb{G}_{{}_{k}}(u)$, 
not included in $F[k,u]$ or, respectively, in $G[k,u]$, drop out. Thus, in our computation
we may replace the indefinite integrals  $\mathbb{F}_{_{k}}(u), \mathbb{G}_{_{k}}(u)$,
with $F[k,u],G[k,u]$. We thus obtain $I(\lambda)$ in terms of $F[k,u],G[k,u]$
and the coefficients of the polynomials $P_l, W_{l-1}$ entering $Q_l$ in (\ref{Al0,10}):
\begin{multline}\label{B[l,lambda]uptoConst}
B_{{}_{l,0}}(\lambda) =
-{\textstyle\frac{i\mathfrak{e}\sqrt{l(l+1)(2l+1)}}{2}} 
\Bigg[
\sum\limits_{k=0}^{l} p_{{}_{l,k}}\big( G[k,u] - G[k+2,u] \big)
\\
+2\sum\limits_{k=0}^{l-1} w_{{}_{l-1,k}} \big( F[k,u] - F[k+2,u] \big)
+ b'[l]
\Bigg],
\,\, u = e^{-2\lambda}.
\end{multline}

To fix $b'[l]$ we compute the limit
\[
\underset{\lambda \rightarrow +\infty}{\textrm{lim}} B_{{}_{l,0}}(\lambda)
= i \mathfrak{e} \sqrt{{\textstyle\frac{8}{3}}} (-1)^{l+1} \int\limits_{0}^\infty 
A_{{}_{l,0 \,\,\, 1,0}}(\lambda) \, d\lambda.
\]
Inserting the last form of the integrand (\ref{Al0,10}), and using the new variable $x$
\[
\lambda = -{\textstyle\frac{1}{2}}\log{\textstyle\frac{x+1}{x-1}},
\]
we get
\[
\underset{\lambda \rightarrow +\infty}{\textrm{lim}} B_{{}_{l,0}}(\lambda)
= i \mathfrak{e} 2 c[l] \sqrt{{\textstyle\frac{8}{3}}} (-1)^{l+1} \int\limits_{-\infty}^{-1} 
Q_{{}_{l}}(x) \, dx =
i \mathfrak{e} 2 \sqrt{{\textstyle\frac{2l+1}{l(l+1)}}},
\]
where we have used the identity
\[
\int\limits_{-\infty}^{-1} 
Q_{{}_{l}}(x) \, dx = {\textstyle\frac{(-1)^{l+1}}{l(l+1)}},
\]
immediately following from the formula (18), p. 324, of \cite{ET}, and the parity
property $Q_{{}_{l}}(-x) = (-1)^{l+1}Q_{{}_{l}}(x)$. 

In our proof we use the fact that the limit of $B(\lambda)$ at $\lambda = +\infty$ is nonzero, 
but we do not need the explict form of the constant $b[l]$, or $b'[l]$. But, concerning $b'[l]$, let us note the following. 
We have the following identities 
\begin{equation}\label{wF}
\sum\limits_{k=0}^{l-1} w_{{}_{l,k}} \left(F[k] - F[k+2] \right)
={\textstyle\frac{-4((-1)^l-1)}{l(l+1)}}
\end{equation}
and 
\begin{equation}\label{pG}
\sum\limits_{k=0}^{l} p_{{}_{l,k}} \left(\mathcal{G}[k] - \mathcal{G}[k+2] \right)
={\textstyle\frac{-4((-1)^l-1)}{l(l+1)}},
\end{equation}
with $\mathcal{G}[k] = \mathcal{G}[k,u=0]$, where $\mathcal{G}[k,u] = G[k,u] - F[k,u] \log u$
is that part of $G[k,u]$ which does not contain the $\log u$-factor.  
From (\ref{pF}) it follows at once that the function 
\[
\sum\limits_{k=0}^{l} p_{{}_{l,k}} \left(F[k,u] - F[k+2,u] \right)
\]
of $u$ is of order greater than zero at $u=0$ in the variable $u$.
From this, and by the L'H\^opital rule, the limit $u\rightarrow 0$ of 
\[
\sum\limits_{k=0}^{l} p_{{}_{l,k}} \left(F[k,u] - F[k+2,u] \right) \log u
\]
is zero and, thus, the 
the limit $u\rightarrow 0$ of the first sum in (\ref{B[l,lambda]uptoConst}) is equal to the l.h.s of 
(\ref{pG}). It is immediately seen that the limit $u\rightarrow 0$ of the second sum 
in (\ref{B[l,lambda]uptoConst}) is equal to the l.h.s of (\ref{wF}). 
Comparing with the limit of $B(\lambda)$ at $\lambda = + \infty$, 
obtained above, we finally get
\begin{multline}\label{B[l,lambda]}
B_{{}_{l,0}}(\lambda) =
-{\textstyle\frac{i\mathfrak{e}\sqrt{l(l+1)(2l+1)}}{2}} 
\Bigg[
\sum\limits_{k=0}^{l} p_{{}_{l,k}}\big( G[k,u] - G[k+2,u] \big)
\\
+2\sum\limits_{k=0}^{l-1} w_{{}_{l-1,k}} \big( F[k,u] - F[k+2,u] \big)
-{\textstyle\frac{4(-1)^l}{l(l+1)}}
\Bigg],
\,\, u = e^{-2\lambda}.
\end{multline}

The formula (\ref{B[l,lambda]}) can be rewritten in the following form
\begin{align}
B_{{}_{2k+1,0}}(\lambda) = {\textstyle\frac{1}{\textrm{sinh}^{2k+2}\lambda}}
\left[
\lambda \sum\limits_{j=0}^{k}\mathfrak{b}_{j} \textrm{cosh}(2j\lambda)
+
\sum\limits_{j=1}^{k+1}\mathfrak{b}_{k+j}\textrm{sinh}(2j\lambda)
\right],
\label{oddB}
\\
B_{{}_{2k,0}}(\lambda) = {\textstyle\frac{1}{\textrm{sinh}^{2k+1}\lambda}}
\left[
\lambda \sum\limits_{j=0}^{k-1}\mathfrak{b'}_{j} \textrm{cosh}^{2j+1}\lambda
+
\sum\limits_{j=0}^{k}\mathfrak{b'}_{k+j}\textrm{sinh}^{2j+1}\lambda
\right],
\label{evenB}
\end{align}
with nonzero constants $(\mathfrak{b}_{0}, \ldots, \mathfrak{b}_{2k+1})$ and $(\mathfrak{b'}_{0}, \ldots, \mathfrak{b'}_{2k})$. 
We have the expansion
\begin{align}
{\textstyle\frac{n}{4\pi\mathfrak{e}}}B_{{}_{l,0}}(-\lambda) = b_{{}_{l,0}}\lambda^l+  b_{{}_{l,2}}\lambda^{l+2} + b_{{}_{l,4}}\lambda^{l+4} + \cdots,
\label{Bexpansion}
\\
b_{{}_{l,0}}= i{\textstyle\frac{n}{4\pi}}\sqrt{{\textstyle\frac{8}{3}}}{\textstyle\frac{(-1)^{l}}{l}}a_{{}_{1, \,\,\, l, \,\, 0}},
\label{bl}
\end{align}
following from (\ref{B(g)}), (\ref{Aexpansion}), (\ref{Aexpansion'}).
Thus, $B_{{}_{l,0}}(\lambda)$  is a function of order $l$ in $\lambda$ at $\lambda=0$
and tends to a nonzero constant $2^l\mathfrak{b}_{l} =2^{2k+1}\mathfrak{b}_{2k+1}$ or, respectively, $\mathfrak{b'}_{l} = \mathfrak{b'}_{2k}$, at infinity:
\begin{equation}\label{BatInfinity}
\underset{\lambda \rightarrow +\infty}{\textrm{lim}} B_{{}_{l,0}}(\lambda) = 
i\mathfrak{e} 2 
{\textstyle\sqrt{\frac{2l+1}{l(l+1)}}} =
 \begin{cases}
  2^l \mathfrak{b}_{{}_{l}},  & \text{if $l$ is odd} \\
  \mathfrak{b'}_{{}_{l}}, & \text{if $l$ is even}
\end{cases}.
\end{equation}

Application of the identities
\begin{equation}\label{helpA}
P_{{}_{l,m}}(y) = (-1)^{m/2}(1-y^2){\textstyle\frac{d^mP_{{}_{l}}(y)}{dy^m}},
\,\,
y^n  = 
\sum\limits_{k=0}^{\lfloor n/2 \rfloor}
{\textstyle\frac{(2n-4k+1)n!}{2^kk!(2n-2k+1)!!}}
P_{{}_{n-2k}}(y),
\end{equation}
to the integrand of (\ref{A(g)}), and the formula (\ref{Al,01,0}), gives
\begin{multline}\label{explicitA}
A_{{}_{l,m \,\, l',m}}(\lambda) 
=  
{\textstyle\frac{1}{\textrm{sinh}^{l+l'+1} \lambda}}
\Bigg[\sum\limits_{\textrm{odd}\, j=0}^{l+l'-||m|-1|}\mathfrak{a}_{{}_{j}}\textrm{sinh}^{{}^{j}} \lambda\textrm{cosh}^{{}^{l+l'-||m|-1|-j}} \lambda
\\
+\lambda \sum\limits_{\textrm{even}\, j=0}^{l+l'-|m|-1}\mathfrak{a}_{{}_{j}}\textrm{sinh}^{{}^{j}} \lambda
\textrm{cosh}^{{}^{l+l'-||m|-1| -j}} \lambda
\Bigg],
\end{multline}
if $l+l'-||m|-1|$ is odd, and the upper limits in the sums are exchanged with each other, if  $l+l'-||m|-1|$ is even. 
Thus, we have the asymptotic $A_{{}_{l,m \,\, l',m}}(\lambda) \sim e^{-(||m|-1|+1)\lambda}$  at infinity. The 
constants $\mathfrak{a}_{{}_{0}}, \ldots, \mathfrak{a}_{{}_{l+l'-||m|-1|}}$ are nonzero.  
From (\ref{A(g)}) it follows that the matrix elements of $A(\lambda)$ are real, 
and $B(\lambda)$ pure imaginary.

In what follows, regarding the constants $\mathfrak{a}_j$,  we use only the fact
that the constant $\mathfrak{a}_0$ is nonzero. To see it, let $m\geq 0$ in (\ref{A(g)}).  
Note that performing the integration (\ref{A(g)}) 
we get $(1-t^2)^{m/2}[W(1/t) \lambda + R(1/t)]$, $t = \tanh \lambda$, with some polynomials $W,R$. 
The constant $\mathfrak{a}_0$ is the highest degree coefficient of the polynomial $W$, which is
of degeree $l+l'+1$. To show $\mathfrak{a}_0 \neq 0$, it is sufficient to compute only the 
contribution proportional to $(1-t^2)^{m/2}\lambda/t^{l+l'+1}$, and show that it is nonzero.
Denoting the constant in front of the integral (\ref{A(g)}) by $c[l,l',m]$,
and using the first formula of (\ref{helpA}) we see that (\ref{A(g)}) can be rewritten as
\begin{multline}\label{A(g)'}
A_{{}_{l,m \,\, l',m}}(\lambda) = (1-t^2)^{m/2} c[l,l',m] \int\limits_{-1}^{1} \left({\textstyle\frac{1-y^2}{1+ty}}\right)^m 
P^{{}^{(m)}}_{{}_{l'}}\left({\textstyle\frac{y+t}{1+ty}}\right) P^{{}^{(m)}}_{{}_{l}}(y) dy
\\
=
(1-t^2)^{m/2} c[l,l',m] \int\limits_{-1}^{1} {\textstyle\frac{P(y)}{(1+ty)^{l'}}} dy
\end{multline}
where $P(y) = p_{{}_{l + l'}}y^{l+l'} + p_{{}_{l+l'-1}}y^{l+l-1} + \ldots$
is a polynomial of degree $l+l'$, with the coefficients $p_{{}_{k}}$ being polynomials in $t$,
and with the highest degree coefficient equal
\[
p_{{}_{l+l'}} = (-1)^m p_{{}_{l,l-m}}^{{}^{(m)}} \sum\limits_{j=0}^{l'-m} p_{{}_{l',j}}^{{}^{(m)}} t^{l'-m-j}, 
\] 
where $p_{{}_{l',j}}^{{}^{(m)}}$ are the coefficients of $P^{{}^{(m)}}_{{}_{l'}}$:
\[
P^{{}^{(m)}}_{{}_{l'}}(t) = \sum\limits_{j=0}^{l'-m} p_{{}_{l',j}}^{{}^{(m)}} t^j.
\]
$P$ can be represented as a linear combination of the Legendre polynomials, 
\begin{equation}\label{polP} 
P(y) =  {\textstyle\frac{p_{{}_{l+l'}}}{p_{{}_{l+l',l+l'}}}} P_{{}_{l+l'}}(y) + \ldots,
\end{equation}
where dots denote linear combination of the Legendre polynomials of degree $< l+l'$. 
Repeated differentiation of (\ref{Al,01,0}) with respect to $t$ gives
\begin{equation}\label{Al,01,0'}
\int\limits_{-1}^{1}{\textstyle\frac{1}{(1+tx)^k}}P_{l}(x) \, dx 
= {\textstyle\frac{1}{(k-1)!}}{\textstyle\frac{1}{t^k}} 
\left[t^2 {\textstyle\frac{d}{dt}} \right]^{k-1} \left( -2Q_l\big(-{\textstyle\frac{1}{t}}\big)\right).
\end{equation}
The operator $t^2 {\textstyle\frac{d}{dt}}$ acting on a polynomial function of the variable $-1/t$
of degree $d$, gives again a polynomial of $-1/t$ of degree $d-1$. Thus, applying (\ref{Al,01,0'}) 
to (\ref{A(g)'}), we see that in order to obtain the contribution proportional to $(1-t^2)^{m/2}\lambda/t^{l+l'+1}$
with the highest possible degree $(1/t)^{l+l'+1}$, we can discard not only the contributions to $P$  
denoted by dots in (\ref{polP}), but also all constributions to the polynomial $p_{{}_{l+l'}}$ of $t$
which are of positive order in $t$,  and compute only 
\[
A_{{}_{l,m \,\, l',m}}(\lambda) 
=
(1-t^2)^{m/2} c[l,l',m] \int\limits_{-1}^{1} 
{\textstyle\frac{(-1)^m p_{{}_{l,l-m}}^{{}^{(m)}}p_{{}_{l',l'-m}}^{{}^{(m)}}P_{{}_{l+l'}}(y) + \ldots}{(1+ty)^{l'} p_{{}_{l+l',l+l'}}}} dy,
\]
discarding remaning terms, denoted by dots. Applying (\ref{Al,01,0'}) we obtain
\begin{multline*}
A_{{}_{l,m \,\, l',m}}(\lambda) = 
{\textstyle\frac{2c[l,l',m] (-1)^{l+m+1} p_{{}_{l,l-m}}^{{}^{(m)}}p_{{}_{l',l'-m}}^{{}^{(m)}} \, p_{{}_{l+l',l+1}}^{{}^{(l'-1)}}}{(l'-1)! \, p_{{}_{l+l',l+l'}}}}
{\textstyle\frac{(1-t^2)^{m/2}\lambda}{t^{l+l'+1}}} 
\\
+ (1-t^2)^{m/2}K(1/t) \lambda + (1-t^2)^{m/2}R(1/t),
\end{multline*}  
where $K$ is a polynomial of degree $< l+l'+1$, and $R$ is a polynomial. Therefore
\[
\mathfrak{a}_0 = {\textstyle\frac{2c[l,l',m] (-1)^{l+m+1} p_{{}_{l,l-m}}^{{}^{(m)}}p_{{}_{l',l'-m}}^{{}^{(m)}} \, p_{{}_{l+l',l+1}}^{{}^{(l'-1)}}}{(l'-1)!\, p_{{}_{l+l',l+l'}}}} \neq 0.
\]
We obtan the same $\mathfrak{a}_0$ for $A_{{}_{l, -m \,\,\,\, l', -m}}(\lambda) = A_{{}_{l, m \,\,\, l', m}}(\lambda)$.
 
Taking into account $\mathfrak{a}_0 \neq 0$, we observe the following.
Because the components of $A(\lambda), B(\lambda)$ 
are analytic also at $\lambda=0$, and we have denominators that have zero
of positive order at $\lambda=0$ in (\ref{explicitA}), (\ref{oddB}), (\ref{evenB}), then it is easily seen that
the sums, say $F_{{}_{0}}(\lambda)$, in the numerators in (\ref{explicitA}), (\ref{oddB}), (\ref{evenB}), which are not multiplied by $\lambda$, 
must be equal zero at $\lambda=0$, and the sums, say $F_{{}_{1}}(\lambda)$, in the numerators, which are multiplied by $\lambda$, 
must be nonzero at $\lambda=0$, because the lower order terms in these numerators must drop out, including the first order term.
Indeed, if we rewrite (\ref{B[l,lambda]}) (resp. (\ref{oddB}), (\ref{evenB})) or (\ref{explicitA}) in the form 
\begin{equation}\label{GeneralAB}
2^q{\textstyle\frac{f_{{}_{0}}(\lambda) +  \lambda f_{{}_{1}}(\lambda)}{(1-e^{-2\lambda})^{q}}}
=
2^q{\textstyle\frac{\sum\limits_{j_0} a_{{}_{0,j_0}} \,\, e^{-j_0\lambda} +  \lambda \sum\limits_{j_1} a_{{}_{1,j_1}} 
e^{-j_1\lambda}}{(1-e^{-2\lambda})^{q}}},
\end{equation}
with $f_{{}_{i}}(\lambda) = e^{-q\lambda}F_{{}_{i}}(\lambda)$, $i=0,1$,
denoting the corresponding sums, and
with a finite linear combination of exponents $e^{j_p\lambda}$,
$j_p \in \mathbb{N}$, in the numerator, then 
\begin{equation}\label{B0odd,B0even,A0}
f_{{}_{0}}(0) 
= \sum\limits_{j_0} a_{{}_{0,j_0}}  = 0,
\end{equation}
and
\begin{multline}\label{B1odd,B1even,A1}
f_{{}_{1}}(0) = \sum\limits_{j=0}^{k}\mathfrak{b}_{j} =  \sum\limits_{j_1} a_{{}_{1,j_1}} \neq 0,
\,\, \textrm{or} \,\,
f_{{}_{1}}(0) = \sum\limits_{j=0}^{k-1}\mathfrak{b'}_{j} = \sum\limits_{j_0} a_{{}_{1,j_1}} \neq 0,
\\
f_{{}_{1}}(0) = \mathfrak{a}_{{}_{0}} = \sum\limits_{j_1} a_{{}_{0,j_1}} \neq 0,
\end{multline} 
respectively, for $B$ and $A$. $q=l+1,l+l'+1$, respectively, for $B,A$.  (\ref{B0odd,B0even,A0}) is obvious. (\ref{B1odd,B1even,A1}) is obvious 
for $A$, because $\mathfrak{a}_0 \neq 0$. For $B$ the statement (\ref{B1odd,B1even,A1}) can be seen
on using (\ref{B1odd,B1even,A1}) for $A_{{}_{l,0 \,\,\, 1,0}}(\lambda)$, and the fact that $A_{{}_{l,0 \,\,\, 1,0}}(\lambda)$
is, up to a nonzero factor, equal to the derivative of $B_{{}_{l,0}}(\lambda)$ with respect to $\lambda$. Moreover,
using this relation between $A$ and $B$, we can easily see that
\begin{multline}\label{sum(b)}
\sum\limits_{j=0}^{(l-1)/2} \mathfrak{b}_{j} \,\,\, \textrm{or, respectively,} \,\,\,\,
\sum\limits_{j=0}^{l/2 -1} \mathfrak{b}'_{j} \,\, = \,\, (-1)^{l} i \mathfrak{e} \, \sqrt{{\textstyle\frac{8}{3}}} {\textstyle\frac{1}{l+1}} \mathfrak{a}_0
\\
=  - i \mathfrak{e} 2^{l+2} 
{\textstyle\binom{l+1/2}{l+1}}
{\textstyle\frac{l(l+1)}{2l+1}}, 
\end{multline}
where the coefficient $\mathfrak{a}_0$ refers to the matrix element $A_{{}_{l,0 \,\,\, 1,0}}(\lambda)$. Because the first order
terms drop out in the numerator of (\ref{GeneralAB}), then from (\ref{B0odd,B0even,A0}) and (\ref{B1odd,B1even,A1}) we get
\begin{equation}
f_{{}_{0}}^{{}^{(1)}}(0) = -\sum\limits_{j_0} a_{{}_{0,j_0}} j_0 = -f_{{}_{1}}(0),
\end{equation}
so that $f_{{}_{1}}$ is a zero-order function and $f_{{}_{0}}$ is a first-order function at $\lambda=0$,
with $f_{{}_{1}}(0) \neq 0, f_{{}_{0}}^{{}^{(1)}}(0) \neq 0$. 

It is easily seen that the natural numbers $j_0, j_1$ in the formula (\ref{GeneralAB}) representing $B_{{}_{l,0}}(\lambda)$,
are always even, independently of the parity of $l$, and have parity the same as $m$, in the formula (\ref{GeneralAB}) representing 
$A_{{}_{l,m \,\,\,\, l',m}}(\lambda)$. We will use these simple facts in Subsection \ref{AsymptoticLaurentCoefficient}.   
 
In order to prove our theorem, in addition to the above-mentioned properties of $A,B$, 
we need to know explicitly the coefficient 
\begin{equation}\label{a0ForA[l,-1,l',-1]}
\mathfrak{a}_0 = (-1)^{l}ll'2^{l+2l'} \textstyle{\frac{\sqrt{(2l+1)(2l'+1)}}{(l+1)!}\binom{l+l'-1/2}{l+l'}\binom{l'-1/2}{l'}}
\prod\limits_{j=0}^{l'-1}{\textstyle\frac{l+j+1}{2(l+j)+1}}\prod\limits_{r=0}^{l'-2}(l+l'-r)
\end{equation}
for the matrix element $A_{{}_{l,-1 \,\,\, l',-1}}(\lambda)$.

\subsection{Explicit formulas for $U^{{}^{(l_0,l_1)}}_{{}_{l,m \,\, l',m'}}\left(g_{{}_{03}}(\lambda)\right)$}\label{explicit(l0,l1)}

In what follows, we will use the following formulas
(let us recall that $l,l' \in\{ l_0, l_0+1, \ldots\}$, $-l \leq m \leq l$, $-l' \leq m' \leq l'$)
\begin{multline}\label{U^l0,l1}
U^{{}^{(l_0,l_1)}}_{{}_{l,m \,\, l',m'}}\left(g_{{}_{03}}(\lambda)\right) = U^{{}^{(l_0,l_1=i\rho)}}_{{}_{l,m \,\, l',m'}}(\lambda) 
= \delta_{{}_{m \, m'}}
{\textstyle\frac{1}{2}} \sqrt{(2l'+1)(2l+1)}(\textrm{cosh}\lambda)^{i\rho-1} \, \times
\\ \times \,\,
\int\limits_{-1}^{1} (1-ty)^{i\rho-1}\overline{P_{l_0 \, m'}^{l'}(y)}P_{l_0 \, m}^{l}({\textstyle\frac{y-t}{1-ty}}) dy.
\end{multline} 
Here $t = \textrm{tanh} \lambda$, and $P_{m \, n}^{l}(\cos \varphi) = e^{im\phi}T^{{}^{l}}_{{}_{m \,\, n}}(a)e^{in\vartheta}$ 
\emph{i.e.} $P_{m \, n}^{l}(\cos \varphi)$ is given by the above formula for $T^{{}^{l}}_{{}_{m \,\, n}}(a)$ in which we substitute
$\cos(\varphi/2)$ for $a_{11}$ and $a_{22}$ and $i\sin(\varphi/2)$ for $a_{12}$ and $a_{21}$. $P_{m \, n}^{l}$
are closely related to the Jacobi polynomials.
These formulas are obtained by changing the coordinate system on $G$ 
in the corresponding formulas for $U^{{}^{(l_0,l_1)}}$ given in \cite{NeumarkLorentzBook}. We need explicit form 
of the matrix elements (\ref{U^l0,l1}) with integer $l_0$. Performing integration (\ref{U^l0,l1}) by parts
and using recurrence rules for the Jacobi polynomials, we get by induction 
\begin{multline}\label{explicitU^(l0,l1)}
U^{{}^{(l_0,l_1=i\rho)}}_{{}_{l,m \,\, l',m}}(\lambda) 
= 
\\
{\textstyle\frac{1}{\rho Q(\rho)\textrm{sinh}^{{}^{l+l'+1}} \lambda}}
\Bigg[
e^{-i\rho\lambda}\sum\limits_{j=0}^{l+l'-||m|-|l_0||}p_{j}^{-}\textrm{sinh}^{{}^{j}} \lambda\textrm{cosh}^{{}^{l+l'-||m|-|l_0||-j}} \lambda
\\
+e^{i\rho\lambda}\sum\limits_{j=0}^{l+l'-|m|-|l_0|}p_{j}^{+}\textrm{sinh}^{{}^{j}} \lambda
\textrm{cosh}^{{}^{l+l'-||m|-|l_0||-j}} \lambda
\Bigg],
\end{multline}
for non-negative $m$. In the formula for non-positive $m$ the upper summation
ranges are exchanged with each other. The matrix elements
(\ref{explicitU^(l0,l1)}) with $l=l'$ and opposite $m$ are mutually complex conjugated.
Here $p_{j}^{-},p_{j}^{+}$ are polynomials in $\rho$ of degree $j$ and parity coinciding with the parity of $j$, 
depending on $l_0,l,l',m$. Thus, $U^{{}^{(l_0,l_1=i\rho)}}_{{}_{l,m \,\, l',m}}(\lambda)\sim e^{-(||m|-|l_0||+1)\lambda}$  at infinity.
$Q$ is a polynomial depending only on $l,l'$.   
We use $Q$ so normalized that the highest-degree coefficient in $Q$ is equal one, \emph{i.e.}
our $Q$ in (\ref{explicitU^(l0,l1)}) is monic: 
\[
Q(\rho) = (-i+\rho)(i+\rho)(-2i+\rho)(2i+\rho)(-3i+\rho) \ldots
\] 
with the number of factors equal to $l+l'$, and with the numbers $-i,i,-2i,2i,\ldots$ in them
increasing in absolute value but alternating in sign, except for the last $|l '-l|$ factors,
in which these numbers continue to increase in absolute value, but with the 
constant sign $+$, when $l>l'$, or with the constant sign $-$, when $l<l'$. For example
for $l=2,l'= 5$ and any $-2\leq m \leq 2$, and any integer $l_0$ 
\[
Q(\rho) = (-i+\rho)(i+\rho)(-2i+\rho)(2i+\rho)(-3i+\rho)(-4i+\rho)(-5i+\rho).
\] 

With this normalization of $Q$, we can fix the coefficients of the polynomials $p_{j}^{-},p_{j}^{+}$. 
We need the highest degree coefficient $p_{j,j}^{\pm}$ of the polynomials 
\[
p_{j}^{\pm}(\rho) = p_{j,0}^{\pm} + p_{j,1}^{\pm} \, \rho + \, \ldots \,
p_{j,j}^{\pm} \, \rho^j
\]
with maximal 
\[
j =j_\textrm{max} = l+l'-||m|-|l_0||
\]
for the matrix elements $U^{{}^{(l_0,i\rho)}}_{{}_{l, m =\pm l_0 \,\, l',m = \pm l_0}}(\lambda)$ and 
$U^{{}^{(l_0=1,i\rho)}}_{{}_{l, m = 0 \,\, l',m = 0}}(\lambda)$. 
The computation of the general
formula for the term of highest order in $\rho$, can be simplified. Indeed, from the formula
(\ref{explicitU^(l0,l1)}) it follows that this term determines the leading contribution 
\begin{equation}\label{MaxDegp,j,j}
{\textstyle\frac{p_{j_\textrm{max},j_\textrm{max}}^{\mp} \, e^{\mp i\rho\lambda}}{\sinh \lambda}} \, {\textstyle\frac{1}{\rho}},
\,\,\,\, \textrm{for} \,\,\, U^{{}^{(l_0,i\rho)}}_{{}_{l, \pm l_0 \,\, l', \pm l_0}}(\lambda)
\end{equation}
\begin{equation}\label{MaxDegp,j,j,0}
\big({\textstyle\frac{p_{j_\textrm{max},j_\textrm{max}}^{+} \, e^{i\rho\lambda}}{\sinh^{l_0+1} \lambda}}
+{\textstyle\frac{p_{j_\textrm{max},j_\textrm{max}}^{-}  \, e^{-i\rho\lambda}}{\sinh^{l_0+1} \lambda}}\big) \, {\textstyle\frac{1}{\rho^{l_0+1}}},
\,\,\,\, \textrm{for} \,\,\, U^{{}^{(l_0,i\rho)}}_{{}_{l, 0 \,\, l', 0}}(\lambda),
\end{equation}
to the asymptotics $\widesim[3]{\rho \rightarrow +\infty}$ of the matrix (\ref{explicitU^(l0,l1)}).
It follows from (\ref{explicitU^(l0,l1)}) that  $p_{j_\textrm{max},j_\textrm{max}}^{+} = 0$, 
if $|m| - |l_0|  \neq 0$, $m>0$, and $p_{j_\textrm{max},j_\textrm{max}}^{-} = 0$, if $|m| - |l_0|  \neq 0$, 
$m<0$. Using the integral (\ref{U^l0,l1}) we can compute the asymptotic $\widesim[3]{\rho \rightarrow +\infty}$ 
expansion by application of the saddle point method. 
In fact, for $l_0=0$, we can use an integral involving only the associated Legendre polynomials:
\begin{multline}\label{U^l0=0,l1}
U_{{}_{l,m \,\, l',m}}^{{}^{(0,i\rho)}}(\lambda) =
\\
(-1)^{l+l'} c_{{}_{l,l',m}} \textrm{cosh}^{i\rho -1} \, \lambda
\int\limits_{-1}^{1} P_{{}_{l',m}}(y) (1+ \, y \, \textrm{tanh} \lambda)^{i\rho -1} 
P_{{}_{l,m}}\big(\textstyle{\frac{y \, + \, \textrm{tanh} \, \lambda}{1 \, + \, y \, \textrm{tanh} \, \lambda}}\big)
\, dy
\end{multline}   
with
\[
c_{{}_{l,l',m}} = 
\sqrt{\textstyle{\frac{(2l+1)(2l'+1)(l-m)!(l'-m)!}{2(l+m)!2(l'+m)!}}}.
\]
Formula (\ref{U^l0=0,l1}) follows from the relation between the Jacobi and the associated Legendre polynomials. 
Independently, (\ref{U^l0=0,l1}) also follows
from the realization $\alpha \mapsto f(\Lambda(\alpha)^{-1}p)$ of the representation
$(l_0=0,l_1=i\rho)$ in the space of homogeneous of degree $i\rho -1$ functions $f$ on the cone, 
thus living effectively on the unit $\mathbb{S}^2$ sphere in the cone, with the inner product
\[
(f,g) = \int\limits_{{}_{\mathbb{S}^2}} \overline{f} g \,\,\, d\mu_{{}_{\mathbb{S}^2}}
\]
invariant under the action naturally induced by the Lorentz transforms in the ambient Minkowski space,
and with ordinary invariant measure $\mu_{{}_{\mathbb{S}^2}}$ on $\mathbb{S}^2$, and
the orthonormal basis $Y_{{}_{lm}}$, $l=0,1, \ldots$, $-l \leq m \leq l$, \cite{GelfandV}. 
The matrix (\ref{U^l0=0,l1}) coincides with (\ref{U^l0,l1}) for $l_0=0$. The integral 
(\ref{U^l0,l1}) and (\ref{U^l0=0,l1}) can be rewritten as
\[
U_{{}_{l,m \,\, l',m}}^{{}^{(l_0,i\rho)}}(\lambda)
\,
= 
{\textstyle\frac{c \, e^{i\rho \, \textrm{ln}(\textrm{cosh} \, \lambda)}}{\textrm{cosh} \, \lambda}}
\,
\int\limits_{-1}^{1} f(y)e^{i\rho S(y)} \, dy  
\]   
with the phase, respectively, equal $S(y) = \textrm{ln}(1 \pm y \, \textrm{tanh} \,  \lambda)$, with 
``$+$'' sign for (\ref{U^l0,l1}) and ``$-$'' sign for (\ref{U^l0=0,l1}), and with 
\[
f(y) = \overline{P_{{}_{l_0,m}}^{l'}(y)}
P_{{}_{l_0,m}}^{l}\big(\textstyle{\frac{y \, - \, \textrm{tanh} \, \lambda}{1 \, - \, y \, \textrm{tanh} \, \lambda}}\big)
{\textstyle\frac{1}{1 \, - \, y \, \textrm{tanh} \,  \lambda}}, 
\]
for (\ref{U^l0,l1}) and
\[
f(y) = P_{{}_{l',m}}(y)
P_{{}_{l,m}}\big(\textstyle{\frac{y \, + \, \textrm{tanh} \, \lambda}{1 \, + \, y \, \textrm{tanh} \, \lambda}}\big)
{\textstyle\frac{1}{1 \, + \, y \, \textrm{tanh} \,  \lambda}}, 
\]
for (\ref{U^l0=0,l1}). $S'(y) \neq 0$ for $ y \in [-1,1]$, and the function $f$ is analytic in an open set containing the 
closed interval $[-1,1]$. This is evident if $m, l_0$ are of the same parity in 
(\ref{U^l0,l1}), or even $m$ in (\ref{U^l0=0,l1}), because in this case each factor $P$ in $f$ separately,
is analytic. But it can be seen that this is generally the case, by noticing that $P_{{}_{l,m}}$ and, respectively, $P^{l'}_{{}_{l_0,m}}$,
can be written as the product of a polynomial and, eventually, the square root factor $(1-y^2)^{1/2}$. Then, it is easily seen that the square 
roots in the product $f$ drop out. The polynomial which arises after extracting the (eventual) square root $(1-y^2)^{1/2}$ in 
$P^{l'}_{{}_{l_0,m}}$ is a Jacobi polynomial multiplied by a natural power of $(1-x)$ and a natural power of $(1+x)$. 
Thus, theorem 1, \S 44.1, p. 410 of \cite{Sidorov}, 
can be applied, and gives the following asymptotic
expansion
\[
U_{{}_{l,m \,\, l',m}}^{{}^{(l_0,i\rho)}}(\lambda) 
\, \widesim[3]{\rho \rightarrow +\infty} \,
{\textstyle\frac{c}{\textrm{cosh} \, \lambda}}
 \sum\limits_{n=0}^{+\infty} \big(e^{-i\rho\lambda} b_n - e^{i\rho\lambda} a_n \big)\, (i\rho)^{-n-1},
\]   
with 
\[
c = \tfrac{1}{2}\sqrt{(2l'+1)(2l+1)} = c_{{}_{l,l',0}},
\]
for (\ref{U^l0,l1}) or, respectively,
$c= (-1)^{l+l'}c_{{}_{l,l',m}}$ for (\ref{U^l0=0,l1}), and with
\[
a_n = (-1)^n M^n \big({\textstyle\frac{f(y)}{S'(y)}}\big)\Big|_{{}_{y=-1}},
\,
b_n = (-1)^n M^n \big({\textstyle\frac{f(y)}{S'(y)}}\big)\Big|_{{}_{y=1}},
\,
M= {\textstyle\frac{1}{S'(y)}}{\textstyle\frac{d}{dy}},
\] 
respectively, in each case. In particular, for $l_0=m=0$, we get the leading order term
\[
U_{{}_{l, 0 \,\, l', 0}}^{{}^{(0,i\rho)}}(\lambda) 
\, \widesim[3]{\rho \rightarrow +\infty} \,  
{\textstyle\frac{(-1)^{l+l'}c_{{}_{l,l',0}}}{\textrm{sinh} \, \lambda}} \big((-1)^{l+l'}e^{i\rho} -e^{-i\rho}\big)  \, {\textstyle\frac{1}{i\rho}}.
\]
For $l_0>0$ and $m=\pm l_0$, respectively, we get
\begin{multline*}
U_{{}_{l, l_0 \,\, l', l_0}}^{{}^{(l_0,i\rho)}}(\lambda) 
\, \widesim[3]{\rho \rightarrow +\infty} \,  
-{\textstyle\frac{c_{{}_{l,l',0}}}{\textrm{sinh} \, \lambda}} e^{-i\rho} \, {\textstyle\frac{1}{i\rho}},
\\
U_{{}_{l, -l_0 \,\, l', -l_0}}^{{}^{(l_0,i\rho)}}(\lambda) 
\, \widesim[3]{\rho \rightarrow +\infty} \,  
(-1)^{l+l'}{\textstyle\frac{c_{{}_{l,l',0}}}{\textrm{sinh} \, \lambda}} e^{i\rho} \, {\textstyle\frac{1}{i\rho}}.
\end{multline*}
For $l_0=1$ and $m=0$ we get
\[
U_{{}_{l, 0 \,\, l', 0}}^{{}^{(1,i\rho)}}(\lambda) 
\, \widesim[3]{\rho \rightarrow +\infty} \,  
- {\textstyle\frac{\sqrt{l(l+1)l'(l'+1)}}{4}}{\textstyle\frac{c_{{}_{l,l',0}}}{\textrm{sinh}^2 \, \lambda}} \big((-1)^{l+l'}e^{i\rho} -e^{-i\rho}\big)  \, {\textstyle\frac{1}{\rho^2}}
\]
Comparing with (\ref{MaxDegp,j,j}) or , respectively, with (\ref{MaxDegp,j,j,0}),
we see that
\begin{align}
p_{{}_{j_\textrm{max},j_\textrm{max}}}^{+} = p_{{}_{l+l',l+l'}}^{+} = {\textstyle\frac{(-1)^{l+l'}c_{{}_{l,l',0}}}{i}}
\,\,\,\,\, \textrm{in} \,\,\,\,\,
U_{{}_{l, -l_0 \,\, l', -l_0}}^{{}^{(l_0,i\rho)}}(\lambda) 
\label{+l0Maxpj+-}
\\
p_{{}_{j_\textrm{max},j_\textrm{max}}}^{-} = p_{{}_{l+l',l+l'}}^{-} = -{\textstyle\frac{c_{{}_{l,l',0}}}{i}}
\,\,\,\,\, \textrm{in} \,\,\,\,\,
U_{{}_{l, l_0 \,\, l', l_0}}^{{}^{(l_0,i\rho)}}(\lambda),
\label{-l0Maxpj+-}
\\
p_{{}_{j_\textrm{max},j_\textrm{max}}}^{+} = p_{{}_{l+l',l+l'-1}}^{+} = -{\textstyle\frac{(-1)^{l+l'}c_{{}_{l,l',0}}\sqrt{l(l+1)l'(l+1)}}{4}}
\,\,\,\,\, \textrm{in} \,\,\,\,\,
U_{{}_{l, 0 \,\, l', 0}}^{{}^{(1,i\rho)}}(\lambda) 
\label{+0Maxpj+-}
\\
p_{{}_{j_\textrm{max},j_\textrm{max}}}^{-} = p_{{}_{l+l',l+l'-1}}^{-} = -{\textstyle\frac{c_{{}_{l,l',0}}\sqrt{l(l+1)l'(l'+1)}}{4}}
\,\,\,\,\, \textrm{in} \,\,\,\,\,
U_{{}_{l, 0 \,\, l', 0}}^{{}^{(1,i\rho)}}(\lambda),
\label{-0Maxpj+-}
\end{align}
for the maximal degree coefficients of the maximal degree polynomials $p_{j}^{\pm}$,
and
\begin{multline}\label{Maxpj+-}
p_{{}_{j_\textrm{max},j_\textrm{max}}}^{+} = p_{{}_{l+l',l+l'}}^{+} = {\textstyle\frac{(-1)c_{{}_{l,l',0}}}{i}},
\\
p_{{}_{j_\textrm{max},j_\textrm{max}}}^{-} = p_{{}_{l+l', l+l'}}^{-} =  \,  {\textstyle\frac{(-1)^{{}^{l'+l+1}}c_{{}_{l,l',0}}}{i}},
\end{multline}
for the maximal degree coefficients of the maximal degree polynomials $p_{j}^{\pm}$ in 
$U_{{}_{l, 0 \,\, l', 0}}^{{}^{(0,i\rho)}}(\lambda)$.
Note that the coefficients (\ref{+l0Maxpj+-}) -- (\ref{-0Maxpj+-}), are always nonzero, with (\ref{+l0Maxpj+-}) and (\ref{-l0Maxpj+-}) 
purely imaginary and with (\ref{+0Maxpj+-}) and (\ref{-0Maxpj+-}) real. Using the saddle point method, as above, one can easily
compute the cofficients (\ref{MaxDegp,j,j,0}) for $l_0>1$, and show that they are all nonzero, but in this paper 
we do not need the explicit values of these coefficients for $l_0>1$.  
We note here, that for fixed $l,l'$, all polynomials $p_j^{\pm}$ in the matrix elements $U_{{}_{l, m \,\, l', m}}^{{}^{(0,i\rho)}}(\lambda)$
with $m \neq 0$, have maximal degree $l+l'-|m|$, which is less than the maximal degree $j_\textrm{max} = l+l'$
of the polynomials $p_{j_\textrm{max}}^{\pm}$ in $U_{{}_{l, 0 \,\, l', 0}}^{{}^{(0,i\rho)}}(\lambda)$.
More generally, for fixed $l,l'$, the matrix elements $U_{{}_{l, \pm l_0 \,\, l', \pm l_0}}^{{}^{(l_0,i\rho)}}(\lambda)$
are precisely these that contain the polynomial $p_{{}_{j}}^{\pm}$ of highest degree $l+l'$ among all matrix elements 
$U_{{}_{l, m \,\, l', m}}^{{}^{(l_0,i\rho)}}(\lambda)$, $-\textrm{min} \, \{l,l'\} \leq m \leq \textrm{min} \, \{l,l'\}$.
All other matrix elements $U_{{}_{l, m \,\, l', m}}^{{}^{(l_0,i\rho)}}(\lambda)$ contain polynomials in $\rho$
of degree strictly smaller. We will use these facts in Subsection \ref{AsymptoticLaurentCoefficient}. 

Let us remark, that the higher order terms in the 
asymptotic $\widesim[3]{\rho \rightarrow +\infty}$ expansion of the matrix elements (\ref{explicitU^(l0,l1)}) are
determined by the lower order coefficients of the polynomials $p_{j}^{\pm}$ of lower order, and
can also be determined by the saddle point method shown above.

General formula for (\ref{explicitU^(l0,l1)}) with $l_0=m=l=0$ and for any $l'$:
\begin{multline}\label{(l0=0,irho)(0,0)x(l,0)}
U^{{}^{(0,i\rho)}}_{{}_{0,0 \,\,\, l,0}}(\lambda)
=
{\textstyle\frac{\sqrt{2l+1}(-1)^{l+1}}{2\prod\limits_{s=0}^{l}(-is+\rho)\textrm{sinh}^{l+1} \lambda}}
\sum\limits_{k=0}^{l}p_{{}_{l,k}} \sum\limits_{j=1}^{k+1}{\textstyle\frac{i^jk!}{(k-j+1)!}}
\Bigg[
\\
\textrm{sinh}^{l+1-j} \lambda (\textrm{cosh} \, \lambda + \textrm{sinh} \, \lambda)^{j-1} \prod\limits_{s=j}^{l}(-is+\rho) e^{i\rho\lambda}
\\
-
\textrm{sinh}^{l+1-j} \lambda (\textrm{cosh} \, \lambda - \textrm{sinh} \, \lambda)^{j-1} (-1)^{k-j+1} \prod\limits_{s=j}^{l}(-is+\rho) e^{-i\rho\lambda}
\Bigg],
\end{multline}
can easily be computed on repeated application of integration by parts in (\ref{U^l0=0,l1}). Here $p_{{}_{l,k}}$ are the coefficients of the Legendre
polynomial $P_l$ (Subsections \ref{Consistency}, \ref{FormulasForA,B}).

If we write $U^{{}^{(l_0,i\rho)}}_{{}_{l,m \,\,\, l',m}}(\lambda)$ in the form
\begin{equation}\label{GeneralU}
{\textstyle\frac{1}{\rho Q(\rho)}}
{\textstyle\frac{\sum\limits_{j\geq-1} \left[a_{{}_{j}}^{{}^{+}} \,\, e^{-j\lambda}e^{i\rho\lambda} +a_{{}_{j}}^{{}^{-}} \,\, 
e^{-j\lambda}e^{-i\rho\lambda}\right]
}{(1-e^{-2\lambda})^{q}}},
\end{equation}
with polynomials $a_{{}_{j}}^{{}^{\pm}}$ of $\rho$, then the natural numbers $j$ have parity the same as the number $-|m|-|l_0|-1$. 

In the sequel we use the fact that the general matrix elements (\ref{explicitU^(l0,l1)})
are quantities of order $|l'-l|$ at $\lambda =0$, with the following expansion at $\lambda=0$ 
\begin{align}
U^{{}^{(l_0,l_1=i\rho)}}_{{}_{l',m \,\, l,m}}(\lambda)  = 
\mathfrak{u}_{{}_{0}} \lambda^{|l-l'|}+\mathfrak{u}_{{}_{1}} \lambda^{|l-l'|+1} 
+ \ldots,
\label{Uexpansion}
\\
\mathfrak{u}_{{}_{0}} = {\textstyle\frac{(-1)^{l-l'}}{(l-l')!}}
\prod\limits_{j=0}^{l-l'-1}C_{l-j,m},
\,\,\, l>l',
\mathfrak{u}_{{}_{0}} = {\textstyle\frac{1}{(l-l')!}}
\prod\limits_{j=0}^{l-l'-1}\big[C_{l+j+1,m}\big],
\,\,\, l'>l,
\\
C_{l,m} =
{\textstyle\frac{\sqrt{l^2-m^2}}{l}}
\sqrt{{\textstyle\frac{(l^2-l_{0}^2)(l^2-l_1^2)}{4l^2-1}}},
\label{Uexpansion'}
\end{align} 
in which $\mathfrak{u}_{{}_{0}} \neq 0$ in the leading order terms. 
One can easily convince himself of the validity of (\ref{Uexpansion}) -- (\ref{Uexpansion'})
by the exponentiation of the generator $M_{{}_{03}}$ of the representation
$(l_0,l_1)$ given in \cite{Geland-Minlos-Shapiro},  \cite{NeumarkLorentzBook}.

\subsection{Decomposition of a cyclic representation. 
Example with $x=c_{{}_{1,0}}^+|u\rangle$}\label{DecompositionGeneralCyclic}

Now we pass to the decomposition of the cyclic representation with the cyclic vector (\ref{x}), and thus to the Fourier
analysis of the associated invariant kernel $\langle g|h\rangle$, and thus, Fourier transform of the associated positive
definite function $\varphi(h) = \langle e|h\rangle$ on $G$. 

Effectiveness of the Fourier analysis
comes from the fact that the matrix elements $U^{{}^{(l_0,l_1=i\rho)}}_{{}_{l,m \,\, l',m'}}$ of the principal series, regarded as functions on $G$, 
compose a complete system (in the generalized sense) of generalized functions on $G$. In general, they are not square-summable on $G$. 
But using them, we can construct square summable functions on $G$ with the required support of their Fourier transform
in the space of unitary irreducible equivalence classes, represented by $(l_0,l_1)$. 
In particular, integrating any matrix function $U^{{}^{(l_0,l_1=i\rho)}}_{{}_{lm \,\, l'm'}}(g)$ with
respect to $\rho$ over any interval $I\subset \mathbb{R}_+$ we get square summable function 
\[
U^{{}^{(l_0,iI)}}_{{}_{lm \,\, l'm'}}(g) = \int\limits_{I} U^{{}^{(l_0,l_1=i\rho)}}_{{}_{lm \,\, l'm'}}(g)d\rho
\]
on $G$ (recall oscillatory character $\cos(\rho \lambda), \sin(\rho \lambda)$ 
of matrix elements), with the Fourier transform supported at $(l_0,l_1)$ with $l_1 \in iI$. Any two such square summable packets
are orthogonal in $L^2(G)$ whenever the corresponding intervals $I$ have empty intersection, or whenever the corresponding
$l_0$ are different.
Moreover, even if the corresponding intervals $I$ and $l_0$ coincide, the packets remain orthogonal if any of their
corresponding matrix coefficients do not coincide:
\begin{equation}\label{Orthogonalityl0l1}
\int\limits_{G}\overline{U^{{}^{(l_0,iI)}}_{{}_{lm \,\, l'm'}}(g)}U^{{}^{(l_0,iI')}}_{{}_{l''m'' \,\, l'''m'''}}(g) \, dg 
= 
\begin{cases}
c(l_0,I,I') \,\, \delta_{{}_{l_0 \,\,\, l'_0}} \, \delta_{{}_{l \, l''}} \delta_{{}_{m \, m''}} \delta_{{}_{l' \, l'''}} \delta_{{}_{m' \, m'''}},
 \\
= 0, \,\, \textrm{if} \,\, I \cap I' = \emptyset
\end{cases}.
\end{equation}        
Orthogonality of the matrix elements corresponding to non-equivalent representations comes from the fact that they compose generalized eigenstates
of the two Casimir operators acting in $L^2(G)$, associated with different generalized eigenvalues. Orthogonality relations of the matrix
functions corresponding to the same representation classes cannot be justified in this manner. But in this case, we use the coordinate system
on $G$ determined by (\ref{g}), representation property of $U^{{}^{(l_0,l_1=i\rho)}}_{{}_{lm \,\, l'm'}}$, and perform
the integration iteratively, in which the integration with respect to $\rho$ must necessary be performed first. Finally using
the orthogonality relations for the matrix elements $T^{{}^{l}}_{{}_{m \, m' }}(a)$, $a \in SU(2, \mathbb{C})$, we obtain
the orthogonality relations (\ref{Orthogonalityl0l1}), compare the analogue orthogonality for the $SL(2,\mathbb{R})$ in \cite{Bargmann}.
The supplementary series does not enter the Plancherel formula for the the group $G$, \emph{i.e.} it is absent in the decomposition of $L^2(G)$. 
Martix elements $U^{{}^{(l_0=0,l_1=1-s)}}_{{}_{lm \,\, l'm'}}(g)$, $0<s<1$, of the supplementary series decrease slower at infinity,
e.g. $U^{{}^{(0,l_1=1-s)}}_{{}_{2,0 \,\, 2,0}}(\lambda) \sim e^{-s\lambda}$ at infinity, and are still not square integrable even after
integration over an interval in the real parameter $s$. But $G=SL(2,\mathbb{C})$ is complex analytic, and 
the matrix elements of the representations $(l_0,l_1)$ are analytic, both as functions of $g \in G$ and as functions of the continuous 
parameter $l_1$ of both series, principal and supplementary, with the matrix elements of the supplementary series $(l_0=0, s)$, $-1 < s < 1$, being the analytic 
continuations of the corresponding matrix elements of the principal sieries $(l_0=0, l_1=i\rho)$. We will use these facts in what follows.  

For $x = |e\rangle = c_\alpha^+ e^{-inS(u)}|0\rangle = c_\alpha^+|u\rangle$ we have
\begin{equation}\label{<g|h>}
\langle g| h\rangle  
= 4\pi\mathfrak{e}^2 \, \left[\overline{A_{{}_{\alpha \,\, \alpha}}(g^{-1}h)}
+ {\textstyle\frac{z}{4}} \overline{{\textstyle\frac{1}{\mathfrak{e}}}B_{{}_{\alpha}}(g^{-1}h)}{\textstyle\frac{1}{\mathfrak{e}}}B_{{}_{\alpha}}(h^{-1}g) \right]
e^{-z(\lambda \textrm{coth} \lambda -1)},
\end{equation}
for $g^{-1}h$ with decomposition (\ref{g}), in which $\lambda$ can be interpreted as the hyperbolic angle between $u$
and $g^{-1}hu$.

Each of the invariant kernels $\langle g |h\rangle$, corresponding to cyclic vectors (\ref{x}), 
can be written in the form of a function depending only on $g,h,z$ (up to a constant factor depending on $\mathfrak{e}$).
Each $\langle g |h\rangle$ depends on $n$ only through $z=n^2\mathfrak{e}^2/\pi$, 
where $n\mathfrak{e}$ is the eigenvalue of the total charge $Q$. 
The simplest examples (\ref{<u,ugh-1>}) and (\ref{<g|h>}) we have already 
seen for $x = |u\rangle, c_{{}_{\alpha}}^+ |u\rangle$. 

The cases with cyclic $x=|u\rangle,c_{{}_{1,0}}^+|u\rangle$ are exceptional, so we will continuously 
be giving explicit formulas for the concrete 
example of representation with cyclic vector $x=c_{{}_{1,0}}^+|u\rangle$.
The exceptional case $x=|u\rangle$ was worked out in \cite{Staruszkiewicz1992ERRATUM}.

The positive definite function $k(h) =\langle e|h\rangle$, corresponding to the left invariant kernel
$\langle g| h\rangle$ on $G$, is in $L^2(G)$ only if $z>1$, and only in this case we can use the Plancherel formula and the inverse Fourier transform
formula for $G$ to decompose $\langle g| h\rangle$. In case $0<z<1$ decomposition will involve additionally the supplementary 
component which cannot be inferred immediately from the Plancherel formula. We use the method \cite{Staruszkiewicz1992ERRATUM}
of analytic continuation to obtain the formula valid in the domain $0<z<1$,
which is based on the fact that $\langle g| h\rangle$ is analytic in $z$.  
In case $z>1$, from Lebesgue's dominated convergence principle, it easily follows that
the integrals
\begin{equation}\label{F[<e,.>]}
\Big[ K(l_0,l_1=i\rho; z)\Big]_{{}_{\beta \,\, \gamma}} 
=\int\limits_{{}_{G}} \langle e| h \rangle \,\, U^{{}^{(l_0,l_1=i\rho)}}_{{}_{\beta \,\, \gamma}}(h) \, dh,
\,\,\, z>1,
\end{equation}
are convergent and represent analytic functions of $z,\rho$. 
Here, in principle, $\beta,\gamma \in \{(l_0,m_0), (l_1,m_1)=(l_0+1, m_{1}), \ldots \}$, $-l_i \leq m_i \leq l_i$ and 
all representations of the principal series are considered in this integral. But for the cyclic (\ref{x})
the integral (\ref{F[<e,.>]}) is nonzero only for $l_0 \leq l(\alpha_1) + \ldots l(\alpha_\mathfrak{q})$ by the product
formula and orthogonality relations of the matrix elements of the standard unitary representations of 
$SU(2,\mathbb{C})$ (Peter-Weyl theorem) and the structure (\ref{SU2Cl0l1}) of representations $(l_0,l_1)$.  
For example, from the formula (\ref{<u,ugh-1>}) it follows that $\langle e|h\rangle = \langle u | hu\rangle$ does not depend 
on the angle coordinates $\left(\theta_{{}_{1}}, \varphi_{{}_{1}}, \vartheta_{{}_{1}}, \vartheta, \varphi \right)$,
thus, for the kernel $\langle g|h\rangle = \langle g u | hu\rangle$ corresponding to the cyclic vector $|u\rangle$ and $z>1$,
the Fourier transform  (\ref{F[<e,.>]}) is nonzero only for $U^{{}^{(l_0=0,l_i=i\rho)}}_{{}_{0,0 \,\,\, 0,0}}$, independent of the
angle coordinates, by the structure (\ref{SU2Cl0l1}) of the restriction of the representations $(l_0,l_1)$ of $G$ to $SU(2,\mathbb{C})$
and orthogonality relations (Peter-Weyl theorem for $SU(2,\mathbb{C})$). 
Because for general $g\in G$ of the form (\ref{g}) 
\[
U^{{}^{(l_0=0,l_1=i\rho)}}_{{}_{0,0 \,\,\, 0,0}}(g)
= {\textstyle\frac{\sin(\rho\lambda)}{\rho \textrm{sinh}(\lambda)}},
\]
then for the kernel $\langle g|h\rangle = \langle g u | hu\rangle$ corresponding to the cyclic vector $|u\rangle$ and $z>1$
the only nonzero matrix element of $K(l_0=0,l_1=i\rho; z)$ is the diagonal element
\begin{equation}\label{ASK}
\Big[ K(l_0=0,l_1=i\rho;z)\Big]_{{}_{0,0 \,\,\,0,0}} =
{\textstyle\frac{4\pi^3e^z}{\rho}} \int\limits_{0}^{+\infty} \textrm{sinh}(\lambda)\sin(\rho\lambda)
e^{-z\lambda\textrm{coth}\lambda} \, d\lambda, 
\end{equation}
and $K(l_0,l_1=i\rho;z) = 0$ for $l_0\neq 0$, which, up to the constant factor $\pi^2$, 
agrees with the weight $K$ of the cyclic representation with the cyclic vector $x=|u\rangle$,
found in \cite{Staruszkiewicz1992ERRATUM}. (The various constant factors come from the free choice in the normalization of the 
invariant measure on $G$ and of the induced invariant measure on the Lobachevsky space
$G/SU(2,\mathbb{C})$, irrelevant in our analysis). The only nonzero element of the 
matrix $K$ for the kernel (\ref{<g|h>}) is the diagonal element 
\begin{multline}\label{Kalphaalpha}
\Big[ K(l_0,i\rho;z)\Big]_{{}_{\alpha \,\, \alpha}} =
\\
\left({\textstyle\frac{4\pi \mathfrak{e}}{2l+1}}\right)^2e^z \bigints\limits_{0}^{+\infty} \Bigg[
\sum\limits_{n=-l}^{l}
\overline{A_{{}_{l,n \,\,\, l,n}}(\lambda)}U^{{}^{(l_0,i\rho)}}_{{}_{l,n \,\,\, l,n}}(\lambda) 
+{\textstyle\frac{z}{4}} (-1)^l \left|{\textstyle\frac{1}{\mathfrak{e}}}B_{{}_{l,0}}(\lambda) \right|^2
U^{{}^{(l_0,i\rho)}}_{{}_{l,0 \,\,\, l,0}}(\lambda)
\\
\Bigg]e^{-z\lambda \textrm{coth} \lambda} \, \pi^2 \textrm{sinh}^2 \lambda \, d\lambda,
\end{multline}
where $\alpha = (l,m)$ is the same which is present in $\langle g| h\rangle$ in (\ref{<g|h>}) 
and in $x= |e\rangle$ defining $\langle g| h\rangle$, and is independent of the azimuthal number $m$. 
To see it, let $\beta = (L,M)$ and $\gamma = (L_1,M_1)$, with $L \leq L_1$ in (\ref{F[<e,.>]}). 
The general argument $h\in G$ of the integrand in (\ref{F[<e,.>]}) we decompose in the form $h = u_1 u_2^* g_{{}_{03}}(\lambda)u_2$,
according to  (\ref{g}), where $u_1 = g_{{}_{12}}(\theta_1)g_{{}_{13}}(\varphi_1)g_{{}_{12}}(\vartheta_1) \in SU(2,\mathbb{C})$,
$u_2 = g_{{}_{13}}(\varphi)g_{{}_{12}}(\vartheta) \in SU(2,\mathbb{C})$.
We use the homomorphism property of the matrices of the representations $(l_0,l_1)$, 
the property (\ref{SU2Cl0l1}) of their restriction to the subgroup $SU(2,\mathbb{C})$, and the
formulas (\ref{B(g)1}), (\ref{B(g)2}), to compute each of the terms of $\langle e, h \rangle$, 
given by (\ref{<g|h>}), with $g=e$ in (\ref{<g|h>}) and $U^{{}^{(l_0,l_1=i\rho)}}_{{}_{L,M \,\, L_1,M_1}}(h)$. Inserting
the integrand computed in this way, and the invariant measure (\ref{dg}), we get, by performing the integration
$du_1$ over $SU(2,\mathbb{}C)$ first, and using the $SU(2,\mathbb{}C)$-invariance of $du_1$: 
\begin{multline*}
K_{{}_{L,M \,\, L_1,M_1}} 
= 4\pi \mathfrak{e}^2 \sum_{\substack{-l \leq m' \leq l \\ -L \leq M' \leq L}} \,\,
\int\limits_{SU(2,\mathbb{C})} \overline{T^{{}^{l}}_{{}_{m \,\, m'}}\big(u_1\big)}T^{{}^{L}}_{{}_{M \,\, M'}}\big(u_1\big) \, du_1
\,\, \times
\\
\times \,\,
\int\limits_{\mathbb{S}^2} \overline{T^{{}^{l}}_{{}_{m' \,\, m}}\big( g_{{}_{13}}(\varphi)g_{{}_{12}}(\vartheta)\big)}
T^{{}^{L_1}}_{{}_{M' \,\, M_1}}\big( g_{{}_{13}}(\varphi)g_{{}_{12}}(\vartheta) \big) \, \sin \varphi d\varphi d\vartheta
\,\, \times
\\
\times \,\,
\int_{0}^{+\infty} \overline{A_{{}_{l,m' \,\, l,m'}}(\lambda)}U^{{}^{(l_0,i\rho)}}_{{}_{L,M' \,\, L_1,M'}}(\lambda) 
e^{-z(\lambda\coth \lambda -1)} \, \pi^2 \sinh^2 \lambda \, d\lambda
\end{multline*}  
\begin{multline*}
+4\pi \mathfrak{e}^2(-1)^l {\textstyle\frac{z}{4}} 
\sum\limits_{-L \leq M' \leq L}
\int\limits_{SU(2,\mathbb{C})} \overline{T^{{}^{l}}_{{}_{m \,\, 0}}\big(u_1\big)}T^{{}^{L}}_{{}_{M \,\, M'}}\big(u_1\big) \, du_1
\,\, \times
\\
\times \,\,
\int\limits_{\mathbb{S}^2} \overline{T^{{}^{l}}_{{}_{0 \,\, m}}\big( g_{{}_{13}}(\varphi)g_{{}_{12}}(\vartheta)\big)}
T^{{}^{L_1}}_{{}_{M' \,\, M_1}}\big( g_{{}_{13}}(\varphi)g_{{}_{12}}(\vartheta) \big) \, \sin \varphi d\varphi d\vartheta
\,\, \times
\\
\times \,\,
\int_{0}^{+\infty} \big| {\textstyle\frac{1}{\mathfrak{e}}}B_{{}_{l,0}}(\lambda)\big|^2 U^{{}^{(l_0,i\rho)}}_{{}_{L,M' \,\, L_1,M'}}(\lambda) 
e^{-z(\lambda\coth \lambda -1)} \, \pi^2 \sinh^2 \lambda \, d\lambda
\end{multline*}
Now we use the first one of the following orthogonality relations: 
\begin{multline}\label{Torthogonality}
\int\limits_{SU(2,\mathbb{C})}\overline{T^{{}^{l_1}}_{{}_{m_1 \,\, n_1}}(a)}T^{{}^{l_2}}_{{}_{m_2 \,\, n_2}}(a) \, da =
{\textstyle\frac{1}{2l_1+1}} \delta_{{}_{l_1 \,\, l_2}}\delta_{{}_{m_1 \,\, m_2}} \delta_{{}_{n_1 \,\, n_2}},
\\
\int\limits_{SU(2,\mathbb{C})/\mathbb{S}^1}
\overline{T^{{}^{l_1}}_{{}_{m_1 \,\, n_1}}\big(g_{{}_{13}}(\varphi)g_{{}_{12}}(\vartheta)\big)}
T^{{}^{l_2}}_{{}_{m_2 \,\, n_2}}\big(g_{{}_{13}}(\varphi)g_{{}_{12}}(\vartheta)\big) \sin \varphi d\varphi d\vartheta
\\
=
{\textstyle\frac{4\pi}{2l_1+1}} I^{{}^{l_1l_2}}_{{}_{m_1m_2}}\, \delta_{{}_{n_1n_2}}, \,\,\, I^{{}^{l_1l_2}}_{{}_{m m}} = \delta^{{}^{l_1l_2}},
\end{multline}
where
\[
I^{{}^{l_1l_2}}_{{}_{m_1m_2}} = {\textstyle\frac{2l_1+1}{2}}
\int\limits_{0}^{\pi}
\overline{P^{{}^{l_1}}_{{}_{m_1 \,\, n_1}}\big(\cos\varphi\big)}
P^{{}^{l_2}}_{{}_{m_2 \,\, n_1}}\big(\cos\varphi\big) \sin \varphi d\varphi.
\]
We obtain 
\begin{multline*}
K_{{}_{L,M \,\, L_1,M_1}} 
= \delta_{{}_{l \,\, L}} \, \delta_{{}_{M \,\, m}} \, 4\pi \mathfrak{e}^2 \sum_{-l \leq m' \leq l}
{\textstyle\frac{1}{2l+1}}
\,\, \times
\\
\times \,\,
\int\limits_{\mathbb{S}^2} \overline{T^{{}^{l}}_{{}_{m' \,\, m}}\big( g_{{}_{13}}(\varphi)g_{{}_{12}}(\vartheta)\big)}
T^{{}^{L_1}}_{{}_{m' \,\, M_1}}\big( g_{{}_{13}}(\varphi)g_{{}_{12}}(\vartheta) \big) \, \sin \varphi d\varphi d\vartheta
\,\, \times
\\
\times \,\,
\int_{0}^{+\infty} \overline{A_{{}_{l,m' \,\, l,m'}}(\lambda)}U^{{}^{(l_0,i\rho)}}_{{}_{L,m' \,\, L_1,m'}}(\lambda) 
e^{-z(\lambda\coth \lambda -1)} \, \pi^2 \sinh^2 \lambda \, d\lambda
\end{multline*}  
\begin{multline*}
+\delta_{{}_{l \,\, L}} \, \delta_{{}_{M \,\, m}} \, 4\pi \mathfrak{e}^2(-1)^l {\textstyle\frac{z}{4}} 
{\textstyle\frac{1}{2l+1}}
\,\, \times
\\
\times \,\,
\int\limits_{\mathbb{S}^2} \overline{T^{{}^{l}}_{{}_{0 \,\, m}}\big( g_{{}_{13}}(\varphi)g_{{}_{12}}(\vartheta)\big)}
T^{{}^{L_1}}_{{}_{0 \,\, M_1}}\big( g_{{}_{13}}(\varphi)g_{{}_{12}}(\vartheta) \big) \, \sin \varphi d\varphi d\vartheta
\,\, \times
\\
\times \,\,
\int_{0}^{+\infty} \big| {\textstyle\frac{1}{\mathfrak{e}}}B_{{}_{l,0}}(\lambda)\big|^2 U^{{}^{(l_0,i\rho)}}_{{}_{l,0 \,\, L_1,0}}(\lambda) 
e^{-z(\lambda\coth \lambda -1)} \, \pi^2 \sinh^2 \lambda \, d\lambda
\end{multline*}
Using the second orthogonality relation (\ref{Torthogonality}) with $m_1 = m_2$, we obtain
\begin{multline*}
K_{{}_{L,M \,\, L_1,M_1}} 
=  \left({\textstyle\frac{4\pi \mathfrak{e}}{2l+1}}\right)^2\, e^z
 \delta_{{}_{l \,\, L}} \, \delta_{{}_{M \,\, m}} \, \delta_{{}_{L_1 \,\, l}} \, \delta_{{}_{M_1 \,\, m}} 
\Bigg[
\\
 \sum_{-l \leq m' \leq l}
\int_{0}^{+\infty} \overline{A_{{}_{l,m' \,\, l,m'}}(\lambda)}U^{{}^{(l_0,i\rho)}}_{{}_{l,m \,\, l,m}}(\lambda) 
e^{-z\lambda\coth \lambda} \, \pi^2 \sinh^2 \lambda \, d\lambda
\\
+(-1)^l {\textstyle\frac{z}{4}} 
\int_{0}^{+\infty} \big| {\textstyle\frac{1}{\mathfrak{e}}}B_{{}_{l,0}}(\lambda)\big|^2 U^{{}^{(l_0,i\rho)}}_{{}_{l,0 \,\, l,0}}(\lambda) 
e^{-z\lambda\coth \lambda} \, \pi^2 \sinh^2 \lambda \, d\lambda
\Bigg],
\end{multline*}
\emph{i.e.} (\ref{Kalphaalpha}) as the only nonzero coefficient of the matrix $K$.
Moreover, from the homomorphism property of $U^{{}^{(l_0,l_1=i\rho)}}$ and orthogonality properties of $T^{{}^{l}}_{{}_{m \,\, m'}}$ it follows
that (\ref{Kalphaalpha}) is nonzero for integer $l_0$ only, and 
(\ref{Kalphaalpha}) is zero for $|l_0|>l(\alpha)$, where for $\alpha = (l,m)$, $l(\alpha) = l$. 
In particular, for $\alpha$ with $l(\alpha) = 1$, the only possible
representations $(l_0,i\rho)$ for which (\ref{Kalphaalpha}) is nonzero, 
are the representations $(l_0=0, l_1=i\rho)$ and $(l_0=\pm 1, l_1= i\rho)=(1,\pm i\rho)$, $\rho>0$, of the principal series.

The integral (\ref{Kalphaalpha}) is convergent for $z>0$ if $|l_0|>0$. It 
follows from the fact that $B_{{}_{l,0}}(\lambda)$ tends to a nonzero constant for $\lambda$ going to infinity,
compare (\ref{BatInfinity}), and by the asymptotic behavior 
$U^{{}^{(l_0,l_1=i\rho)}}_{{}_{l,0 \,\,\, l',0}}\sim e^{-(|l_0|+1)\lambda}$ at infinity. This is generally
true for the integral (\ref{F[<e,.>]}).

As we have already mentioned, $K$ is nothing else but the Hermitian conjugation of the Fourier 
transform of the complex conjugation $\overline{k}$ of the 
positive definite function $k(h)=\langle e|h\rangle$ 
corresponding to the positive definie kernel $\langle g | h \rangle$ on $G$.  
From the invariance it easily follows 
\[
\int\limits_{{}_{G}} \langle g| h \rangle \,\, U^{{}^{(l_0,l_1)}}(h) \, dh = U^{{}^{(l_0,l_1)}}(g) \,\, \int\limits_{{}_{G}} \langle e| h \rangle \,\, U^{{}^{(l_0,l_1)}}(h) \, dh, 
\,\,\, g\in G.
\] 
Let $f\in L^2(G)$. We have the following Fourier transform and its inverse on $G$, \cite{NeumarkLorentzBook}:
\begin{multline*}
\mathcal{F}f(l_0,l_1=i\rho) = \int\limits_{{}_{G}} f(g)U^{{}^{(l_0,l_1=i\rho) \, *}}(g) \, dg,
\\
f(g) = \sum\limits_{l_0=-\infty}^{+\infty} \int\limits_{0}^{+\infty}\textrm{Tr} \, \left[\mathcal{F}f(l_0,l_1=i\rho)U^{{}^{(l_0,l_1=i\rho)}}(g) \right]{\textstyle\frac{l_0^2+\rho^2}{2\pi^4}} \, d\rho
\end{multline*}
\begin{multline*}
= \int\limits_{0}^{+\infty}\textrm{Tr} \, \left[\mathcal{F}f(0,l_1=i\rho)U^{{}^{(l_0,l_1=i\rho)}}(g) \right]{\textstyle\frac{\rho^2}{2\pi^4}} \, d\rho
\\
+ \sum\limits_{l_0>0} \,\, \int\limits_{-\infty}^{+\infty}\textrm{Tr} \, \left[\mathcal{F}f(l_0,l_1=i\rho)U^{{}^{(l_0,l_1=i\rho)}}(g) \right]{\textstyle\frac{l_0^2+\rho^2}{2\pi^4}} \, d\rho,
\end{multline*}
where only the representations of the principal series are present here (recall $U^{{}^{(l_0,l_1)}} = U^{{}^{(-l_0,-l_1)}}$). 
Star denotes the Hermitian conjugation.
Because of the last three formulas and because the integral (\ref{F[<e,.>]}) is convergent ($z>1$), changing the order of
integration is legitimate in deriving the following formula obtained by inserting the inverse Fourier formula 
\begin{multline}\label{decomositionz>1}
\langle f|f'\rangle =  \int\limits_{{}_{G \times G}} dg \, dh \,\, \langle g | h \rangle f(h) \overline{f'(g)}
 =
\\
\sum\limits_{l_0=-l_{{}_{\textrm{max}}}}^{l_{{}_{\textrm{max}}}}\int\limits_{0}^{+\infty} {\textstyle\frac{l_0^2+\rho^2}{2\pi^4}} \, d\rho
\textrm{Tr} \, \left[\mathcal{F}f(l_0,l_1=i\rho)K(l_0,l_1=i\rho; z)\mathcal{F}f'(l_0,l_1=i\rho)^* \right],
\end{multline}
As we will show, only the contribution $l_0=0$ is nonzero for the kernel (\ref{<g|h>}) with $l(\alpha) = 1$.  
The formula (\ref{decomositionz>1}), valid if $z>1$, gives the decomposition of the cyclic representation
with the cyclic (\ref{x}) into the direct integral of the principal series representations $(l_0,l_1)$,
each entering with multiplicity equal to the rank of $K(l_0,l_1=i\rho; z)$, which in this case is equal $1$.
This is the case for cyclic (\ref{x}) with $\mathfrak{q} =1$ and, generally for  $\mathfrak{q} > 1$, it follows from 
the relations between the cyclic spaces proved in Section \ref{CyclicDomains}.
The multiplicity of $(l_0,l_1)$, equal to the rank of $K(l_0,l_1=i\rho; z)$, is equal $1$, because the corresponding representation is cyclic. 
Positivity of the operator matrix $K(l_0,l_1=i\rho; z)$ (for almost
all $\rho$) follows from the positivity of the kernel $\langle g|h\rangle$. 

In passing to the domain $0<z<1$ we observe first that also in this domain the formula (\ref{decomositionz>1}) remains meaningful, except
the term with $l_0=0$, as the integral (\ref{F[<e,.>]})  is convergent for $|l_0|>0$. The integral (\ref{F[<e,.>]}) with $l_0=0$
becomes divergent for $0<z<1$, and so the term with $l_0=0$ in (\ref{decomositionz>1})  
becomes divergent for $0<z<1$. Using the symmetry $U^{{}^{(l_0=0,l_1)}} = U^{{}^{(l_0=0,-l_1)}}$, the contribution with $l_0=0$ in 
(\ref{decomositionz>1}) can be written as
\begin{multline*}
{\textstyle\frac{1}{2}}\int\limits_{-\infty}^{+\infty} d\rho {\textstyle\frac{\rho^2}{2\pi^4}} 
\textrm{Tr} \, \left[\mathcal{F}f(0,i\rho)K(0,i\rho; z)\mathcal{F}f'(0,i\rho)^* \right] =
\\
\int\limits_{{}_{G \times G}} dg \, dh \,\, \langle g | h \rangle f(h) \overline{f'(g)}
\, {\textstyle\frac{1}{2}}\int\limits_{-\infty}^{+\infty} d\rho {\textstyle\frac{\rho^2}{2\pi^4}} 
\textrm{Tr} \, \left[U^{{}^{(0,i\rho) \, *}}(h)K(0,i\rho; z)U^{{}^{(0,i\rho)}}(g)\right],
\end{multline*}
where the last integral $1/2 \int d\rho \ldots = \int d\rho F(\rho,z)$ we treat (compare \cite{Staruszkiewicz1992ERRATUM}) as a  contour integral in the 
complex $\rho$-plane. In passing to the domain $0<z<1$ we are using the analytic continuation in $z$ of the $l_0=0$
contribution. Its correct construction is summarized in the following prescription \cite{Staruszkiewicz1992ERRATUM}: we replace the contour $d\rho$
integral with the $d\rho$-integral in which the integrand $F(\rho,z)$ is replaced by its analytic continuation 
in the variables $\rho,z$ (\emph{i.e.} we replace $K(0,l_1=i\rho; z)$ with its analytic continuation), then add to it the sum of all residues
of the analytic continuation of $F(\rho,z)$ mutlipled by $2\pi i$, which cross the contour going ``top down'' when $z$ is passing from the domain $z>1$ to the 
domain $0<z<1$, and finally subtract the sum of all residues
of the analytic continuation of $F(\rho,z)$ multiplied by $2\pi i$, which cross the contour going ``down top'' when $z$ is passing from the domain $z>1$ to the 
domain $0<z<1$. For general $\alpha = (l,m_l)$, $-l\leq m_l\leq l$, $l\geq 1$, in (\ref{<g|h>}), the only residues in complex $\rho$-plane 
of $F(\rho,z)$ crossing the contour are located at $\pm i(1-z)$. At the same time, the only poles in complex $\rho$-plane
of the analytic continuation of  $K(0,l_1=i\rho; z)$ which cross the contour, when $z$ changes the domains $z>1$  and $0<z<1$, 
are the simple poles, \emph{i.e.} the residues at  $\pm i(1-z)$.
This is generally the case for the kernel $\langle g|h\rangle$ corresponding to the general cyclic vector (\ref{x}), and follows from the fact that the
functions $U^{{}^{l_0, i\rho}}_{{}_{l',0 \,\, l,0}}$ behave at infinity as $\sim e^{-(|l_0|+1)\lambda}$ in the coordinates given by decomposition (\ref{g}),
so that the integrands of the particular contributions to the matrix element of $K(l_0,l_1=i\rho; z)$, in the $d\lambda$-integral, 
behave as $\sim e^{-s\lambda}$ at infinity, with $s= 1-z+|l_0|, 1-z+|l_0|+1, 1-z+|l_0|+2, \ldots$, for the essential contribution
which does not contain the $A$-factors, and
$s=z, z+1, z+2, \ldots$, for the part of contributions containing $A$-factors, compare our discussion below. 
Taking these circumstances into account, the analytic continuation 
of the $l_0=0$ contribution in (\ref{decomositionz>1}) has the form 
\begin{multline}\label{ResDec}
\int\limits_{0}^{+\infty} d\rho {\textstyle\frac{\rho^2}{2\pi^4}} 
\textrm{Tr} \, \left[\mathcal{F}f(l_0=0,l_1=i\rho)K(0,i\rho; z)\mathcal{F}f'(l_0=0,l_1=i\rho)^* \right] 
\\
+\textrm{Tr} \, \left[\mathcal{F}f(l_0=0,l_1=1-z)\kappa(z) \mathcal{F}f'(l_0=0,l_1=1-z)^* \right]
\end{multline}
where the second term coming from the residues is present only if $0<z<1$, and where $K(0,i\rho; z)$ is the analytic continuation of 
$K(l_0=0,i\rho; z)$ given by (\ref{F[<e,.>]}) or, in more explicit form, by (\ref{Kalphaalpha}), 
\emph{e.t.c.}. 
$\kappa$ is a finite dimensional positive definite matrix $\kappa_{{}_{\beta \,\, \gamma}}$ which, in case of the kernel
(\ref{<g|h>}), has only one nonzero matrix element $\kappa_{{}_{\alpha \,\, \alpha}}$, 
with $\alpha$ being the same as the $\alpha$ in definition of the cyclic vector $x= c_\alpha^+e^{-inS(u)}|0\rangle$ and in the invariant kernel
(\ref{<g|h>}). In general, we have (\ref{ResDec}) and
\begin{multline}\label{resK}
\kappa(z) = {\textstyle\frac{1}{2}}2\pi i \,\, \textrm{res}_{{}_{\rho=-i(1-z)}} \Big[{\textstyle\frac{\rho^2}{2\pi^4}} 
K(0,i\rho; z) \Big]
\\
-
{\textstyle\frac{1}{2}}2\pi i \,\, \textrm{res}_{{}_{\rho=i(1-z)}} \Big[{\textstyle\frac{\rho^2}{2\pi^4}} 
K(0,i\rho; z) \Big],
\end{multline}
In (\ref{ResDec}) and (\ref{resK}) we have used the fact that the only poles of $K(0,l_1=i\rho; z)$ crossing the contour, are the simple poles at
$\rho=\mp i(1-z)$.

Concerning analytic continuation, we perhaps should make the following remark. 
Suppose we have an analytic function $f$ of complex $z$ in a  
domain $D \subset \mathbb{C}$. Let $\Omega$ be an open, simply connected, subset of $\mathbb{C}$ with compact closure 
$\overline{\Omega}$ and with a regular boundary $\Gamma = \partial \Omega$.
Let moreover $\Gamma$ be an oriented cycle, inducing orientation on $\Omega$ through the Cauchy integral formula,
so that $\Omega \cup \Gamma$ comopse an oriented chain.   
Let $W$ be an analytic function of complex $\rho,z \in D$, 
except for poles, which has no poles in the $\rho$-variable which cross the closed
contour $\Gamma$ when $z$ varies within a subdomain $D_1 \subset D$.
Suppose that for $z \in D_1$, $f$ has a contour integral form 
\begin{equation}\label{ResContinuation}
f(z) =  {\textstyle\frac{1}{2\pi i}} \int\limits_\Gamma W(\rho,z) d\rho
= \sum\limits_{{}_{i}} \underset{{}_{\Gamma}}{\textrm{ind}}\rho_i \,\,\, \textrm{res}_{{}_{\rho_i}} W_{{}_{z}}, 
\,\,\,\,\, W_{{}_{z}}(\rho) = W(\rho,z),
\end{equation}
equal, by the Cauchy theorem, to the sum of all residues in $\Omega$, counted with the indices
induced by the orientation of $\Gamma$ (with residues, which should be understood 
as residues of the one-form $W_{{}_{z}} d\rho$ rather than of the function $W_{{}_{z}}$, 
in order for this construction to make sense on any Riemann surface).
It is obvious that the same integral is in general not analytic in the whole $D$, and cannot represent $f$ in the whole domain $D$, 
as for $z$ going outside $D_1$, some residues of the one-form $W_{{}_{z}} d\rho$, 
will move outside $\Gamma$, and others will move inside $\Gamma$,
causing the jump precisely equal to the difference between the sum of the residues going inside and the sum of residues, which goes outside. 
By definition, $\rho$ lies inside or outside $\Gamma$, iff $\rho \in \Omega$ or $\rho \notin \overline{\Omega}$, respectively.
In order to restore the analytic continuation of $f(z)$ for $z$ outside $D_1$, we need to add to the integral (\ref{ResContinuation})
all residues that come out, and subtract all that come in, of course all weighted with one and the same index 
fixed by the chosen $\Omega\cup \Gamma$-orientation. 
In this way we recover the initial sum (\ref{ResContinuation}) of residues, which represented the 
integral (\ref{ResContinuation}) for $z$  inside $D_1$, but now with the argument $z$ placed outside $D_1$. 
That this sum represents analytic continuation of $f$ immediately follows from the fact that the residua
are, by assumption, analytic functions of $z$.  
These considerations are valid for any Riemann surface, 
not just for $\mathbb{C}$. Passing to our case, our contour can be considered 
as a closed contour on the Riemann sphere $\mathbb{C} \cup \infty$. Next, we use the fact that only a finite number of poles
cross our contour when $z$ goes outside the domain $D_1: \textrm{Re} \, z >1$, $-\epsilon < \textrm{Im} \, z < \epsilon$. We separate our analytic
function $F: \rho,z \mapsto F(\rho,z)$ into two analytic parts: $F = W_1 + W$, with $W$ containing a finite number of poles, 
including all poles crossing the contour, and $W_1$, subsuming the rest of poles. 
$W$ can be chosen in the form of a finite sum of rational functions of the variables $\rho,z$ 
(the reader will find specific formulas below). Finally, we observe that the integral
\[
\int\limits_{-\infty}^{+\infty} W_1(\rho,z) d\rho
\]
with the poles of $W_1$ separated from our contour, represents an analytic function of $z$ in the full domain $\textrm{Re} \, z>0$,
$-\epsilon < \textrm{Im} \, z < \epsilon$.
To the part
\[
\int\limits_{-\infty}^{+\infty} W(\rho,z) d\rho
\]
we apply the same reasoning as to (\ref{ResContinuation}), because $W$, being a rational function, 
can be lifted to a meromorphic function on the Riemann sphere. As for the rule of signs, they are determined 
by the orientation of our contour and by the Cauchy theorem. In turn, the orientation of our contour 
follows from the construction of the integral $\int d\rho F(\rho,z)$ through a one-dimensional version of Stokes' theorem, 
in which $d\rho$ becomes the one-form, determined by the coordinate (or variavle) $\rho$, on the one-dimensional cycle (after adjusting $\infty$) with
orientation of the cycle directed from $-\infty$ to $+\infty$. This fixes the signs in the invariant form independent
of the choice of the open set $\Omega$, which, when regarded as a subset of the Riemann sphere $\mathbb{C}\cup \infty$,
can be chosen as the upper open hemisphere (with $\textrm{Im} \, \rho >0$) or as the lower open hemisphere 
(with $\textrm{Im} \, \rho < 0$). For the upper hemisphere $\Omega$ the residues in $\Omega$ get the index $+1$,
and thus all residues have to be counted with the weight $+1$ in the corresponding sum (\ref{ResContinuation}). 
Therefore, the residues coming out of $\Omega$ (``top down''), have to be added. 
If we choose the lower hemisphere for $\Omega$, the residues in $\Omega$ will get the index $-1$,
and thus all residues in the corresponding sum (\ref{ResContinuation}) have to be counted with the weight $-1$. 
Thus, the residues coming in $\Omega$ (``top down''), have to be added, as before.  We see that changing the hemisphere
does not change the rule of signs, given above.

We base our analysis of the cyclic representations on the general properties of the matrix elements of the kernel 
$K$ of decomposition (\ref{decomositionz>1}) and its analytic 
continuation to the domain $0<z<1$ for the kernel $\langle g|h\rangle$ associated to the cyclic vector of the general form (\ref{x}). These properties
are deeply related to their analyticity.  Before we formulate them, let us give explicit formulas for the 
particular kernel (\ref{<g|h>}) with $\alpha = (1,0)$ in (\ref{<g|h>}). We compute explicitly (\ref{Kalphaalpha}) 
for $l_0=0$, $\alpha =(1,0)$ and $z>1$, and then compute the analytic continuation of such $K(l_0=0,l_1;z)$. 
We show that $K(l_0,l_1;z) = 0$, for $|l_0|>0$, $z>0$, $\alpha=(1,0)$ in (\ref{<g|h>}). As we have already mentioned, the case $|l_0|>1$, $z>0$, 
follows from the orthogonality properties of the matrix elements of the irreducible representations 
of the $SU(2,\mathbb{C})$ group (Peter-Weyl theorem). 
It remains to show that in case $\alpha=(1,0)$, $K(l_0,l_1;z) = 0$ also for $|l_0|=1$, $z>0$. 

We compute the analytic continuation first.
Up to the overall factor $\mathfrak{e}^2$ in front of (\ref{<g|h>}) and (\ref{Kalphaalpha}) we have
\begin{equation}\label{K10}
\big[K(0,i\rho,z)\big]_{{}_{\alpha \,\,\, \alpha}} = 
{\textstyle\frac{8\pi^4 e^z}{\rho^2+1}} \, f_{{}_{1}}(\rho,z) 
+ {\textstyle\frac{16\pi^4 e^z}{\rho^2+1}} \, f_{{}_{2}}(\rho,z) 
+ {\textstyle\frac{2\pi^4 z e^z}{\rho^2+1}} \, f_{{}_{3}}(\rho,z), 
\end{equation}
for each $\alpha$ with $l(\alpha)=1$, where
\begin{multline*}
f_{{}_{1}}(\rho,z)  = 
\int\limits_{0}^{+\infty} \left( {\textstyle\frac{\textrm{sinh}(2\lambda)-2\lambda}{\textrm{sinh} \lambda}} \right)
\left(
{\textstyle\frac{\textrm{cosh} \lambda\textrm{sin}(\rho\lambda)}{\rho \textrm{sinh} \lambda}} - \textrm{cos} (\rho \lambda)
\right)
{\textstyle\frac{e^{-z\lambda\textrm{coth} \lambda}}{\textrm{sinh}^2 \lambda}}
\, d\lambda,
\\
f_{{}_{2}}(\rho,z)  = 
\,\,\,\,\,\,\,\,\,\,\,\,\,\,\,\,\,\,\,\,\,\,\,\,\,\,\,\,\,\,\,\,\,\,\,\,\,\,\,\,\,\,\,\,\,\,\,\,\,\,\,\,\,\,\,\,\,\,\,\,\,\,\,\,
\,\,\,\,\,\,\,\,\,\,\,\,\,\,\,\,\,\,\,\,\,\,\,\,\,\,\,\,\,\,\,\,\,\,\,\,\,\,\,\,\,\,\,\,\,\,\,\,\,\,\,\,\,\,\,\,\,\,\,\,\,\,\,\,
\,\,\,\,\,\,\,\,\,\,\,\,\,\,\,\,\,\,\,\,\,\,\,\,\,\,\,\,\,\,\,\,\,\,\,\,\,\,\,\,\,\,\,\,\,\,\,\,\,\,\,\,\,\,\,\,\,\,\,\,\,\,\,\,
\,\,\,\,\,\,\,\,\,\,\,\,\,\,\,\,\,\,\,\,\,\,\,\,\,\,\,\,\,\,\,\,\,\,\,\,\,\,\,\,\,\,\,\,\,\,\,\,\,\,\,\,\,\,\,\,\,\,\,\,\,\,\,\,
\\
\int\limits_{0}^{+\infty} {\textstyle\frac{\left[(\rho^2+1)\textrm{sinh}\lambda \textrm{sin} (\rho \lambda)
+2\rho \textrm{cosh}\lambda \textrm{cos} (\rho \lambda) 
- 2\textrm{cosh}\lambda \textrm{coth}\lambda \textrm{sin} (\rho \lambda)
\right](\lambda\textrm{coth} \lambda-1)e^{-z\lambda\textrm{coth} \lambda}}{\rho \textrm{sinh}^2 \lambda}} \,  d\lambda
\\
f_{{}_{3}}(\rho,z)  = 
\,\,\,\,\,\,\,\,\,\,\,\,\,\,\,\,\,\,\,\,\,\,\,\,\,\,\,\,\,\,\,\,\,\,\,\,\,\,\,\,\,\,\,\,\,\,\,\,\,\,\,\,\,\,\,\,\,\,\,\,\,\,\,\,
\,\,\,\,\,\,\,\,\,\,\,\,\,\,\,\,\,\,\,\,\,\,\,\,\,\,\,\,\,\,\,\,\,\,\,\,\,\,\,\,\,\,\,\,\,\,\,\,\,\,\,\,\,\,\,\,\,\,\,\,\,\,\,\,
\,\,\,\,\,\,\,\,\,\,\,\,\,\,\,\,\,\,\,\,\,\,\,\,\,\,\,\,\,\,\,\,\,\,\,\,\,\,\,\,\,\,\,\,\,\,\,\,\,\,\,\,\,\,\,\,\,\,\,\,\,\,\,\,
\,\,\,\,\,\,\,\,\,\,\,\,\,\,\,\,\,\,\,\,\,\,\,\,\,\,\,\,\,\,\,\,\,\,\,\,\,\,\,\,\,\,\,\,\,\,\,\,\,\,\,\,\,\,\,\,\,\,\,\,\,\,\,\,
\\
\int\limits_{0}^{+\infty}
 {\textstyle\frac{\left[(\rho^2+1)\textrm{sinh}\lambda \textrm{sin} (\rho \lambda)
+2\rho \textrm{cosh}\lambda \textrm{cos} (\rho \lambda) 
- 2\textrm{cosh}\lambda \textrm{coth}\lambda \textrm{sin} (\rho \lambda)
\right](\textrm{sinh}(2\lambda)-2\lambda)^2e^{-z\lambda\textrm{coth} \lambda}}{\rho\textrm{sinh}^4\lambda}} \, d\lambda
\end{multline*}
We construct the analytic continuation in two steps. In the first step, we construct a series of simple fractions which
pointwisely (in the variables $\rho,z$, within the domain $|\textrm{Im} \, \rho|<\epsilon$, $z>1+\epsilon$, for any positive $\epsilon$) 
converges to $K$. In the second step, we prove the absolute and almost uniform
convergence of this series for $(\rho,z) \in \mathbb{C}\times\mathbb{C}$, which proves that indeed it represents the required analytic continuation of $K$, 
regarded as an analytic function of the two complex variables $\rho,z$. We do it separately for each integral $f_1,f_2,f_3$ using a method
which can be applied to more general $K$ associated with cyclic (\ref{x}). 
In order to illustrate the general method used in the first step, let us consider the simple integral
\[
f_{{}_{0}}(z) = \int\limits_{0}^{+\infty} e^{-z\lambda\textrm{coth} \lambda} \, d\lambda.
\]   
Then we use the new variable $t=e^{-z\lambda}$, which converts the integral into the following
\[
{\textstyle\frac{1}{z}}\int\limits_{0}^{1}\exp\left[{\textstyle\frac{2t^{2/z}}{1-t^{2/z}}} \textrm{ln}t \right] \, dt,
\]
in which the function of $t$ in the argument of $\textrm{exp}$ is bounded over the closed integration interval $[0,1]$. Because the 
exponential series is uniformly convergent over any bounded domain, and the integration domain is bounded, then insertion 
of the exponential series and changing the order of summation and integration operations is allowed:
\[
f_{{}_{0}}(z) = {\textstyle\frac{1}{z}} \sum\limits_{n=0}^{+\infty} 
{\textstyle\frac{1}{n!}} \int\limits_{0}^{1}\left[{\textstyle\frac{2t^{2/z}}{1-t^{2/z}}} \textrm{ln}t\right]^n \, dt 
\]
In the last integrals we use the new variable $s=-\textrm{ln} t$, which, eventually, after simple rescaling of the variable,  
converts these integrals into
\[
\int\limits_{0}^{1}\left[{\textstyle\frac{2t^{2/z}}{1-t^{2/z}}}\textrm{ln}t\right]^n \, dt 
= (-2)^nz^{n+1} \int\limits_{0}^{+\infty} s^n e^{-(2n+z)s} {\textstyle\frac{1}{(1-e^{-2s})^n}} \, ds.
\]
In this integral we use the convergence
\[
{\textstyle\frac{1}{(1-e^{-2s})^n}} = \sum\limits_{k=0}^{\infty} {\textstyle\binom{k+n-1}{n-1}}e^{-2ks}
\]
for  all $s>0$. Finally the Lebesgue dominated convergence principle is applicable to the last integral with
the above series representation of the last factor in it, which again allows changing the integration and summation
over $k$. Therefore, using the identity
\[
n! = \int\limits_{0}^{+\infty} s^n e^{-s} \, ds
\]
we finally get
\[
f_{{}_{0}}(z) = {\textstyle\frac{1}{z}} + \sum\limits_{n=1}^{+\infty}\sum\limits_{k=0}^{+\infty} {\textstyle\binom{k+n-1}{n-1}}
{\textstyle\frac{(-2z)^n}{(2k+2n+z)^{n+1}}} = {\textstyle\frac{1}{z}} 
-2z\sum\limits_{m=1}^{+\infty} {\textstyle\frac{(2m-z)^{m-1}}{(2m+z)^{m+1}}}, 
\]
where in this case the absolute and almost uniform convergence of the double series is easily seen, and which allows reorganized summation 
(first with respect to all pairs $(n,k)$ for which $n+k=m$ and then with respect to $m$) giving the last equality. Exactly the same changing of variables,
with the same argument, allowing changing the order of integration over compact set and summation for uniformly convergent series, and dominated
convergence principle, is applicable to our general integrals (including $f_{{}_{k}}$, $k=0,1,2,3$)
\begin{equation}\label{GeneralInt}
f(\rho,z) = \int\limits_{0}^{+\infty} g(\rho,z,\lambda)e^{-z\lambda\textrm{coth} \lambda} \, d\lambda,
\end{equation}  
representing matrix elements of $K$, and thus, $\textrm{Tr} \,K$, associated with general cyclic (\ref{x}) in which the function 
$g$ can be represented as 
\begin{equation}\label{Generalgk}
g(\rho,z,\lambda) 
= {\textstyle\frac{\sum\limits_{p\geq 0, j_p\geq-1} \left[a_{{}_{p,j_p}}^{{}^{+}} \,\, \lambda^p e^{-j_p\lambda}e^{i\rho\lambda} +a_{{}_{p,j_p}}^{{}^{-}} \,\, \lambda^p 
e^{-j_p\lambda}e^{-i\rho\lambda}\right]
}{(1-e^{-2\lambda})^{q+1}}},
\end{equation}
with integer $q\geq -1$ and finite sum over integer $p\geq 0$ and $j_p\geq -1$, 
and with $a_{{}_{p,j_p}}^{{}^{\pm}}$ being polynomial functions of $\rho,z$. 
The coefficients $a_{{}_{p,j_p}}^{{}^{\pm}}$ and powers of the exponent $e^{-\lambda}$ 
are such that $g$ is analytic in $\rho,\lambda$, which, in some nonempty 
domain $|\textrm{Im} \, \rho| \leq \epsilon$, $\textrm{Re} \, z>1+\epsilon$ of the convergence of the integral, it
has no singular points and the asymptotic $\sim e^{s\lambda}$, $s \leq 1+\epsilon$ in $\lambda$ at infinity, 
so that the zero of order $q+1$ in $\lambda$ at $\lambda=0$ in the denominator cancels out with the zero of order $\geq q+1$ of the numerator, 
and the above construction of the series expansion for (\ref{GeneralInt}) can be applied. Indeed, (\ref{oddB}), (\ref{evenB}) and the matrix elements 
(\ref{explicitA}),(\ref{explicitU^(l0,l1)}) all have the form (\ref{Generalgk}) with $a_{{}_{p,j_p}}^{{}^{\pm}}$ being polynomial functions of 
$\rho,z$, divided by the common polynomial $\rho Q(\rho)$ in $\rho$, coming from the matrix 
elements (\ref{explicitU^(l0,l1)}). Their products, containing at most one factor (\ref{explicitU^(l0,l1)}), 
again have the same form (\ref{Generalgk}), with $a_{{}_{p,j_p}}^{{}^{\pm}}$ being polynomial functions of $\rho,z$, divided
by the polynomial $\rho Q(\rho)$, coming from the matrix elements (\ref{explicitU^(l0,l1)}). 
But any matrix element of $K$ corresponding to a cyclic (\ref{x}) is given by the integral (\ref{GeneralInt}) 
with $g$ equal to a finite sum of products
of the functions (\ref{oddB}), (\ref{evenB}),(\ref{explicitA}),(\ref{explicitU^(l0,l1)}) and $\textrm{sinh}^2 \lambda$, with the factor of the type
(\ref{explicitU^(l0,l1)}) entering each summand exactly once. 
Thus,  $g$ in (\ref{GeneralInt})  representing matrix element of $K$, has the form (\ref{Generalgk}), divided by the polynomial $\rho Q(\rho)$. 
For (\ref{Generalgk}) the integral (\ref{GeneralInt}) has the following series representation
\begin{multline}\label{GeneralSeries}
f(\rho,z) = \sum\limits_{n=0}^{+\infty}\sum\limits_{k=0}^{+\infty} b_{{}_{n,k}}(\rho,z) =
\sum\limits_{n=0}^{+\infty}\sum\limits_{k=0}^{+\infty}
(-2z)^n  {\textstyle\binom{k+n+q}{n+q}} \sum\limits_{p\geq 0, j_p\geq-1} \Bigg[
\\
{\textstyle\frac{a_{{}_{p,j_p}}^{{}^{+}} \,\,\, (n+1)\ldots(n+p)}{(2k+2n+j_p+z-i\rho)^{n+p+1}}} 
+ {\textstyle\frac{a_{{}_{p,j_p}}^{{}^{-}} \,\,\, (n+1)\ldots(n+p)}{(2k+2n+j_p+z+i\rho)^{n+p+1}}}
\Bigg],
\end{multline}
where the product $(n+1)\ldots(n+p)$ is by definition equal $1$ for $p=0$.

Returning to (\ref{<g|h>}) with $l(\alpha)=1$, we observe
that the integrals $f_1,f_2$ are convergent for all $z>0$, 
so they will have no contribution to the
residue crossing the contour. 

\begin{lem}
Each of the double series (\ref{GeneralSeries}), 
representing matrix elements of $K$, is absolutely and uniformly convergent on any closed domain 
in $\mathbb{C}\times\mathbb{C}$ of the complex variables $\rho,z$ 
in which $|j_p+z \mp i \rho|\leq R$, for any finite $R>0$, except
eventually the finite number of terms of the series, which have 
a pole in this domain or at its boundary. In particular, the series (\ref{GeneralSeries})
representing $K$ are analytic functions of $\rho,z$, having poles of finite order, 
without essential singularities.
\label{Lconvergence}
\end{lem} 
\qedsymbol \,
We ignore in the proof the common factor $\tfrac{1}{\rho Q(\rho)}$, coming from $U^{{}^{l_0,i\rho}}$, in $ b_{{}_{n,k}}(\rho,z)$,
which is irrelevant for the convergence claimed in the lemma. 
General term $b_{{}_{n,k}}$ of the series is of the form
\begin{equation}\label{ess-bn,k}
(-2z)^n {\textstyle\binom{k+n+q}{n+q}} \big[ \ldots \big], 
\end{equation}
where dots denote the corresponding finite sum of simple fractions with denominators $(2k+2n+j_p+z \mp i \rho)^{n+p}$, $p=1,2, \ldots, p_{{}_{\textrm{max}}}$. 
Let us replace the expression $2k+2n$ in each denominator by a variable $x$. Then, after reducing the sum in $[\ldots]$ 
to a common denominator $D$,  $b_{{}_{n,k}}$ can be written as
\[
(-2z)^n {\textstyle\binom{k+n+q}{n+q}} \big[{\textstyle\frac{N(x=2k+2n)}{D(x=2k+2n)}}\big]
\]  
with some polynomials $N$ and $D$. If 
\begin{equation}\label{DegreeCondition}
\textrm{degree} \, [D(x)] - \textrm{degree}[x^{n+q}N(x)] \geq 2,
\end{equation}
with $x^{n+q}$ in the second term coming from the estimation 
\begin{equation}\label{binomEstim}
\left({\textstyle\frac{n}{k}}\right)^k < {\textstyle\binom{n}{k}} < \left({\textstyle\frac{ne}{k}}\right)^k,
\end{equation}
then the remainder, counting all terms of the series $|b_{{}_{n,k}}|$ with $n+k\geq R$, is easily seen to be bounded by a remainder of a convergent
series with the bound independent of the particular choice of $\rho,z$ in the assumed domain. The degree condition (\ref{DegreeCondition}) 
can be checked ``by hand on paper'' only for the simplest double series, e.g. the first three simplest subseries in $f_3$. It would be difficult 
to check it ``by hand'' even for some  of the  subseries in $f_3$, and for the general term of the series (\ref{GeneralSeries}). 
But we can use the observation: the condition (\ref{DegreeCondition}) follows from the pointwise convergence of the series 
in the domain $|\textrm{Im} \, \rho| <\epsilon$, 
$z>1$, where the integral, represented by the series, is convergent. In this domain of the variables $\rho,z$,
$|b_{{}_{n,k}}(\rho,z)|$ goes to zero faster than $\tfrac{1}{n^2}\tfrac{1}{(n+k+q)^3}$. 
Indeed, let $\mathfrak{l}(\rho,z,\lambda)$ be the numerator of $g$ in the integral 
representation (\ref{GeneralInt}) of the series (\ref{GeneralSeries}).
Then $b_{{}_{n,k}}(\rho,z)$ is given by the Laplace integral
\[
b_{{}_{n,k}}(\rho,z) = {\textstyle\frac{(-2z)^n}{n!}}{\textstyle\binom{k+n+q}{n+q}} 
\int\limits_{0}^{+\infty} \mathfrak{l}(\rho,z,\lambda) \lambda^n e^{-(z+2n+2k)\lambda} \, d\lambda,
\]
in which $\mathfrak{l}$ has zero of order $\geq q+3$ at $\lambda=0$, and, thus, 
$\mathfrak{l} \cdot \lambda^n$ has zero of order $\geq n+q+3$. Therefore, we have the asymptotic
(\cite{Sidorov}, Theorem 1, p. 398)
\[
b_{{}_{n,k}}(\rho,z) \widesim[3]{n,k \rightarrow +\infty} 
{\textstyle\frac{1}{(z+2n+2k)^{n+q+3}}}{\textstyle\frac{(-2z)^n}{n!}}{\textstyle\binom{k+n+q}{n+q}},
\] 
if $\rho,z$ are in the domain of pointwise convergence of the integral (\ref{GeneralInt}).
Using this asymptotic and the estimation (\ref{binomEstim}),
it is easily seen that $|b_{{}_{n,k}}(\rho,z)|$ 
goes to zero faster than $\tfrac{1}{n^2}\tfrac{1}{(n+k+q)^3}$, if $\rho,z$ 
are in the domain of pointwise convergence of the integral (\ref{GeneralInt}). 
Using this, we proceed as follows. Let $L_{{}_{d}}$ be the coefficient 
of degree $d$ of the polynomial $L(x) = x^{n+q}N(x)$, with $d\geq \textrm{degree} \, [D(x)]-1$. 
We observe that if $L_{{}_{d}}\neq 0$, the components $b_{{}_{n,k}}$
would be going to zero slower than $1/k^3$ for each fixed $n$ -- contradiction.
Therefore, $L_{{}_{d}}=0$ for all $z>1$, in the domain of convergence of the integral. 
$L_{{}_{d}}$ has the form of a polynomial in $\rho,z$. For each fixed $\rho$,
$L_{{}_{d}}$ is a polynomial in $z$, with coefficients polynomially depending on $\rho$. 
A polynomial can have a continuum many zeros only if it is identically zero. Thus, for each $|\textrm{Im} \, \rho| <\epsilon$, 
all coefficients of the polynomial $L_{{}_{d}}$ in $z$, depending on 
$\rho$, are zero. Therefore, $L_{{}_{d}}(\rho,z)=0$ identically for all $\rho,z \in \mathbb{C}$.    
\qed

We have finished the construction of the analytic continuation of $K(0,i\rho;z)$ applicable generally to all $K(l_0,i\rho;z)$ corresponding
to cyclic (\ref{x}).

Now we show that in case $l(\alpha)=1$, $K(l_0,l_1;z) = 0$ also for $|l_0|=1$, and  for all $z>0$. 
Up to the overall factor $\mathfrak{e}^2$ in front of (\ref{<g|h>}) and (\ref{Kalphaalpha}) we have
\[
\big[K(1,i\rho,z)\big]_{{}_{\alpha \,\,\, \alpha}} = \int\limits_{0}^{+\infty}g(\rho,z;\lambda) \, d\lambda, \,\,\, l(\alpha)=1,
\]
convergent for all $z>0$, with the integrand
\begin{multline*}
g(\rho,z;\lambda)= {\textstyle\frac{2\pi^4e^z}{\rho(\rho^2+1)}}\Big[
{\textstyle\frac{(\textrm{sinh}(2\lambda)-2\lambda)\left(
-2\textrm{sin} (\rho \lambda) + 2\rho^2\textrm{sinh}^2\lambda \textrm{sin} (\rho \lambda)
+ \rho\textrm{sinh} (2\lambda)\textrm{cos} (\rho \lambda)
\right)}{\textrm{sinh}^2\lambda}}
\\
+ \left(
8(\lambda\textrm{coth}\lambda-1)-z\left({\textstyle\frac{\textrm{sinh}(2\lambda)-2\lambda}{\textrm{sinh}\lambda}}\right)^2
\right)
\left(
\textrm{coth}\lambda\textrm{sin} (\rho \lambda) -\rho\textrm{cos} (\rho \lambda)
\right)
\Big]
{\textstyle\frac{e^{-z\lambda\textrm{coth}\lambda}}{\textrm{sinh}^2\lambda}}
\end{multline*}
which has the antiderivative of the form
\begin{multline*}
G(\rho,z;\lambda) = {\textstyle\frac{8\pi^4e^z}{\rho(\rho^2+1)}} e^{-z\lambda\textrm{coth}\lambda}
\Big[
\textrm{sin} (\rho \lambda) +
{\textstyle\frac{\rho\lambda\textrm{cos} (\rho \lambda)+\textrm{sin} (\rho \lambda)}{\textrm{sinh}^2\lambda}}
\\
-\textrm{coth}\lambda
\left(
\rho\textrm{cos} (\rho \lambda) +{\textstyle\frac{\lambda\textrm{sin} (\rho \lambda)}{\textrm{sinh}^2\lambda}}
\right)
\Big] + C,
\end{multline*}
with $C$ being arbitrary constant. Because
\[
\underset{\lambda\rightarrow 0}{\textrm{lim}} G(\rho,z;\lambda) = \underset{\lambda\rightarrow +\infty}{\textrm{lim}} G(\rho,z;\lambda)= C, 
\]
then
\[
\big[K(1,i\rho,z)\big]_{{}_{\alpha \,\,\, \alpha}} = \int\limits_{0}^{+\infty}g(\rho,z;\lambda) \, d\lambda =0,
\]
for all $z>0$, $\rho\in \mathbb{R}$ and $l(\alpha)=1$. Because this holds both, for non-negative and negative $\rho$, and because
$U^{{}^{(1,i\rho)}} = U^{{}^{(-1,-i\rho)}}$, we have therefore shown that the Fourier transform
$K(l_0,i\rho,z)$, \emph{i.e.} (\ref{Kalphaalpha}), of the kernel (\ref{<g|h>}) with $l(\alpha)=1$,
is equal zero for all $z>0$, $\rho \geq 0$, $|l_0|>0$. Now we are ready to formulate 
\begin{lem} 
Let $\langle g|h\rangle$ be the kernel (\ref{<g|h>}) with $l(\alpha)=1$ and $f,f'$ smooth functions of compact support on $G$. 
We have the following decomposition valid for all $z>0$
\begin{multline}\label{10decompositionJW}
\langle f|f'\rangle =  \int\limits_{{}_{G \times G}} dg \, dh \,\, \langle g | h \rangle f(h) \overline{f'(g)}
 =
\\
\int\limits_{0}^{+\infty} {\textstyle\frac{d\rho \, \mathfrak{e}^2\rho^2}{2\pi^4}} \,
\textrm{Tr} \, \left[\mathcal{F}f(0, i\rho) \,\, K(0,i\rho; z) \,\, \mathcal{F}f'(0,i\rho)^* \right]
\\
+{\textstyle\frac{2\mathfrak{e}^2\pi e^z(1-z)(3-z)}{2-z}} 
\textrm{Tr} \, \left[\mathcal{F}f(0,1-z) \,\, \kappa \,\, \mathcal{F}f'(0,1-z)^* \right],
\end{multline}
where the last term is present only if $0<z<1$, and where the only nonzero component of the matrix $\kappa$
is the component $\kappa_{{}_{\alpha \,\, \alpha}}=1$, and similarly the only nonzero component
of the positive matrix $K(0,i\rho; z)$ is the component (\ref{K10}) with $f{{}_{1}}$, $f_{{}_{2}}$, $f_{{}_{3}}$ given by the above
sums of absolutely and almost uniformly convergent, in $\rho,z\in \mathbb{C}$, series.
\label{L1} 
\end{lem} 

We have also proved a lemma more general than lemma \ref{L1}, 
valid for the representations with cyclic vectors (\ref{x}), 
but in general, decomposition will also contain non-spherical representations:
\begin{lem}
Let $\langle g|h\rangle$ be the invariant kernel of the representation with cyclic vector (\ref{x}) and $q=1$,
$\alpha_1 = \alpha$. Then 
\begin{multline}\label{decomositionz>1'}
\langle f|f'\rangle =  \int\limits_{{}_{G \times G}} dg \, dh \,\, \langle g | h \rangle f(h) \overline{f'(g)}
 =
\\
  \sum\limits_{l_0=-l(\alpha)}^{l(\alpha)}\int\limits_{0}^{+\infty} \, d\rho \, {\textstyle\frac{l_0^2+\rho^2}{2\pi^4}} 
\textrm{Tr} \, \left[\mathcal{F}f(l_0,i\rho)K(l_0,i\rho; z)\mathcal{F}f'(l_0,i\rho)^* \right]
\\
+\textrm{Tr} \, \left[\mathcal{F}f(0,1-z) \,\, \kappa(z) \,\, \mathcal{F}f'(0,1-z)^* \right],
\end{multline}
with the sum over integer $l_0$. The last term is present only if $0<z<1$.
$K(l_0,l_1=i\rho;z), \kappa(z)$ are finite dimensional positive definite matrices with the only nonzero
coefficient $K_{{}_{\alpha \,\, \alpha}}, \kappa_{{}_{\alpha \,\, \alpha}}$, analytic in $\rho,z$. 
The nonzero matrix element $\left[K(l_0,l_1=i\rho;z)\right]_{{}_{\alpha \,\,\, \alpha}}$  
is given by the absolutely and almost uniformly convergent, in $\rho,z\in \mathbb{C}$,
series of the general form (\ref{GeneralSeries}), divided by $\rho Q(\rho)$, where $Q(\rho)$ 
is the polynomial in $\rho$, coming from the denominator in the formula (\ref{explicitU^(l0,l1)})
with $l=l(\beta), l' = l(\gamma)$.  $\kappa(z)$ is given by the residue formula (\ref{resK}).  
\label{L3}
\end{lem}

\section{Relation between the cyclic representations. Example of cyclic $x=|u\rangle$ and $x=c_{1,0}^+|u\rangle$}\label{RelationCycRep}

From lemma \ref{L1} it follows that the representation with the cyclic vector $x = |e\rangle =c_{\alpha}^+e^{-inS(u)}|0\rangle$, and with $\alpha=(1,0)$
decomposes into direct integral of the spherical representations $(l_0=0, l_1=i\rho)$, $\rho>0$, of the principal series if $z>1$,
and in addition it has the supplementary component $(l_0=0, l_1 = 1-z)$ entering as a discrete direct summand, if $0<z<1$. 
The absence of non-spherical representations
in lemma \ref{L1} independently follows from 
\begin{lem}
\[
c_{{}_{1,0}}^+|u\rangle \in \mathcal{H}_{{}_{|u\rangle}}.
\]
\label{Hc10+|u>InH|u>}
\end{lem}


\qedsymbol \, 
Let $U\big(g_{{}_{03}}(\lambda)\big) = U(\lambda)$. 
From the Baker-Campbell-Hausdorff formula, commutation rules (\ref{CommutationRules}), 
(\ref{Vacuum}), and the second transformation
rule in (\ref{c'S'}), we get
\begin{multline}\label{U(lambda)|u>}
U(\lambda)|u\rangle = U(\lambda)e^{-inS(u)}|0\rangle 
\\
= e^{-{\textstyle\frac{n^2}{8\pi}}||B(\lambda)||^2}e^{-inS(u)} 
\textrm{exp}\left[{\textstyle\frac{n}{4\pi \mathfrak{e}}\sum\limits_{l}B_{{}_{l,0}}(-\lambda)}c_{{}_{l,0}}^+\right]|0\rangle.
\end{multline}
Differentiating (\ref{U(lambda)|u>}) at $\lambda =0$, we obtain a non-zero 
vector proportional to $c_{{}_{1,0}}^+|u\rangle \in \mathcal{H}_{{}_{|u\rangle}}$, becuse of
the expansion (\ref{Bexpansion}).
\qed

Lemma \ref{Hc10+|u>InH|u>} can still be strengthened 
using orthogonality relations of the matrix elements of the representations $(l_0,l_1)$.

\begin{lem}
The representations with cyclic $x=|u\rangle$ and $x=c_{1,0}^+|u\rangle$ coincide.
\label{LCyc|u>andc10+|u>}
\end{lem}
\qedsymbol \,
We ``smear out''
the Lorentz transformed cyclic state $|g\rangle= U(g)c_{{}_{1,0}}^+|u\rangle$ with a function $f$, Fourier transform 
$\mathcal{F}f$ of which has prescribed support.
In fact in case $z>1$ we can use generalized state $f$ with a single point support $\{(l_0,l_1)\}$. Namely, we consider first
\[
|f\rangle = \int f(g) |g\rangle dg,
\,\,\,\,
f(g) = \int\limits_{I_{{}_{\epsilon}}} U^{{}^{(l_0=0,l_1=i\rho)}}_{{}_{0,0 \,\,\, 1,0}}(g) \, 
d\rho, \,\,\,\ I_{{}_{\epsilon}}=[\rho-\epsilon/2, \rho+\epsilon/2]
\]
This state, decomposed accordingly to the last lemma, has nonzero decomposition components only at
$(l_0,l_1) = (0,i\rho)$, $\rho \in I_{{}_{\epsilon}}$.  Next we compute 
\begin{equation}\label{CyclicProj1}
\underset{\epsilon \rightarrow 0}{\textrm{lim}} \,\, {\textstyle\frac{1}{\epsilon}}\langle u |f\rangle
={\textstyle\frac{4\pi^{7/2}\sqrt{z}e^z}{3}} \int\limits_{0}^{+\infty} \overline{B_{{}_{1,0}}(\lambda)} U^{{}^{(l_0=0,i\rho)}}_{{}_{0,0 \,\,\, 1,0}}(\lambda)
\,\, e^{-z\lambda \textrm{coth} \lambda} \, \textrm{sinh}^2\,\lambda d\lambda,  
\end{equation}
obtaining analytic function of $\rho,z$ of the same kind as the matrix elemets of $K$, if we ignore the, irrelevant here, factor
$4\pi^{7/2}\sqrt{z}e^z/5$ in front of the integral (\ref{CyclicProj1}). 
We have used $z>1$, allowing echanging the $\epsilon$-limit
with integration (dominated convergence principle), commutation rules (\ref{CommutationRules}), (\ref{Vacuum}), transformation
rule (\ref{c'S'}) and orthogonality (\ref{Torthogonality}).
(\ref{CyclicProj}) can be represented by almost uniformly and absolutely
convergent double series of finite sums of simple fractions of the general form (\ref{GeneralSeries}), 
divided by the polynomial $\rho Q(\rho)$ in $\rho$,
and which can be explicitly computed with the general method given above. This function has the property that 
it is not identically equal to zero as a function of $\rho$ for each $z\neq 0$, in particulat for each $z>0$,  
understood as the analytic continuation for $0<z<1$. This can immediately be seen from the value
of the residues of (\ref{CyclicProj}) at $\rho = \pm i(z-1)$, which, in turn, can easily be read off from the general series 
(\ref{GeneralSeries}) representing (\ref{CyclicProj}), with the irrelevant factor in front of the integral
in  (\ref{CyclicProj}) ignored. Being analytic, it is, for each $z>0$, nonzero for almost all $\rho$.
Therefore, decomposition component of the projection of the state $|u\rangle$ on the domain 
$\mathcal{H}_{{}_{c_{{}_{1,0}}^+|u\rangle}}$ of the cyclic representation with cyclic vector
$c_{{}_{1,0}}^+|u\rangle$ is nonzero for $(l_0=0,l_1=i\rho)$ with almost all $\rho$. 
It means that the subspace spanned by the spherical representations in $\mathcal{H}_{{}_{c_{{}_{1,0}}^+|u\rangle}}$ is contained in
$\mathcal{H}_{{}_{|u\rangle}}$, if $z>1$. In the more general case $z>0$, it means that the subspace of $\mathcal{H}_{{}_{c_{{}_{1,0}}^+|u\rangle}}$ 
spanned by the spherical representations, which is orthogonal to the subspace spanned by the supplementary series, is contained in the subspace
of $\mathcal{H}_{{}_{|u\rangle}}$, orthogonal to the subspace spanned by the supplementary series.  
Analogously, we``smear out'' the state $|g\rangle = U(g)|u\rangle$, and consider the ``smeared'' state
\[
|f\rangle = \int f(g) |g\rangle dg,
\,\,\,\,
f(g) = \int\limits_{I_{{}_{\epsilon}}} U^{{}^{(l_0=0,l_1=i\rho)}}_{{}_{1,0 \,\,\, 0,0}}(g) \, 
d\rho, \,\,\,\ I_{{}_{\epsilon}}=[\rho-\epsilon/2, \rho+\epsilon/2]
\]
with the corresponding projection function
\begin{equation}\label{1CyclicProj}
\underset{\epsilon \rightarrow 0}{\textrm{lim}} \,\, {\textstyle\frac{1}{\epsilon}}\langle u|c_{{}_{1,0}}|f\rangle
=-{\textstyle\frac{4\pi^{7/2}\sqrt{z}e^z}{3}} \int\limits_{0}^{+\infty} B_{{}_{1,0}}(\lambda) 
U^{{}^{(l_0=0,i\rho)}}_{{}_{1,0 \,\,\, 0,0}}(\lambda)
\,\, e^{-z\lambda \textrm{coth} \lambda} \, \textrm{sinh}^2\,\lambda d\lambda,  
\end{equation}
and show that decomposition components of the projection of the state $c_{{}_{1,0}}^+|u\rangle$ on the domain $\mathcal{H}_{{}_{|u\rangle}}$
of the cyclic representation with the cyclic vector $|u\rangle$ are nonzero for almost all $\rho$ in the decomposition 
found in  \cite{Staruszkiewicz1992ERRATUM} (with decomposition parameter $\rho$ denoted by $\nu$ in \cite{Staruszkiewicz1992ERRATUM}).
It means that the subspace of $\mathcal{H}_{{}_{|u\rangle}}$ 
spanned by the spherical representations, which is orthogonal to the subspace spanned by the supplementary series, is contained in the subspace
of $\mathcal{H}_{{}_{c_{1,0}^+|u\rangle}}$, orthogonal to the subspace spanned by the supplementary series. 
Therefore, the spherical parts of the representations with cyclic vectors $c_{{}_{1,0}}^+|u\rangle$ and $|u\rangle$, orthogonal to
the subspaces spanned by the supplementary series, do actually coincide for all $z>0$.
Because the non-spherical components of these cyclic representations are zero,
these representations do actually coincide if $z>1$. From  $\mathcal{H}_{{}_{c_{{}_{1,0}}^+|u\rangle}} \subset \mathcal{H}_{{}_{|u\rangle}}$
and lemma \ref{L1} it follows a still stronger statement, that they coincide for
all $z>0$. 
\qed

This is the method that we apply to the analysis of cyclic representations
with cyclic vectors (\ref{x}) and for their comparison.

Let us remark that, in principle, we can use the analytic continuations of the projection functions 
(\ref{CyclicProj1}), (\ref{1CyclicProj}) to compare also the supplementary components, 
without adhering to the expansions (\ref{U(lambda)|u>}) and (\ref{Bexpansion}). To this end we continue analytically 
(\ref{CyclicProj1}), (\ref{1CyclicProj}) to the value $\rho = -i(z-1)$ or $\rho = i(z-1)$, with $0<z<1$. 
The subtle point is that these values of $\rho$
are in general among the places where the projection functions (\ref{CyclicProj1}), (\ref{1CyclicProj}) (and their analogues with
$c_{{}_{1,0}}$ replaced with $c_{{}_{l,0}}$) have residues. This reflects the fact that the matrix elements of the supplementary series are
not square summable on $G$, even after integration over the continuous parameter of the series. Equivalently, the ``smeared out'' 
state, $|f\rangle$, as it stands in (\ref{CyclicProj1}) and (\ref{1CyclicProj}), is not normalizable for the values of $\rho$ 
lying in neighborhoods of the points $\pm i(z-1)$, with $0<z<1$, uderstood as the analytic continuation of the above-given formula for $|f\rangle$. 
This shortcoming can be, as can easily be seen, repaired by multiplying 
the state $|f\rangle$, or the ``smearing'' function $f$, by $(\rho + i(z-1))$ or, respectively, by  $(\rho -i(z-1))$, 
which removes the corresponding singularity and gives a normalizable state $ |(\rho \pm i(z-1))f\rangle$ 
in a vicinity of the corresponding $\rho = \mp i(z-1)$. In other words
\begin{rem}
For the cyclic representations with cyclic vectors, respectively, $x = |u\rangle$ and $x= c_{{}_{l,0}}^+|u\rangle$, 
the mutual projections of the supplementary components are nonzero, 
and thus these components coincide, iff the residue at $\rho = -i(z-1)$, or at $\rho = i(z-1)$, with $0<z<1$, of the analytic 
continuation of the projection functions, (\ref{CyclicProj}), (\ref{lCyclicProj}), is nonzero.
\label{ResidueComponentsComparison}      
\end{rem}
In the next Section we show that these residues are all nonzero.

Instead of calculating these residues, we can still use yet another method. 
Namely, in \cite{Staruszkiewicz2009} a spherically symmetric normalized state 
\[
|u,\infty\rangle = \underset{\lambda\rightarrow +\infty}{\textrm{lim}} \, 
c_{{}_{\lambda}} \, \int\limits_{{}_{SU(2,\mathbb{C})}} da_1 \,\, U(a_1)U(\lambda)|u\rangle, 
\]
was constructed, well-defined for real $z$ if and only if $0<z<1$, and lying in the supplementary component of 
$\mathcal{H}_{{}_{|u\rangle}}$, with the normalization factor $c_{{}_{\lambda}}$ equal
\[
c_{{}_{\lambda}} = {\textstyle\frac{\sqrt{2} \, \textrm{sinh} \, \lambda}
{\left[\int\limits_{0}^{2\lambda}dx \,\, \textrm{sinh}\, x e^{-z(x\textrm{coth}\, x-1)}\right]^{1/2}}}.
\]
Instead of calculating the residues of remark \ref{ResidueComponentsComparison}, we can calculate
projections of the Lorentz transforms $U(\sigma)c_{{}_{l,0}}^+|u\rangle \in \mathcal{H}_{{}_{c_{{}_{l,0}}^+|u\rangle}}$ 
of the cyclic vectors $c_{{}_{l,0}}^+|u\rangle$ on the normalized bound state $|u,\infty\rangle$.
Here $U(\sigma) = U\big(g_{{}_{03}}(\sigma)\big)$.

\begin{lem}
\[
\langle u, \infty|U(\sigma)c_{{}_{l,0}}^+|u\rangle, \,\,\,\,\, l=1,2, \ldots,
\,\,\, 0 < z < 1,
\] 
is nonzero for almost all hyperbolic angles $\sigma$.  
\label{BoundStateComponentsComparison}  
\end{lem}
\qedsymbol \,
It is not difficult to show that the one-parameter group $\sigma \mapsto U(\sigma)$ is strongly continuous.
Because by construction $|0\rangle$, $e^{iS(u)}|0\rangle$, belong to the domains of polynomials in $c_{{}_{l',0}}^+$,
$l'=1,2, \ldots$, then also $c_{{}_{l,0}}^+|u\rangle$ belongs to the domain of $M_{{}_{03}}^n$, $n \in \mathbb{N}$, 
compare \cite{Staruszkiewicz1995}, \cite{obata-book}, and beginning of Subsection \ref{Consistency}. 
Using (\ref{CommutationRules}), (\ref{Vacuum}), (\ref{c'S'}) and the exponentiation $U(\sigma)$ of $i\sigma M_{{}_{03}}$, we show that 
$h_{{}_{\infty,l}}(\sigma) = \langle u, \infty|U(\sigma)c_{{}_{l,0}}^+|u\rangle$ 
is an analytic function of $\sigma$, given by a power series around $\sigma =0$ with the covergence radius $R >0$. 
Let $|u_l\rangle = c_{{}_{l,0}}^+|u \rangle$.
Using (\ref{CommutationRules}), (\ref{Vacuum}), (\ref{c'S'}), we obtain analytc function
\begin{equation}\label{hll}
 h_{{}_{l,l}}(\sigma) = \langle u_l |U(\sigma)|u_l\rangle = \pi z \, e^{-z(\sigma \coth \sigma -1)} 
\left[
4\pi + (-1)^{l+1} B_{{}_{l,0}}(\sigma)^2
\right],
\end{equation}
represented by a power series around $\sigma =0$, with the convergence radius equal $\pi$. 
By theorem 13.35 of \cite{Rudin}, $d^nh_{_{l,l}}/d\sigma^n(0) = i^n \langle u_l| M_{{}_{03}}^n |u_l\rangle$. Because
the power series $h_{{}_{l,l}}$ has convergence radius $\pi$, 
we see that $\underset{n\rightarrow +\infty}{\textrm{lim sup}} \, (c'_{{}_{n}}/n!)^{1/n} = 1/\pi$, 
on using the Cauchy-Hadamard formula for convergence radius of a power series around 
$\sigma=0$, where $c'_{{}_{n}} = |\langle u_l| M_{{}_{03}}^n u_l \rangle|$. By the Cauchy-Schwartz inequality 
\[
c_n {:}= |\langle u, \infty |M_{{}_{03}}^n u_l \rangle| \leq || u, \infty|| \cdot | \langle u_l |M_{{}_{03}}^{2n} u_l \rangle|^{1/2}
= |\langle u_l| M_{{}_{03}}^{2n} u_l \rangle|^{1/2} = \sqrt{c'_{{}_{2n}}}.
\]
 From this,  and by Stirling's approximation formula,
\begin{multline*}
\underset{n\rightarrow +\infty}{\textrm{lim sup}} \, (c_{{}_{n}}/n!)^{1/n}
\leq \underset{n\rightarrow +\infty}{\textrm{lim sup}} \, \left(\sqrt{c'_{{}_{2n}}}/n!\right)^{1/n}
\\
= 2 \, \underset{n\rightarrow +\infty}{\textrm{lim sup}} \, (c'_{{}_{2n}}/(2n)!)^{1/(2n)} = 2/\pi.
\end{multline*} 
Thus, by the Cauchy-Hadamard formula, the convergence radius $R$ of the power series around $\sigma=0$ for $h_{{}_{\infty,l}}$ is 
greater than $\pi/2$. 
 
To show our lemma, it is sufficient to determine the lowest-order terms in the expansion of
$h_{{}_{\infty,l}}= \langle u, \infty|U(\sigma)c_{{}_{l,0}}^+|u\rangle$ with respect to $\sigma$ at $\sigma=0$, and show that they are nonzero.
In fact, it is sufficient to determine the lowest-order term and the behavior of $h_{{}_{\infty,l}}$ at $\sigma = \infty$.
Using the expansions (\ref{Aexpansion}), (\ref{Bexpansion}), (\ref{U(lambda)|u>}), the first transformation 
rule in (\ref{c'S'}), (\ref{Torthogonality}),  and (\ref{BatInfinity}), and the product formula 
\begin{equation*}
T^{{}^{l_1}}_{{}_{m_1 \,\, n_1}}(a)T^{{}^{l_2}}_{{}_{m_2 \,\, n_2}}(a)
= \sum\limits_{l=|l_1-l_2|}^{l_1+l_2}\sum\limits_{m,n=-l}^{l}
C^{{}^{l,m}}_{{}_{l_1,m_1 \,\,\, l_2,m_2}}T^{{}^{l}}_{{}_{m \,\, n}}(a)C^{{}^{l,n}}_{{}_{l_1,n_1 \,\,\, l_2,n_2}},
\end{equation*} 
with the Clebsch-Gordan coefficients $C^{{}^{l,n}}_{{}_{l_1,n_1 \,\,\, l_2,n_2}}$, and (\ref{Aexpansion}), (\ref{Bexpansion}),
it is not difficult to see that the contribution of the lowest $l$-order term is equal
\begin{equation}\label{l-orderBoundStateProj''}
\langle u, \infty|U(\sigma)c_{{}_{l,0}}^+|u\rangle 
= \textrm{sgn}(n)\sqrt{\pi}\sqrt{z}(1-z)
b_{{}_{l,0}}\left[ 1
+ r_{{}_{1}}(l) z \ldots + r_{{}_{l}}(l) z^l
\right] \, \sigma^l  + \ldots
\end{equation}
where dots denote terms of order $> l$ in $\sigma$, and where
\begin{multline*}
r_{{}_{1}}(l) 
= \sum\limits_{j=1}^{l}\textstyle{\binom{l}{j}}{\textstyle\frac{(-1)^j }{j(j+1)}}
 = 1 -\gamma - \psi(l+2), \,\,\,\,\,\,\,\,\,\,\,\,\, \,\,\,\,\,\,\,\,\,\,\,\,\,\,\,\,\,\,\,\,\,\,\,\,\,\,\,\,\,\,\,\,\,\,\,\,\,\,\,\,\,\,\,\,\,\,\,\,\,\,
\,\,\,\,\,\,\,\,\,\,\,\,\,\,\,\,\,\,\,\,\,\,\,\,\,\,\,\,\,\,\,\,\,\,\,\,\,\,\,\,\,\,\,\,\,\,\,\,\,\,
\,\,\,\,\,\,\,\,\,\,\,\,\,\,\,\,\,\,\,\,\,\,\,\,\,\,\,\,\,\,\,\,\,\,\,\,\,\,\,\,\,\,\,\,\,\,\,\,\,\,
\,\,\,\,\,\,\,\,\,\,\,\,\,\,\,\,\,\,\,\,\,\,\,\,\,\,\,\,\,\,\,\,\,\,\,\,\,\,\,\,\,\,\,\,\,\,\,\,\,\,
\,\,\,\,\,\,\,\,\,\,\,\,\,\,\,\,\,\,\,\,\,\,\,\,\,\,\,\,\,\,\,\,\,\,\,\,\,\,\,\,\,\,\,\,\,\,\,\,\,\,
\\
r_{{}_{2}}(l) = {\textstyle\frac{(-1)^l}{2!3!4^2}}\sum\limits_{j=0}^{l-2}\sum\limits_{l_2=1}^{l-j-1}
{\textstyle\frac{(-1)^j \frac{1}{i\mathfrak{e}}b_{{}_{l_2}}\frac{1}{i\mathfrak{e}}b_{{}_{l-j-l_2}} a_{{}_{l-j, l,0}} }{\frac{1}{i\mathfrak{e}}b_{{}_{l}}}} 
\, \times \,\,\,\,\,\,\,\,\,\,\,\,\,\,\,\,\,\,\,\,\,\,\,\,\,\,\,\,\,\,\,\,\,\,\,\,\,\,\,\,\,\,\,\,\,\,\,\,\,\,
\,\,\,\,\,\,\,\,\,\,\,\,\,\,\,\,\,\,\,\,\,\,\,\,\,\,\,\,\,\,\,\,\,\,\,\,\,\,\,\,\,\,\,\,\,\, \\
\times \,\, 
{\textstyle 2\sqrt{\frac{2(l-j)+1}{(l-j)(l-j+1)}}} {\textstyle 2\sqrt{\frac{2l_2+1}{l_2(l_2+1)}}}{\textstyle 2\sqrt{\frac{2(l-j-l_2)+1}{(l-j-l_2)(l-j-l_2+1)}}}  
\, \times \\
\times \,\, 
{\textstyle\frac
{\left\langle u |c_{{}_{l-j,0}} c_{{}_{l_2,0}}c_{{}_{l-j -l_2,0}} c_{{}_{l-j,0}}^+ c_{{}_{l_2,0}}^+c_{{}_{l-j -l_2,0}}^+ |u\right\rangle}{(4\pi\mathfrak{e}^2)^3}}
\int\limits_{\mathbb{S}^2} {\textstyle\frac{T^{{}^{l-j}}_{{}_{0 \,\, 0}}(a)T^{{}^{l_2}}_{{}_{0 \,\, 0}}(a)T^{{}^{l-j - l_2}}_{{}_{0 \,\, 0}}(a)}{4\pi}} \, da,
\\
\vdots \,\,\,\,\,\,\,\,\,\,\,\,\,\,\,\,\,\,\,\,\,\,\,\,\,\,\,\,\,\,\,\,\,\,\,\,\,\,\,\,\,\,\,\,\,\,\,\,\,\,
\,\,\,\,\,\,\,\,\,\,\,\,\,\,\,\,\,\,\,\,\,\,\,\,\,\,\,\,\,\,\,\,\,\,\,\,\,\,\,\,\,\,\,\,\,\,
\,\,\,\,\,\,\,\,\,\,\,\,\,\,\,\,\,\,\,\,\,\,\,\,\,\,\,\,\,\,\,\,\,\,\,\,\,\,\,\,\,\,\,\,\,\,\,\,\,\,
\,\,\,\,\,\,\,\,\,\,\,\,\,\,\,\,\,\,\,\,\,\,\,\,\,\,\,\,\,\,\,\,\,\,\,\,\,\,\,\,\,\,\,\,\,\,
\end{multline*}
\begin{multline*}
r_{{}_{l-1}}(l) = 
{\textstyle\frac{(-1)^l}{(l-2)!^24^{l-1}}}
{\textstyle\frac{\frac{1}{i\mathfrak{e}}b_{{}_{2}}(\frac{1}{i\mathfrak{e}}b_{{}_{1}})^{l-2} }{\frac{1}{i\mathfrak{e}}b_{{}_{l}}}} 
{\textstyle 2\sqrt{\frac{2l+1}{l(l+1)}}} 
{\textstyle 2\sqrt{\frac{5}{6}}}\big({\textstyle 2\sqrt{\frac{3}{2}}}\big)^{l-2}
\, \times \\
\times \,\, 
{\textstyle\frac
{\left\langle u |c_{{}_{l,0}} c_{{}_{2,0}}(c_{{}_{1,0}})^{l-2} c_{{}_{l,0}}^+ c_{{}_{2,0}}^+(c_{{}_{1,0}}^+)^{l-2} |u\right\rangle}{(4\pi\mathfrak{e}^2)^l}}
\int\limits_{\mathbb{S}^2} {\textstyle\frac{T^{{}^{2}}_{{}_{0 \,\, 0}}(a)(T^{{}^{1}}_{{}_{0 \,\, 0}}(a))^{l-2}T^{{}^{l}}_{{}_{0 \,\, 0}}(a)}{4\pi}} \, da,
\end{multline*}
\begin{multline*}
+ {\textstyle\frac{(-1)^{l-1}}{(l-1)!^24^{l-1}}}
{\textstyle\frac{(\frac{1}{i\mathfrak{e}}b_{{}_{1}})^{l-1} a_{{}_{l-1,l,0}}}{\frac{1}{i\mathfrak{e}}b_{{}_{l}}}} 
{\textstyle 2\sqrt{\frac{2(l-1)+1}{(l-1)l}}} 
\big({\textstyle 2\sqrt{\frac{3}{2}}}\big)^{l-1}
\, \times \\
\times \,\, 
{\textstyle\frac
{\left\langle u |c_{{}_{l-1,0}} (c_{{}_{1,0}})^{l-1} c_{{}_{l-1,0}}^+(c_{{}_{1,0}}^+)^{l-1} |u\right\rangle}{(4\pi\mathfrak{e}^2)^l}}
\int\limits_{\mathbb{S}^2} {\textstyle\frac{(T^{{}^{1}}_{{}_{0 \,\, 0}}(a))^{l-1}T^{{}^{l-1}}_{{}_{0 \,\, 0}}(a)}{4\pi}} \, da,
\end{multline*}
\begin{multline*}
r_{{}_{l}}(l) = 
{\textstyle\frac{(-1)^l}{l!^24^{l}}}
{\textstyle\frac{(\frac{1}{i\mathfrak{e}}b_{{}_{1}})^{l} }{\frac{1}{i\mathfrak{e}}b_{{}_{l}}}} 
{\textstyle 2\sqrt{\frac{2l+1}{l(l+1)}}} 
\big({\textstyle 2\sqrt{\frac{3}{2}}}\big)^{l}
\, \times \\
\times \,\, 
{\textstyle\frac
{\left\langle u |c_{{}_{l,0}}(c_{{}_{1,0}})^{l} c_{{}_{l,0}}^+(c_{{}_{1,0}}^+)^{l} |u\right\rangle}{(4\pi\mathfrak{e}^2)^{l+1}}}
\int\limits_{\mathbb{S}^2} {\textstyle\frac{(T^{{}^{1}}_{{}_{0 \,\, 0}}(a))^{l}T^{{}^{l}}_{{}_{0 \,\, 0}}(a)}{4\pi}} \, da,
\end{multline*}
with $\gamma$ -- the Euler gamma constant, $\psi$ -- digamma function. Here $b_{{}_{l,0}}$ is the coefficient
of the lowest order in the expansion of $B_{{}_{l,0}}(\lambda)$ at zero, and thus is equal
to (\ref{bl}) multiplied by $\tfrac{4\pi\mathfrak{e}(-1)^{l}}{n}$, 
$a = g_{{}_{12}}(\vartheta)g_{{}_{13}}(\varphi), da = \sin \varphi d\varphi d\vartheta$.
A closer look at $r_{{}_{k}}(l)$ shows that $r_{{}_{k}}(l)$ are all negative and $r_{{}_{1}}(l)  \rightarrow - \infty$
if $l \rightarrow \infty$. For $l=1, \ldots, 4$, we have:  
\begin{center}
\begin{tabular}{ c | c|c}
 $l$ & $R_{{}_{l}}(z) = 1+ r_{{}_{1}}(l) z \ldots + r_{{}_{l}}(l) z^l$ & roots $\in (0,1)$   \\
\hline
     $1$  &   $1 - z/2$  & no roots   \\
     $2$  &   $1 - 5z/6 - 3z^2/8 -z^3/6$   &  no roots \\
     $3$  &   $1 - 13z/12 - 3z^2/8 -z^3/24$  & $\approx 0.726$ \\ 
     $4$  &   $1 - 77z/60 - 71z^2/120 - 7z^3/60 - z^4/120$ & $\approx 0.596$  \\
\end{tabular}
\end{center}
Starting with $l=3$, $R_{{}_{l}}$ has exactly one root in this interval, approximately equal to $-1/r_{{}_{1}}(l)$ for
large $l$, and converging to zero if $l \rightarrow \infty$.
  
To prove our lemma, it is sufficient to show, moreover, that
\begin{multline*}
\underset{\sigma \rightarrow \infty}{\textrm{lim}} c_{{}_{\sigma}} \langle u, \infty|U(\sigma)c_{{}_{l,0}}^+|u\rangle 
=
\textrm{sgn}(n)\sqrt{\pi}\sqrt{z}(1-z)^2 2\textstyle{\sqrt{\frac{2l+1}{l(l+1)}}} \Bigg[
\\
1 + \sum\limits_{k=2}^{\infty} {\textstyle\frac{(-z)^k}{k!^2}} \sum\limits_{l_1, \ldots, l_k =1}^{\infty} 
\prod\limits_{j=1}^{k}\big[{\textstyle\frac{2l_j+1}{l_j(l_j+1)}}  \big]
{\textstyle\frac{\langle u| \prod\limits_{j=1}^{k} [c_{{}_{l_j,0}}] \prod\limits_{s=1}^{k} [c_{{}_{l_s,0}}^+]  |u\rangle}{(4\pi\mathfrak{e}^2)^k}}
\int\limits_{\mathbb{S}^2}{\textstyle\frac{\prod\limits_{s=1}^{k} T^{{}^{l_s}}_{{}_{0 \,\, 0}}(a)}{4\pi}} \, da
\Bigg]
\end{multline*}
\begin{equation}\label{sigma->Infinity}
= \sqrt{\pi}\sqrt{z}(1-z)^2 2\textstyle{\sqrt{\frac{2l+1}{l(l+1)}}} \Bigg[
1 + z^2/2 + \sum\limits_{k=3}^{\infty} (-1)^k  w_{{}_{k}} z^k 
\Bigg]
\end{equation}
is nonzero in the interval $z \in (0,0.73)$. Using the asymptotic properties of the Clebsch-Gordan coefficients
it can be shown that $0<w_{{}_{k}} < 1$ in the series (\ref{sigma->Infinity}), so that (\ref{sigma->Infinity})
is absolutely convergent for $0 < z < 1$, which legitimates exchanging the summation and transition to the limit, in obtaining the formulas 
(\ref{l-orderBoundStateProj''}) and (\ref{sigma->Infinity}) (averaging over $SU(2,\mathbb{C})$ intervenes substantially).
The coefficient $w_{{}_{2}} = 1/2$ can be easily exactly computed, and we have estimated numerically more accurately the coefficients 
$w_{{}_{3}} \approx 32/375, w_{{}_{4}} \approx 259/4000$, so that
\begin{multline*}
1 +  \sum\limits_{k=2}^{\infty} (-1)^k  w_{{}_{k}} z^k  >
1 + z^2/2 - w_{{}_{3}} z^3 + w_{{}_{4}}z^4 - {\textstyle\frac{z^5}{1-z}}
\\
>
1 + z^2/2 - 32z^3/375 + 259z^4/4000 - {\textstyle\frac{z^5}{1-z}} > 0, \,\,\, \textrm{for} \,\, 0 < z < 0.77.
\end{multline*}
\qed

From lemma \ref{BoundStateComponentsComparison}
and lemma \ref{L3}, it follows that the supplementary component 
of the representation with cyclic vector $c_{{}_{l,0}}^+|u\rangle$
is nonzero for each $l=1,2, \ldots$, and each of these supplementary components 
coincides with the supplementary component of the representation with the cyclic vector $|u\rangle$.

To implement generally the method used in the proof of lemma \ref{LCyc|u>andc10+|u>}, 
we prove, in Section \ref{PolesAsymptotic}, that the functions (\ref{Kalphaalpha}), and the analogues of the projection functions 
(\ref{CyclicProj1}), (\ref{1CyclicProj}), joining the cyclic representations with cyclic vectors 
$c_{{}_{\alpha}}^{+}|u\rangle, c_{{}_{\alpha'}}^{+}|u\rangle$, are nonzero functions of $\rho$ for each positive $z$, if 
\[
\textrm{min}\big(l(\alpha),l(\alpha')\big)>1.
\] 
The case $l(\alpha) = 1, l_0=1$, in (\ref{Kalphaalpha}) is exceptional.
The cyclic representation with the cyclic vector $c_{{}_{\alpha}}^+|u\rangle$ and with
$l(\alpha) = 1$, coincides with the representation with the cyclic vector $|u\rangle$. This makes it exceptional
among the cyclic representations with cyclic vectors $c_{{}_{\alpha}}^+|u\rangle$.
As we have seen (\ref{Kalphaalpha}) is identically zero function for $l(\alpha) = 1$ and $l_0=1$ in (\ref{Kalphaalpha}), 
and the representation with the cyclic vector $c_{{}_{\alpha}}^+|u\rangle$ and $l(\alpha)=1$ 
decomposes into purely spherical representations, which is an exceptional property coming from the 
theory \cite{Staruszkiewicz}. This is why we have treated the case with $l(\alpha) = 1$ in (\ref{Kalphaalpha}) explicitly.
To investigate the relation between the representation spaces of representations with cyclic (\ref{x}) in which $\mathfrak{q}>1$, to the representations
with cyclic (\ref{x}) in which $\mathfrak{q}=1$, we use the expansions (\ref{U(lambda)|u>}) and (\ref{Bexpansion}),
as in the proof that $c_{{}_{1,0}}^+|u\rangle \in \mathcal{H}_{{}_{|u\rangle}}$, given above, 
which allows us to reduce the problem to the mutual relations of the representations with cyclic  
(\ref{x}) in which $\mathfrak{q}=1$, \emph{i.e.} with cyclic vector of the form $c_{{}_{\alpha}}^+|u\rangle$.

\section{Poles and asymptotic of $\textrm{Tr} \, K$ and projection functions}\label{PolesAsymptotic}

For the proof of our theorem, we need the 
projection functions of the cyclic vectors -- the analogues of (\ref{CyclicProj}) in which we replace the cyclic vectors
$|u\rangle$ and $c_{{}_{1,0}}^+|u\rangle$ with the more general cyclic vectors of the form $c_{{}_{\alpha}}^+|u\rangle$ 
and the components $0,0 \times 0,1$ in the ``smearing'' with arbitrary components.  

Namely, the first class of projective functions we need, connect the cyclic representations
with cyclic vectors, respectively, equal $x=|u\rangle$ and $x=c_{{}_{l,0}}^+|u\rangle$, in the sense explained in the proof
of lemma \ref{LCyc|u>andc10+|u>} .
Thus, they arise by replacing the state 
$|g\rangle= U(g)c_{{}_{1,0}}^+|u\rangle$ with $|g\rangle= U(g)c_{{}_{l,0}}^+|u\rangle$ in (\ref{CyclicProj1}) 
and with the ``smeared'' state
\[
|f\rangle = \int f(g) |g\rangle dg,
\,\,\,\,
f(g) = \int\limits_{I_{{}_{\epsilon}}} U^{{}^{(l_0=0,l_1=i\rho)}}_{{}_{0,0 \,\,\, l,0}}(g) \, d\rho, 
\,\,\,\ I_{{}_{\epsilon}}=[\rho-\epsilon/2, \rho+\epsilon/2]
\]
and we introduce the following projection function
\begin{equation}\label{CyclicProj}
\underset{\epsilon \rightarrow 0}{\textrm{lim}} \,\, {\textstyle\frac{1}{\epsilon}}\langle u |f\rangle
={\textstyle\frac{4\pi^{7/2}\sqrt{z}e^z}{2l+1}} \int\limits_{0}^{+\infty} \overline{B_{{}_{l,0}}(\lambda)} U^{{}^{(l_0=0,i\rho)}}_{{}_{0,0 \,\,\, l,0}}(\lambda)
\,\, e^{-z\lambda \textrm{coth} \lambda} \, \textrm{sinh}^2\,\lambda d\lambda.  
\end{equation}
For the comparison of representations, we also need the corresponding projection function with the cyclic vectors reversed. Namely
with $|g\rangle = U(g)|u\rangle$, and with the ``smeared'' state
\[
|f\rangle = \int f(g) |g\rangle dg,
\,\,\,\,
f(g) = \int\limits_{I_{{}_{\epsilon}}} U^{{}^{(l_0=0,l_1=i\rho)}}_{{}_{l,0 \,\,\, 0,0}}(g) \, d\rho, 
\,\,\,\ I_{{}_{\epsilon}}=[\rho-\epsilon/2, \rho+\epsilon/2]
\]
and the projection function
\begin{equation}\label{lCyclicProj}
\underset{\epsilon \rightarrow 0}{\textrm{lim}} \,\, {\textstyle\frac{1}{\epsilon}}\langle u|c_{{}_{l,0}}|f\rangle
={\textstyle\frac{4\pi^{7/2}\sqrt{z}e^z}{2l+1}} (-1)^l \int\limits_{0}^{+\infty} B_{{}_{l,0}}(\lambda) U^{{}^{(l_0=0,i\rho)}}_{{}_{l,0 \,\,\, 0,0}}(\lambda)
\,\, e^{-z\lambda \textrm{coth} \lambda} \, \textrm{sinh}^2\,\lambda d\lambda.  
\end{equation}
But for each fixed $z$, (\ref{CyclicProj}) is a nonzero function of $\rho$ if and only if  (\ref{lCyclicProj}) is a nonzero function
of $\rho$, and we need to analyze only one of the two (\ref{CyclicProj}) and (\ref{lCyclicProj}). The same
is true of the residues of (\ref{CyclicProj}) and (\ref{lCyclicProj}) at $\rho = -i(z-1)$, or at $\rho = i(z-1)$, with $0<z<1$:
the residue is nonzero for (\ref{CyclicProj}) if and only if it is nonzero for (\ref{lCyclicProj}).
We need the second class of projection functions, which connect the cyclic representations
with cyclic vectors, respectively, equal $x=c_{{}_{\alpha'}}^+|u\rangle$ and $x=c_{{}_{\alpha}}^+|u\rangle$
defined in the analogous way. 
Namely, let $|g\rangle = U(g)c_{{}_{\alpha'}}^+|u\rangle$ and $z>1$. Let us consider 
\[
|f\rangle = \int f(g) |g\rangle dg,
\,\,\,\,
f(g) = \int\limits_{I_{{}_{\epsilon}}} U^{{}^{(l_0,i\rho)}}_{{}_{\beta \,\,\, \beta'}}(g) \, d\rho, 
\,\,\,\ I_{{}_{\epsilon}}=[\rho-\epsilon/2, \rho+\epsilon/2]
\]
and the following projection functions
\begin{multline}\label{ProjF}
\underset{\epsilon \rightarrow 0}{\textrm{lim}} \,\, {\textstyle\frac{1}{\epsilon}}\langle u|c_{{}_{\alpha}} |f\rangle 
=  
{\textstyle\frac{\delta_{{}_{\beta \,\, \alpha}} \delta_{{}_{\beta' \,\, \alpha'}} (4\pi \mathfrak{e})^2e^z}{(2l+1)(2l'+1)}} \bigints\limits_{0}^{+\infty} \Bigg[
\sum\limits_{n=-l}^{l}
\overline{A_{{}_{l,n \,\,\, l',n}}(\lambda)}U^{{}^{(l_0,i\rho)}}_{{}_{l,n \,\,\, l',n}}(\lambda) 
\\
+{\textstyle\frac{z}{4}} (-1)^{l} \overline{{\textstyle\frac{1}{\mathfrak{e}}}B_{{}_{l',0}}(\lambda)}
{\textstyle\frac{1}{\mathfrak{e}}}B_{{}_{l,0}}(\lambda)
U^{{}^{(l_0,i\rho)}}_{{}_{l,0 \,\,\, l',0}}(\lambda)
\Bigg]e^{-z\lambda \textrm{coth} \lambda} \, \pi^2 \textrm{sinh}^2 \lambda \, d\lambda,
\end{multline}
where $\alpha=(l,m), \alpha'=(l',m') $, $l\leq l'$.
Again, we have used in (\ref{CyclicProj}), (\ref{lCyclicProj}) and (\ref{ProjF}), 
the fact that the integration and passing to the $\epsilon\rightarrow 0$ limit operations
can be exchanged if $z>1$, by the dominated convergence principle, and where we have used (\ref{Torthogonality}),
commutation rules (\ref{CommutationRules}), (\ref{Vacuum}), and transformation
rule (\ref{c'S'}). (\ref{CyclicProj}), (\ref{lCyclicProj}) and (\ref{Kalphaalpha}), (\ref{ProjF}), 
are analytic functions of $(\rho,z) \in \mathbb{C}^2$ and (up to the factors in front of the integrals)
can be represented by almost uniformly and absolutely convergent double series of finite sums 
of simple fractions of the general form (\ref{GeneralSeries}). 

We note first the following
\begin{lem}
The residue at $\rho = -i(z-1)$, or at $\rho = i(z-1)$, with $0<z<1$, of the analytic 
continuation of the projection functions, (\ref{CyclicProj}), (\ref{lCyclicProj}), is nonzero
for each $l=1,2, \ldots$.
\label{ResiduePart}
\end{lem}
\qedsymbol \,
We need to analyze only (\ref{CyclicProj}).
Having given (\ref{(l0=0,irho)(0,0)x(l,0)}), (\ref{oddB}), (\ref{evenB}), we represent
the function $g$ of the integrand $g e^{-z\lambda\textrm{coth} \lambda}$ in 
(\ref{CyclicProj}) in the general form (\ref{Generalgk}) with the analytic continuation
of the integral in (\ref{CyclicProj}) given by the corresponding series (\ref{GeneralSeries}). 
It is easily seen that the residue  of this series 
at $\rho = -i(z-1)$ or, respectively, at $\rho = i(z-1)$ is equal
\[
i a_{{}_{0,-1}}^{{}^{+}}(\rho= -i(z-1)) \,\,\,\, \textrm{or, respectively}, \,\,\,\,
i a_{{}_{0,-1}}^{{}^{-}}(\rho= i(z-1)),
\]  
where $a_{{}_{0,-1}}^{{}^{\pm}}(\rho)$ are the polynomials in (\ref{Generalgk}) and in the 
series (\ref{GeneralSeries}) equal to the analytic continuation of the integral in (\ref{CyclicProj}). In these formulas for the residues
we ignore the factor in front of the integral (\ref{CyclicProj}). We'll bring it back at the very end of the calculation.
Using the formulas (\ref{BatInfinity}) and (\ref{(l0=0,irho)(0,0)x(l,0)}), we get the following values
\begin{multline*}
{\textstyle\frac{\mathfrak{e} 32 \pi^{{}^{7/2}}\sqrt{z}e^zi^{l+1}}{\sqrt{l(l+1)}\prod\limits_{s=0}^{l}(-i(s + z-1))}}
 \prod\limits_{j=2}^{l+1}(z-j)
\,\,\, 
\textrm{or, respectively},
\\
-{\textstyle\frac{\mathfrak{e}8\pi^{{}^{7/2}}\sqrt{z}e^zi^{l}}{\sqrt{l(l+1)}\prod\limits_{s=0}^{l}(-i(s - z +1))}}   
\prod\limits_{j=2}^{l+1}(z-j),
\end{multline*}
for the residues. It is evident that they are nonzero for each $0 < z < 1$ and $l=1,2, \ldots$.
\qed

\begin{lem}
(\ref{Kalphaalpha}) is identically zero function of $(z,\rho)$ if $l=l_0=1$ in (\ref{Kalphaalpha}). 
(\ref{ProjF}) is identically zero function of $(z,\rho)$,
if $l=l_0=1$ or $l'=l_0=1$ in (\ref{ProjF}).
\label{LzeroProjF}
\end{lem}
\qedsymbol \,
The first statement we have already proved in Subsection \ref{DecompositionGeneralCyclic}.
Because (\ref{Kalphaalpha}) is identically zero function of $(z,\rho)$, if $l=l_0=1$ in (\ref{Kalphaalpha}),
then by definition, the projection functions (\ref{ProjF}) must be identically zero functions of $(z,\rho)$,
if $l=l=l_0=1$ or $l'=l_0=1$ in (\ref{ProjF}). (Compare the proof of lemma \ref{LCyc|u>andc10+|u>}.)  
\qed

One can also convince himself of the validity of this lemma for particular values of $l$ or $l'$ with $l_0=1$ and $l=1$ 
or $l'=1$, by explicit calculation. In these cases, the integrals
(\ref{Kalphaalpha}), (\ref{ProjF}), are elementary and have the primitives which are in the same
class (\ref{Generalgk}), multiplied by $e^{-z\lambda\textrm{coth} \lambda}$, as the integrands themselves in  
(\ref{Kalphaalpha}), (\ref{ProjF}), with the primitives which have the same limit value at zero 
and at infinity, thus giving identically zero values for (\ref{Kalphaalpha}), (\ref{ProjF}).    

We show that for all remaining cases, (\ref{Kalphaalpha}) and (\ref{ProjF}), 
with $\alpha=\beta$, $\alpha'=\beta'$ in (\ref{ProjF}),
are, for each fixed $z>0$, nonzero functions of $\rho$, \emph{i.e.} in this Section we give 
a proof that (\ref{Kalphaalpha}), (\ref{ProjF}), 
(\ref{CyclicProj}) are, for each positive $z$, nonzero functions of $\rho$,
except for $l=l'=l_0=1$ in (\ref{Kalphaalpha}) and except for $l_0=l=1$, or 
$l_0=l'=1$ in  (\ref{ProjF}). 
Because the common factor, equal to the inverse of the polynomial $\rho Q(\rho)$, and coming from the denominator
in the formula (\ref{explicitU^(l0,l1)}), is nonzero for all real $\rho$, we ignore this common factor
$\tfrac{1}{\rho Q(\rho)}$ in the proof, and the factors in front of the integrals 
in (\ref{Kalphaalpha}), (\ref{ProjF}), (\ref{CyclicProj}). We give here a short outline
of the proof. First, we show the statement that, under these restrictions, and any finite positive $z$,
(analytic continuation of) (\ref{Kalphaalpha}), (\ref{ProjF}), 
(\ref{CyclicProj}) will have nontrivial poles, \emph{i.e.}
with nonzero coefficients $L_{{}_{m,s}}(z)$ in the Laurent series, for which the distance $m+z$ of the pole from the real axis 
in the complex $\rho$-plane is arbitrarily large, the order $s$ of the pole is arbitrarily large, and the
ratio $m/s$ of the distance $m$ to the order $s$ is arbitrarily large. In fact, we put $m=s^2$ with the order $s$ 
going to infinity. We prove it in Subsection
\ref{AsymptoticLaurentCoefficient}.
Because we have ignored the factor $\tfrac{1}{\rho Q(\rho)}$, and the factors in front of the integrals,
the Laurent coefficients $L_{{}_{m,s}}(z)$ are polynomial functions of $z$. 
We compute the asymptotic form of the polynomials  $L_{{}_{m,s}}(z)$ equal to the Laurent coefficients 
for large $m,s,m/s$, and show that for each $z>0$ they are nonzero for large $m,s,m/s$, if $\textrm{min}(l,l')>1$ or $l_0 \neq 1$. 
This proves that (\ref{Kalphaalpha}), (\ref{ProjF}) and (\ref{CyclicProj}), are nonzero functions for each $z>0$,
except for $l=l' =l_0 = 1$ in (\ref{Kalphaalpha}) and except $l_0=l=1$ or $l_0=l'=1$ in  (\ref{ProjF}).
In these exceptional cases, (\ref{Kalphaalpha}), (\ref{ProjF}), are identically zero functions of
$z, \rho$.

\subsection{Asymptotic of the Laurent coefficient $L_{{}_{m,s}}(z)$}\label{AsymptoticLaurentCoefficient}

The integrals in (\ref{Kalphaalpha}), (\ref{CyclicProj}) or (\ref{ProjF}), regarded as functions of $\rho,z$
have the asymptotic expansions of the general form (\cite{Sidorov}, Corollary 1, p. 408) 
\begin{equation}\label{asymtoticTrK}
\widesim[3]{z \rightarrow +\infty} e^{-z} \sum\limits_{n= n_0}^{+\infty} c_n z^{-(n+1)/2}
\end{equation}
with $c_n$ depending rationally on $\rho$,  and with $n_0$ greather than or equal to the order of zero of $g_0$ at $\lambda=0$,
where $g(\rho,z,\lambda) =\Sigma_{i=0}^{{\mathfrak{q}}}z^ig_i(\rho,\lambda)$ in the integrand in the general form (\ref{GeneralInt}) 
of these integrals. More generally, the integrals (\ref{GeneralInt}), 
giving, up to the factor $e^z$, the matrix elements of the Fourier transform (\ref{F[<e,.>]}),
have the asymptotic (\ref{asymtoticTrK}).  Here we ignore the factors, respectively,
\begin{equation}\label{factors}
\left({\textstyle\frac{4\pi \mathfrak{e}}{2l+1}}\right)^2e^z, \,\,\, {\textstyle\frac{4\pi^{7/2}\sqrt{z}e^z}{2l+1}}
\,\,\, \textrm{or} \,\,\,
{\textstyle\frac{\delta_{{}_{\beta \,\, \alpha}} \delta_{{}_{\beta' \,\, \alpha'}} (4\pi \mathfrak{e})^2e^z}{(2l+1)(2l'+1)}}, 
\end{equation}
standing before the integrals in (\ref{Kalphaalpha}), (\ref{CyclicProj}) or (\ref{ProjF}). Therefore in order 
to obtain the asymptotic of the full expressions (\ref{Kalphaalpha}), (\ref{CyclicProj}) or (\ref{ProjF}),
we need to multiply (\ref{asymtoticTrK}), respectively, by (\ref{factors}). Similarly, we need to multiply
(\ref{asymtoticTrK}) by $e^z$ in order to obtain asymptotics of the matrix elements of $K$. 

Applying Corollary 1, pp. 407, 408 of \cite{Sidorov}, to each integrand $g_i(\rho,\lambda)$ separately, and using the
expansions (\ref{Uexpansion})-(\ref{Uexpansion'}) and (\ref{Bexpansion}) it is easily seen,
that for appropriately large $z$, (\ref{Kalphaalpha}), (\ref{CyclicProj}) and (\ref{ProjF}), are nonzero functions of $\rho$, if 
$l>1$ in (\ref{Kalphaalpha}) and  $\textrm{min}\{l,l'\}>1$ in  (\ref{ProjF}). Therefore, only in the mentioned
exceptional cases these functions can be identically zero. 

We represent (\ref{Kalphaalpha}) or (\ref{ProjF}) or, respectively, (\ref{CyclicProj}), in the form
(\ref{GeneralInt}) with the integrand (\ref{Generalgk}), with common denominator $(1-e^{-2\lambda})^{q+1}$, with $q$ equal $4l,2(l+l')$, or 
$2l-1$, respectively, for (\ref{Kalphaalpha}), (\ref{ProjF}) or, respectively, (\ref{CyclicProj}).
Using the parities of $j_p,j$ in (\ref{GeneralAB}) and (\ref{GeneralU}), we can easily see that the 
parities of the integer numbers $j_p \geq -1$ in the formula (\ref{Generalgk}) representing
(\ref{Kalphaalpha}), (\ref{ProjF}) or (\ref{CyclicProj}), is the same as the parity of the number
$l_0-1$, where $l_0$ is the number in $U^{{}^{(l_0,i\rho)}}$ in (\ref{Kalphaalpha}), (\ref{ProjF}) or (\ref{CyclicProj}).
In what follows, we ignore the common factor $\tfrac{1}{\rho Q(\rho)}$, coming from $U^{{}^{(l_0,i\rho)}}$, 
and the factors (\ref{factors}) in the formula (\ref{Generalgk}) representing  
(\ref{Kalphaalpha}), (\ref{ProjF}) or (\ref{CyclicProj}).

Let us introduce the coefficients $a^{{}^{\pm}}_{{}_{p,j_p,k}}$ of the polynomials
\[
a^{{}^{\pm}}_{{}_{p,j_p}}(\rho) = \sum\limits_{k} a^{{}^{\pm}}_{{}_{p,j_p,k}} \rho^{{}^{k}}
\]
in (\ref{Generalgk}). In the formula of the integrand (\ref{Generalgk}) in (\ref{Kalphaalpha}) or in (\ref{ProjF}), 
we collect all terms $a^{{}^{\pm}}_{{}_{p,j_p,k}}$, containing linearly the parameter $z$, which represent the second
summand of the integrand in (\ref{Kalphaalpha}) or (\ref{ProjF}), containing $z$ as a factor, and denote them with
the prime:
\[
a^{{}^{\pm}}_{{}_{p,j_p,k}} = z {a'}^{{}^{\pm}}_{{}_{p,j_p,k}}
\]
and all remaining terms $a^{{}^{\pm}}_{{}_{p,j_p,k}}$ without $z$, and coming from the first summand of the integrand in 
(\ref{Kalphaalpha}) or (\ref{ProjF}), let us denote with double prime:
\[
a^{{}^{\pm}}_{{}_{p,j_p,k}} = {a''}^{{}^{\pm}}_{{}_{p,j_p,k}}.
\]
The coefficients will be denoted simply by $a^{{}^{\pm}}_{{}_{p,j_p,k}}$ without any primes, 
in the formula for the integrand (\ref{Generalgk}) in the integral (\ref{GeneralInt}) 
in  (\ref{CyclicProj}). Note that $p =0,1$ or $2$ for
the single primed coefficients, and $p=0$ or $1$ for the double primed in (\ref{Generalgk}), representing 
(\ref{Kalphaalpha}) or (\ref{ProjF}), and $p=0$ or $1$ for $a^{{}^{\pm}}_{{}_{p,j_p}}$ in (\ref{Generalgk}) 
representing (\ref{CyclicProj}).  

Using the fact that $f_{{}_{0}}$ is of first-order, and $f_{{}_{1}}$ is of zero-order at $\lambda=0$ in the formula
(\ref{GeneralAB}) for $B_{{}_{l,0}}(\lambda)$ and $A_{{}_{l,m \,\,\, l',m}}(\lambda)$, and the formulas
(\ref{MaxDegp,j,j}), (\ref{MaxDegp,j,j,0}) for the coefficents of the maximal order in $\rho$ in the numerator
of (\ref{explicitU^(l0,l1)}), we easily see the following order behavior of the function 
in the numerator of (\ref{Generalgk}). 
The sum of exponents of $\lambda$, regarded as a function $f(\lambda)$ of $\lambda$, multiplying the monomial 
$\lambda^0\rho^{{}^{j_\textrm{max}}}e^{i\rho\lambda}$ of the maximal degree in $\rho$, which collects all single primed terms, 
is of $j_\textrm{max}+2 = l+l'-l_0+2$ -order
at $\lambda=0$ for (\ref{ProjF}), with $l=l'$ for (\ref{Kalphaalpha}). 
The function of $\lambda$ multiplying the monomial 
$\lambda^1\rho^{{}^{j_\textrm{max}}}e^{i\rho\lambda}$ of the maximal degree in $\rho$, which collects all single primed terms, 
is of $j_\textrm{max}+1 = l+l'-l_0+1$ -order for (\ref{ProjF}), with $l=l'$ for (\ref{Kalphaalpha}).  
The function of $\lambda$ multiplying the monomial 
$\lambda^2\rho^{{}^{j_\textrm{max}}}e^{i\rho\lambda}$ of the maximal degree in $\rho$, which collects all single primed terms, 
is of $j_\textrm{max} = l+l'-l_0$ -order for (\ref{ProjF}), with $l=l'$ for (\ref{Kalphaalpha}).  
The function of $\lambda$ multiplying the monomial 
$\lambda^0\rho^{{}^{j_\textrm{max}}}e^{i\rho\lambda}$ of the maximal degree in $\rho$, which collects all double primed terms, 
is of $j_\textrm{max}+2 = l+l'+2$ -order for (\ref{ProjF}) with $l=l'$ for (\ref{Kalphaalpha}).  
The function of $\lambda$ multiplying the monomial 
$\lambda^1\rho^{{}^{j_\textrm{max}}}e^{i\rho\lambda}$ of the maximal degree in $\rho$, which collects all double primed terms, 
is of $j_\textrm{max}+1 = l+l'+1$ -order for (\ref{ProjF}), with $l=l'$ for (\ref{Kalphaalpha}).  
The function of $\lambda$ multiplying the monomial 
$\lambda^0\rho^{{}^{j_\textrm{max}}}e^{i\rho\lambda}$ of the maximal degree in $\rho$, which collects all unprimed terms, 
is of $j_\textrm{max}+1 = l+1$ -order for (\ref{CyclicProj}).  
The function of $\lambda$ multiplying the monomial 
$\lambda^1\rho^{{}^{j_\textrm{max}}}e^{i\rho\lambda}$ of the maximal degree in $\rho$, which collects all unprimed terms, 
is of $j_\textrm{max} = l$ -order for (\ref{CyclicProj}). The function(s) $f(\lambda)$ multiplying the monomial
$\lambda^p\rho^{{}^{j_\textrm{max}}}e^{i\rho\lambda}$ has the $k$-order Taylor coefficient at $\lambda=0$
equal
\[
{\textstyle\frac{1}{k!}}f^{{}^{(k)}}(0) = {\textstyle\frac{(-1)^k}{k!}} \sum\limits_{j_p} {a'}^{{}^{+}}_{{}_{p,j_p,j_\textrm{max}}}j_{p}^{k},
\]
for the primed contribution, and analogously for the double primed contribution in the integrand (\ref{Generalgk}) in
(\ref{ProjF}) or in (\ref{Kalphaalpha}), and the unprimed ${a}^{{}^{+}}_{{}_{p,j_p,j_\textrm{max}}}$ in the integrand (\ref{Generalgk})
of the integral (\ref{GeneralInt}) in  (\ref{CyclicProj}).
Therefore, this order behavior can be summarized by the following formulas
\begin{multline}\label{a'order}
\sum\limits_{j_0} {a'}^{{}^{+}}_{{}_{0,j_0,l+l'-l_0}}j_{0}^{l+l'-l_0+2} \neq 0,
\,\,\, \sum\limits_{j_0} {a'}^{{}^{+}}_{{}_{0,j_0,l+l'-l_0}}j_{0}^{k} = 0, \,\, k=0, \ldots, l+l'-l_0+1,
\\
\sum\limits_{j_1} {a'}^{{}^{+}}_{{}_{1,j_1,l+l'-l_0}}j_{0}^{l+l'-l_0+1} \neq 0,
\,\,\, \sum\limits_{j_0} {a'}^{{}^{+}}_{{}_{1,j_1,l+l'-l_0}}j_{1}^{k} = 0, \,\, k=0, \ldots, l+l'-l_0,
\\
\sum\limits_{j_2} {a'}^{{}^{+}}_{{}_{2,j_2,l+l'-l_0}}j_{2}^{l+l'-l_0} \neq 0,
\,\,\, \sum\limits_{j_0} {a'}^{{}^{+}}_{{}_{2,j_2,l+l'-l_0}}j_{2}^{k} = 0, \,\, k=0, \ldots, l+l'-l_0-1,
\end{multline}
for the single primed coefficients.
\begin{multline}\label{a''order}
\sum\limits_{j_0} {a''}^{{}^{+}}_{{}_{0,j_0,l+l'}}j_{0}^{l+l'+2} \neq 0,
\,\,\, \sum\limits_{j_0} {a''}^{{}^{+}}_{{}_{1,j_1,l+l'}}j_{0}^{k} = 0, \,\, k=0, \ldots, l+l'+1,
\\
\sum\limits_{j_1} {a''}^{{}^{+}}_{{}_{1,j_1,l+l'}}j_{1}^{l+l'+1} \neq 0,
\,\,\, \sum\limits_{j_1} {a''}^{{}^{+}}_{{}_{1,j_1,l+l'}}j_{1}^{k} = 0, \,\, k=0, \ldots, l+l',
\end{multline}
for the double primed coefficients.  
\begin{multline}\label{aorder}
\sum\limits_{j_0} {a}^{{}^{+}}_{{}_{0,j_0,l}}j_{0}^{l+1} \neq 0,
\,\,\, \sum\limits_{j_0} {a}^{{}^{+}}_{{}_{0,j_0,l}}j_{0}^{k} = 0, \,\, k=0, \ldots, l+1,
\\
\sum\limits_{j_1} {a}^{{}^{+}}_{{}_{1,j_1,l}}j_{1}^{l} \neq 0,
\,\,\, \sum\limits_{j_1} {a}^{{}^{+}}_{{}_{1,j_1,l}}j_{1}^{k} = 0, \,\, k=0, \ldots, l,
\end{multline} 
for the unprimed coefficients. In particular, using (\ref{a0ForA[l,-1,l',-1]}) and (\ref{+l0Maxpj+-}),
we obtain
\begin{multline}\label{suma''1}
{\textstyle\frac{(-1)^{l+l'+1}}{(l+l'+1)!}}\sum\limits_{j_1} {a''}^{{}^{+}}_{{}_{1,j_1,l+l'}}j_{1}^{l+l'+1}
\\
= {\textstyle\frac{1}{i}}2^{3l+4l'}(-1)^{2l+l'}ll'{\textstyle\frac{(2l+1)(2l'+1)}{(l'+1)!}}
{\textstyle\binom{l+l'-1/2}{l+l'}\binom{l'-1/2}{l'}}\prod\limits_{j=0}^{l'-1}\textstyle{\frac{l+j+1}{2(l+j)+1}}
\prod\limits_{r=0}^{l'-2}(l+l'-r), 
\end{multline}
for $l_0=1$. Using (\ref{sum(b)}) and (\ref{+0Maxpj+-}), we get
\begin{multline}\label{suma'2}
{\textstyle\frac{(-1)^{l+l'-l_0}}{(l+l'-l_0)!}}
\sum\limits_{j_2} {a'}^{{}^{+}}_{{}_{2,j_2,l+l'-l_0}}j_{2}^{l+l'-l_0}
=
\\
(-1)^{2l+l'+1}2^{4l+2l'+1}l(l+1)l'(l'+1){\textstyle\binom{l'+1/2}{l'+1}\binom{l+1/2}{l+1}}
\end{multline}
for $l_0=1$.

Next, we consider the Laurent coefficient $L_{{}_{m,s}}(z)$ of the Laurent expansion  
of the series (\ref{GeneralSeries}), representing (\ref{Kalphaalpha}), (\ref{ProjF}) or (\ref{CyclicProj}), 
at $\rho = -i(m+z)$, of order $s$, with the factor $\tfrac{1}{\rho Q(\rho)}$ and the factors (\ref{factors})
ignored. Therefore, in order to compute $L_{{}_{m,s}}(z)$ we need to consider only the simple fractions 
in the series (\ref{GeneralSeries}) in which $2k+2n+j_p = m$ is fixed and equal $m$, 
with the parity of $m$ the same as the parity of $l_0-1$. We compute $L_{{}_{m,s}}(z)$ and its asymptotic for $s \rightarrow +\infty$,
with $m=s^2$, for the general (\ref{GeneralSeries}), with the restriction that the range of values of $p$ is $0,1,2$,  and with 
${a}^{{}^{\pm}}_{{}_{p,j_p}}$ being polynomials in $\rho$, which is the case for (\ref{Kalphaalpha}), (\ref{ProjF}) 
or (\ref{CyclicProj}), because we have ignored the factor $\tfrac{1}{\rho Q(\rho)}$
and the factors (\ref{factors}). We compute separately the contributions to  $L_{{}_{m,s}}(z)$ coming from the simple 
fractions containing, respectively, the polynomials ${a}^{{}^{\pm}}_{{}_{0,j_0}}$,  ${a}^{{}^{\pm}}_{{}_{1,j_1}}$ and separately 
${a}^{{}^{\pm}}_{{}_{2,j_2}}$. Let $\mathfrak{d}_0, \mathfrak{d}_1, \mathfrak{d}_2$ be, respectively, the degrees of these polynomials. 

Then the contribution to the Laurent coefficient
$L_{{}_{m,s}}(z)$, coming from the simple fractions containing the polynomials ${a}^{{}^{\pm}}_{{}_{0,j_0}}$,  
${a}^{{}^{\pm}}_{{}_{1,j_1}}$ and, respectively, ${a}^{{}^{\pm}}_{{}_{2,j_2}}$,   
is equal, for large $m,s$, to
\begin{equation}\label{L[m,s,z,0]}
\sum\limits_{\ell}^{\mathfrak{d}_0}
i{a}^{{}^{+}}_{{}_{0,j_0,\ell}} (-2iz)^{s-1}
\sum\limits_{r=0}^{\ell}
{\textstyle\binom{\ell}{r}} \big(-i(m+z) \big)^{\ell-r}(-2iz)^r
{\textstyle\binom{(m-j_0)/2+q}{s+r+q-1}},
\end{equation}
\begin{equation}\label{L[m,s,z,1]}
\sum\limits_{\ell}^{\mathfrak{d}_1}
i^2{a}^{{}^{+}}_{{}_{1,j_1,\ell}} (-2iz)^{s-2}
\sum\limits_{r=0}^{\ell}
{\textstyle\binom{\ell}{r}} \big(-i(m+z) \big)^{\ell-r}(-2iz)^r
{\textstyle\binom{(m-j_1)/2+q}{s+r+q-2}},
\end{equation}
and
\begin{equation}\label{L[m,s,z,2]}
\sum\limits_{\ell}^{\mathfrak{d}_2}
i^3{a}^{{}^{+}}_{{}_{2,j_2,\ell}} (-2iz)^{s-3}
\sum\limits_{r=0}^{\ell}
{\textstyle\binom{\ell}{r}} \big(-i(m+z) \big)^{\ell-r}(-2iz)^r
{\textstyle\binom{(m-j_2)/2+q}{s+r+q-3}},
\end{equation}
so that $L_{{}_{m,s}}(z)$ is equal to the sum of (\ref{L[m,s,z,0]}), (\ref{L[m,s,z,1]})
and (\ref{L[m,s,z,2]}).
The asymptotic expansion $\widesim[3]{s \rightarrow +\infty}$  of the contribution to the Laurent coefficient
$L_{{}_{m=s^2,s}}(z)$, coming from the simple fractions containing the polynomials ${a}^{{}^{\pm}}_{{}_{0,j_0}}$,  
${a}^{{}^{\pm}}_{{}_{1,j_1}}$ and, respectively, ${a}^{{}^{\pm}}_{{}_{2,j_2}}$,   
is equal
\begin{multline}\label{assL[m,s,z,0]}
\widesim[3]{s \rightarrow +\infty}
{\textstyle\frac{i(-2iz)^{s-1}}{e\sqrt{2\pi s}}}e^{{}^{-{\textstyle\frac{2}{3s}}-{\textstyle\frac{2}{3s^2}}
-{\textstyle\frac{4}{5s^3}} -{\textstyle\frac{16}{15s^4}} - \ldots}} \big(\textstyle\frac{es}{2}\big)^s \big(\textstyle\frac{s}{2}\big)^{q-1}
\sum\limits_{\ell}^{\mathfrak{d}_0}\sum\limits_{r=0}^{\ell}\sum\limits_{j_0} \Bigg[
\\
{\textstyle\binom{\ell}{r}}\big(-i(s^2+z)\big)^{\ell-r}\big(-{\textstyle\frac{2izs}{2}}\big)^r
{a}^{{}^{+}}_{{}_{0,j_0,\ell}} \Big[
1 + \Big( {\textstyle\frac{(-1)^1 j_0^1}{1!}} + \ldots \Big){\textstyle\frac{1}{s^1}}
\\
+\Big( {\textstyle\frac{(-1)^2 j_0^2}{2!}} + \ldots \Big){\textstyle\frac{1}{s^2}}
+\Big( {\textstyle\frac{(-1)^3 j_0^3}{3!}} + \ldots \Big){\textstyle\frac{1}{s^3}}
+\Big( {\textstyle\frac{(-1)^4 j_0^4}{4!}} + \ldots \Big){\textstyle\frac{1}{s^4}}
+ \ldots
\Big]
\Bigg],
\end{multline}
\begin{multline}\label{assL[m,s,z,1]}
\widesim[3]{s \rightarrow +\infty}
{\textstyle\frac{i^2(-2iz)^{s-2}}{e\sqrt{2\pi s}}}e^{{}^{-{\textstyle\frac{2}{3s}}-{\textstyle\frac{2}{3s^2}}
-{\textstyle\frac{4}{5s^3}} -{\textstyle\frac{16}{15s^4}} - \ldots}} \big(\textstyle\frac{es}{2}\big)^s \big(\textstyle\frac{s}{2}\big)^{q-2}
\sum\limits_{\ell}^{\mathfrak{d}_1}\sum\limits_{r=0}^{\ell}\sum\limits_{j_1} \Bigg[
\\
{\textstyle\binom{\ell}{r}}\big(-i(s^2+z)\big)^{\ell-r}\big(-{\textstyle\frac{2izs}{2}}\big)^r
{a}^{{}^{+}}_{{}_{1,j_1,\ell}}(r+s-1) \Big[
1 + \Big( {\textstyle\frac{(-1)^1 j_1^1}{1!}} + \ldots \Big){\textstyle\frac{1}{s^1}}
\\
+\Big( {\textstyle\frac{(-1)^2 j_1^2}{2!}} + \ldots \Big){\textstyle\frac{1}{s^2}}
+\Big( {\textstyle\frac{(-1)^3 j_1^3}{3!}} + \ldots \Big){\textstyle\frac{1}{s^3}}
+\Big( {\textstyle\frac{(-1)^4 j_1^4}{4!}} + \ldots \Big){\textstyle\frac{1}{s^4}}
+ \ldots
\Big]
\Bigg]
\end{multline}
and
\begin{multline}\label{assL[m,s,z,2]}
\widesim[3]{s \rightarrow +\infty}
{\textstyle\frac{i^3(-2iz)^{s-3}}{e\sqrt{2\pi s}}}e^{{}^{-{\textstyle\frac{2}{3s}}-{\textstyle\frac{2}{3s^2}}
-{\textstyle\frac{4}{5s^3}} -{\textstyle\frac{16}{15s^4}} - \ldots}} \big(\textstyle\frac{es}{2}\big)^s \big(\textstyle\frac{s}{2}\big)^{q-3}
\sum\limits_{\ell}^{\mathfrak{d}_2}\sum\limits_{r=0}^{\ell}\sum\limits_{j_2} \Bigg[
\\
{\textstyle\binom{\ell}{r}}\big(-i(s^2+z)\big)^{\ell-r}\big(-{\textstyle\frac{2izs}{2}}\big)^r
{a}^{{}^{+}}_{{}_{2,j_2,\ell}}(r+s-1)(r+s-2) \Big[
1 + \Big( {\textstyle\frac{(-1)^1 j_2^1}{1!}} + \ldots \Big){\textstyle\frac{1}{s^1}}
\\
+\Big( {\textstyle\frac{(-1)^2 j_2^2}{2!}} + \ldots \Big){\textstyle\frac{1}{s^2}}
+\Big( {\textstyle\frac{(-1)^3 j_2^3}{3!}} + \ldots \Big){\textstyle\frac{1}{s^3}}
+\Big( {\textstyle\frac{(-1)^4 j_2^4}{4!}} + \ldots \Big){\textstyle\frac{1}{s^4}}
+ \ldots
\Big]
\Bigg],
\end{multline}
so that the total asymptotic expansion of $L_{{}_{m=s^2,s}}(z)$ is equal to the sum of (\ref{assL[m,s,z,0]}), (\ref{assL[m,s,z,1]})
and (\ref{assL[m,s,z,2]}). Dots in each term
\[
\Big( {\textstyle\frac{(-1)^k j_p^k}{k!}} + \ldots \Big){\textstyle\frac{1}{s^k}}
\]
denote polynomials in $j_p,r,q$, which are of degree in $j_p$ stricly
less than $k$. These polynomials are inessential  for us because of the order identities 
(\ref{a'order}) -- (\ref{aorder}). In the derivation of the asymptotic expansion
(\ref{assL[m,s,z,0]}) -- (\ref{assL[m,s,z,2]}) we have used the asymptotic expansion $\widesim[3]{s \rightarrow +\infty}$ of the 
binomials
\[
{\textstyle\binom{(s^2-j_0)/2+q}{s+r+q-1}}, {\textstyle\binom{(s^2-j_1)/2+q}{s+r+q-2}}, {\textstyle\binom{(s^2-j_2)/2+q}{s+r+q-3}}
\] 
which are contained in the contributions to $L_{{}_{m=s^2,s}}(z)$, coming from the simple fractions
containing ${a}^{{}^{\pm}}_{{}_{0,j_0}}$,  ${a}^{{}^{\pm}}_{{}_{1,j_1}}$ and, respectively, ${a}^{{}^{\pm}}_{{}_{2,j_2}}$.

\begin{lem}
Let $\alpha=\beta$ and $\alpha'=\beta'$ in (\ref{ProjF}).
For each $z \neq 0$,  analytic continuations of (\ref{Kalphaalpha}), (\ref{ProjF}) and (\ref{CyclicProj}), are nonzero
functions of $\rho$, except for they are identically zero, which is the case if and only if
$l=l_0 =1$ in  (\ref{Kalphaalpha}) and $l_0=l=1$ or $l_0=l'=1$ in (\ref{ProjF}).
\label{ProjF=/=0}
\end{lem}
\qedsymbol \,
We apply the formulas (\ref{assL[m,s,z,0]}) -- (\ref{assL[m,s,z,2]}) to our series 
(\ref{GeneralSeries}), representing (\ref{Kalphaalpha}), (\ref{ProjF}) or (\ref{CyclicProj}), 
remembering that the singly primed coefficients ${a'}^{{}^{\pm}}_{{}_{p,j_p}}$, representing the
contribution from the second integrand in (\ref{Kalphaalpha}) or (\ref{ProjF}), are in addition
multiplied by $z$. We compute the leading order contribution to $L_{{}_{m=s^2,s}}(z)$, with respect to $s$
going to infinity. Taking into account the order identities (\ref{a'order}) -- (\ref{a''order}), it is easily seen 
that for (\ref{GeneralSeries}), representing (\ref{ProjF}), 
the leading order contributions to $L_{{}_{m=s^2,s}}(z)$, in the asymptotic $\widesim[3]{s \rightarrow +\infty}$, 
are equal
\begin{multline}\label{assProjF}
{\textstyle\frac{i^2(-2iz)^{s-2}}{e\sqrt{2\pi s}}}
\big(\textstyle\frac{es}{2}\big)^s \big(\textstyle\frac{s}{2}\big)^{q-2}
(-is^2)^{l+l'}s\sum\limits_{j_1}{a''}^{{}^{+}}_{{}_{1,j_1,l+l'}}{\textstyle\frac{j_{1}^{l+l'+1}(-1)^{l+l'+1}}{(l+l'+1)! \,\, s^{l+l'+1}}}
\\
+
{\textstyle\frac{i^3(-2iz)^{s-3}z}{e\sqrt{2\pi s}}}
\big(\textstyle\frac{es}{2}\big)^s \big(\textstyle\frac{s}{2}\big)^{q-3}
(-is^2)^{l+l'-1}s^2\sum\limits_{j_2}{a'}^{{}^{+}}_{{}_{2,j_2,l+l'-l_0}}{\textstyle\frac{j_{2}^{l+l'-l_0}(-1)^{l+l'-l_0}}{(l+l'+1)! \,\, s^{l+l'-l_0}}}
\end{multline}
with $q=2(l+l')$, and with $l=l'$ in this formula for the leading order contributions to $L_{{}_{m=s^2,s}}(z)$ for (\ref{Kalphaalpha}).
Similarly, using the identities (\ref{a'order}), we see that for (\ref{CyclicProj}) the leading order contribution 
to $L_{{}_{m=s^2,s}}(z)$, in the asymptotic $\widesim[3]{s \rightarrow +\infty}$, is equal
\begin{equation}\label{assCyclicProj}
{\textstyle\frac{i^2(-2iz)^{s-2}}{e\sqrt{2\pi s}}}
\big(\textstyle\frac{es}{2}\big)^s \big(\textstyle\frac{s}{2}\big)^{q-2}
(-is^2)^{l}s\sum\limits_{j_1}{a}^{{}^{+}}_{{}_{1,j_1,l}}{\textstyle\frac{j_{1}^{l}(-1)^{l}}{l! \,\, s^{l}}}, \,\,\ q=2l-1.
\end{equation}

From (\ref{a'order}), it follows that for any fixed and nonzero $z$, (\ref{assCyclicProj}) is nonzero for appropriately large $s$.
Therefore, $L_{{}_{m=s^2,s}}(z) \neq 0$ for all sufficiently large $s$. Thus (\ref{CyclicProj}) is a nonzero function of $\rho$
for each nonzero value of $z$. In particular, for each $z>0$, (\ref{CyclicProj}) is a nonzero function of $\rho$.

Let $z\neq 0$ in (\ref{assProjF}). From (\ref{a'order}) -- (\ref{a''order}) it follows that
the contributions in  (\ref{assProjF}), can cancel out for all $s \rightarrow +\infty$, if and only if they are of the 
same order in $s$. This cancellation holds if and only if $l_0=1$ and
\begin{equation}\label{assProjF=0}
-i {\textstyle\frac{(-1)^{l+l'+1}}{(l+l'+1)!}}\sum\limits_{j_1}{a''}^{{}^{+}}_{{}_{1,j_1,l+l'}}j_{1}^{l+l'+1}
= {\textstyle\frac{(-1)^{l+l'-1}}{(l+l'-1)!}}\sum\limits_{j_1}{a'}^{{}^{+}}_{{}_{2,j_2,l+l'-1}}j_{2}^{l+l'-1}
\end{equation}
with $l=l'$ for (\ref{Kalphaalpha}).  Therefore, if $l_0 \neq 1$, we are left with one of the two terms in (\ref{assProjF}),
as the leading term. Therefore, if $l_0 \neq 1$, then for each nonzero $z$, in particular for each $z>0$, 
(\ref{ProjF}) and (\ref{Kalphaalpha}) are nonzero functions of $\rho$.
It remains to investigate the case $l_0=1$ for  (\ref{ProjF}) and (\ref{Kalphaalpha}). 
Inserting (\ref{suma''1}) and (\ref{suma'2}) into (\ref{assProjF=0}),
and using standard binomial identities, we easily see that (\ref{assProjF=0}) is equivalent to
\[
{\textstyle\frac{2}{(l+l'+2)(l+l'+1)}\binom{l+l'+2}{l+1}} =1
\]
which is equivalent to the assertion that $l=1$ and $l'$ is any natural $\geq 1$, or 
$l'=1$ and $l$ is any natural $\geq 1$. Therefore, if $l_0=1$, then the leading order term in the asymptotic
$\widesim[3]{s \rightarrow +\infty}$ of  $L_{{}_{m=s^2,s}}(z)$ is zero for  
(\ref{ProjF}) or, respectively, for (\ref{Kalphaalpha}), if and only if $l$ or $l'=1$ in 
(\ref{ProjF}), respectively, in (\ref{Kalphaalpha}). 
Thus, $L_{{}_{m=s^2,s}}(z) \neq 0$ for (\ref{ProjF}) and (\ref{Kalphaalpha})
if $l_0 =1$ and $\textrm{min}(l,l') >1$, which shows that for each $z \neq 0$, 
(\ref{ProjF}) and (\ref{Kalphaalpha}) are nonzero functions of $\rho$, if $l_0=1$
and $\textrm{min}(l,l') >1$, accordingly with Lemma \ref{LzeroProjF}.
\qed

Note that lemma \ref{ProjF=/=0} for (\ref{CyclicProj}) follows already from lemma \ref{ResiduePart},
so the essential part of lemma \ref{ProjF=/=0} is the assertion concerning 
(\ref{Kalphaalpha}) and (\ref{ProjF}). In lemmas \ref{ResiduePart} -- \ref{ProjF=/=0}, 
we have used the representations $U^{{}^{(l_0,i\rho)}}$ with non-negative integer $l_0$, but
with any real $\rho$. With the convention $l_0 \in \mathbb{Z}$ and $\rho \geq 0$, numbering the 
equivalence classes of representations $(l_0,l_1=i\rho)$ (used in lemma \ref{L3}), lemma \ref{ProjF=/=0} says, 
that for each positive $z$, (\ref{Kalphaalpha}), (\ref{ProjF}) and (\ref{CyclicProj}) are nonzero functions of $\rho$
except $|l_0|=l=1$ or $|l_0|=l'=1$ in (\ref{ProjF}), (\ref{Kalphaalpha}). With this convention
lemma \ref{LzeroProjF} implies that (\ref{ProjF}), (\ref{Kalphaalpha}) are identically zero functions
of $z,\rho$, whenever $|l_0|=l=1$ or $|l_0|=l'=1$ in (\ref{ProjF}), (\ref{Kalphaalpha}).

\section{Relation between the subspaces with cyclic vectors}\label{CyclicDomains}

\begin{lem}
\begin{align*}
\mathcal{H}_{{}_{c_{{}_{\alpha'}}^+|u\rangle}} \subset
\mathcal{H}_{{}_{c_{{}_{\alpha''}}^+|u\rangle}}, &
\,\,\,
\textrm{whenever}
\,\,
l(\alpha') \leq l(\alpha''),
\\
\textrm{and} \,\, \mathcal{H}_{{}_{|u\rangle}} \subset
\mathcal{H}_{{}_{c_{{}_{\alpha}}^+|u\rangle}}. &
\end{align*}
\label{Hcalpha'|u>,Hcalpha''|u>}
\end{lem}
\qedsymbol \,
This is a corollary of lemmas \ref{L3}, \ref{ResiduePart}, \ref{ProjF=/=0}. Let $l(\alpha') \leq l(\alpha'')$. 
We apply the same proof as that of lemma \ref{LCyc|u>andc10+|u>} and remark \ref{ResidueComponentsComparison}
(instead of remark \ref{ResidueComponentsComparison} and lemma \ref{ResiduePart}, 
one can use lemma \ref{BoundStateComponentsComparison}). 
From lemmas  \ref{ResiduePart}, \ref{ProjF=/=0} it follows that the corresponding 
direct sum $l_0$-components, for $-l(\alpha') \leq l_0 \leq l(\alpha')$,
in the decomposition of lemma \ref{L3} of the cyclic spaces
$\mathcal{H}_{{}_{c_{{}_{\alpha'}}^+|u\rangle}},
\mathcal{H}_{{}_{c_{{}_{\alpha''}}^+|u\rangle}}$, coincide, 
including the supplementary component. By lemma \ref{L3}, decomposition of $\mathcal{H}_{{}_{c_{{}_{\alpha'}}^+|u\rangle}}$
does not contain direct sum components with $l_0 < -l(\alpha')$ and $l(\alpha') < l_0$, whence the inclusion.
In particular, it follows from lemma \ref{ResiduePart} that $\mathcal{H}_{{}_{|u\rangle}} \subset
\mathcal{H}_{{}_{c_{{}_{\alpha}}^+|u\rangle}}$ for each $l(\alpha) =1,2, \ldots$. 
In particular, the supplementary component of $\mathcal{H}_{{}_{c_{{}_{\alpha}}^+|u\rangle}}$ is nonzero
and is common for all $\mathcal{H}_{{}_{c_{{}_{\alpha}}^+|u\rangle}}$ and $\mathcal{H}_{{}_{|u\rangle}}$.     
\qed

Let $U\big(g_{{}_{03}}(\lambda)\big) = U(\lambda)$. Using (\ref{CommutationRules}), the first transformation rule in (\ref{c'S'}) 
and  (\ref{U(lambda)|u>}), we obtain the following generalization of the expansion
(\ref{U(lambda)|u>}) 
\begin{multline}\label{U(lambda)cl0+|u>}
U(\lambda)c_{{}_{\ell_1,m_1}}^+|u\rangle  
\\
= \sum\limits_{\ell'_1}\overline{A_{{}_{\ell'_1,m_1 \,\,\, \ell_1,m_1}}(\lambda)}e^{-{\textstyle\frac{n^2}{8\pi}}||B(\lambda)||^2} 
\textrm{exp}\left[{\textstyle\frac{n}{4\pi \mathfrak{e}}\sum\limits_{l}B_{{}_{l,0}}(-\lambda)}c_{{}_{l,0}}^+\right]c_{{}_{\ell'_1,m_1}}^+|u\rangle
\\
+ \delta_{{}_{m_1 \, 0}} \,  n\mathfrak{e}\overline{B_{{}_{\ell_1,0}}(\lambda)} e^{-{\textstyle\frac{n^2}{8\pi}}||B(\lambda)||^2}
\textrm{exp}\left[{\textstyle\frac{n}{4\pi \mathfrak{e}}\sum\limits_{l}B_{{}_{l,0}}(-\lambda)}c_{{}_{l,0}}^+\right]|u\rangle,
\end{multline}
and  
\begin{multline}\label{U(lambda)cli10+...clq+|u>}
U(\lambda)c_{{}_{\ell_1,m_1}}^+ \ldots c_{{}_{\ell_\mathfrak{q},m_\mathfrak{q}}}^+|u\rangle  
\\
= \sum\limits_{{}_{\ell'_1, \ldots, \ell'_\mathfrak{q}}} \prod\limits_{i}^{\mathfrak{q}} \left[ \overline{A_{{}_{\ell'_i,m_i \,\,\, \ell_i,m_i}}(\lambda)}\right]
e^{-{\textstyle\frac{n^2}{8\pi}}||B(\lambda)||^2} 
\textrm{exp}\left[{\textstyle\frac{n}{4\pi \mathfrak{e}}\sum\limits_{l}B_{{}_{l,0}}(-\lambda)}c_{{}_{l,0}}^+\right]
\prod\limits_{i}^{\mathfrak{q}} c_{{}_{\ell'_i,m_i}}^+ |u\rangle \,\,\, 
\\
+ \,\,\, \ldots
\\
+ \prod\limits_{i}^{\mathfrak{q}} \left[\delta_{{}_{m_i \, 0}} \, n\mathfrak{e}\overline{B_{{}_{\ell_i,0}}(\lambda)}\right]
e^{-{\textstyle\frac{n^2}{8\pi}}||B(\lambda)||^2}
\textrm{exp}\left[{\textstyle\frac{n}{4\pi \mathfrak{e}}\sum\limits_{l}B_{{}_{l,0}}(-\lambda)}c_{{}_{l,0}}^+\right]|u\rangle,
\end{multline}
where dots denote terms in which one of the factors $\overline{A_{{}_{\ell'_i,m_i \,\,\, \ell_i,m_i}}(\lambda)} c_{{}_{\ell'_i,m_i}}^+$
in the first sum is replaced with $\delta_{{}_{m_i \, 0}} \,n\mathfrak{e}\overline{B_{{}_{\ell_i,0}}(\lambda)}$, with the summation over the corresponding
$\ell'_i$ withdrawn, and further terms, which arise from the first sum in which two 
\[
\overline{A_{{}_{\ell'_i,m_i \,\,\, \ell_i,m_i}}(\lambda)}c_{{}_{\ell'_i,m_i}}^+, 
\,\, \overline{A_{{}_{\ell'_j,m_j \,\,\, \ell_j,m_j}}(\lambda)} c_{{}_{\ell'_j,m_j}}^+,
\]
of the factors in the first sum are replaced with
\[
\left(\delta_{{}_{m_i \, 0}} \, n\mathfrak{e}\overline{B_{{}_{\ell_i,0}}(\lambda)}\right), 
\,\, \left(n\mathfrak{e}\overline{\delta_{{}_{m_j \, 0}} \, B_{{}_{\ell_j,0}}(\lambda)}\right),
\]
with the summation over the corresponding
$\ell'_i, \ell'_j$, withdrawn, and so on, up to the last term written explicitly in (\ref{U(lambda)cli10+...clq+|u>}). 
We use the expansions (\ref{U(lambda)cl0+|u>}), (\ref{U(lambda)cli10+...clq+|u>})
in the same way as the expansion (\ref{U(lambda)|u>}) in Subsection \ref{RelationCycRep}, utilizing the fact that $B_{{}_{l,0}}(\lambda)$
is of $l$-order at zero, and $A_{{}_{l,m \,\,\, l',m}}(\lambda) = U^{{}^{(1,0)}}_{{}_{l,m \,\,\, l',m}}(\lambda)$ 
is of $|l-l'|$-order at zero, compare (\ref{Bexpansion}) and 
(\ref{Uexpansion}), with $l_0=1, \rho=0$, in (\ref{Uexpansion}).

\begin{lem}
\[
c_{{}_{\ell_1,m_1}}^+c_{{}_{\ell_2,m_2}}^+ \ldots c_{{}_{\ell_\mathfrak{q},m_\mathfrak{q}}}^+|u\rangle 
\in \mathcal{H}_{{}_{c_{{}_{\ell,0}}^+|u\rangle}},
\,\,
\textrm{whenever} 
\,\,
\sum\limits_{j=1}^{\mathfrak{q}} \ell_j \leq \ell,
\,\,
-\ell_i \leq m_i \leq \ell_i.
\]
\label{azimuth=0}
\end{lem}
\qedsymbol \,
We proceed by induction with respect to $\mathfrak{q}$. Let $\mathfrak{q}=1$.
The cyclic subspace of the lemma is by definition invariant under $U$, and thus invariant under the generators
$M_{23},M_{13},M_{23}$ of the action of the subgroup $SU(2,\mathbb{C})\subset G$. The vector $|u\rangle$
is invariant under the action of this subgroup. From this and from the first transformation rule in (\ref{c'S'}),
with $B(a)=0$, $a \in SU(2,\mathbb{C})$, it follows that this subgroup acts on the vectors 
$\xi_{\ell_1,m_1} = c_{{}_{\ell_1,m_1}}^+|u\rangle$, 
in the cyclic subspace of the lemma with $\mathfrak{q}=1$, by the standard representation, given by right multiplication by the matrix
$\overline{A_{{}_{\ell'_1,m'_1 \,\,\, \ell_1,m_1}}(a)} = \delta_{{}_{\ell_1 \,\, \ell'_1}} \overline{T^{{}^{\ell_1}}_{{}_{m'_1 \,\, m_1}}(a)}$,
$a \in SU(2,\mathbb{C})$, with
\begin{multline}\label{H-+xq=1}
H_+\xi_{{}_{\ell_1,m_1}} = \sqrt{(\ell_1+m_1+1)(\ell_1-m_1)} \,\, \xi_{{}_{\ell_1,m_1+1}}, 
\\
H_-\xi_{{}_{\ell_1,m_1}} = \sqrt{(\ell_1+m_1)(\ell_1-m_1+1)} \,\,  \xi_{{}_{\ell_1,m_1-1}},
\end{multline}
where $H_{\pm} = M_{23} \pm i M_{13}$. This representation reduces to the standard representation 
of weight $\ell_1$, \cite{Geland-Minlos-Shapiro}, on each invariant subspace, 
spanned by $\xi_{\ell_1,m_1}$, $-\ell_1 \leq m_1 \leq \ell_1$.
The lemma \ref{azimuth=0} for $\mathfrak{q}=1$ follows from (\ref{H-+xq=1}) 
and from lemma \ref{Hcalpha'|u>,Hcalpha''|u>}. 

Now we proceed to the proof of the inductive step: assuming the assertion of lemma \ref{azimuth=0} for $\mathfrak{q}\leq \mathfrak{q'}$, 
we show the assertion of lemma \ref{azimuth=0} for $\mathfrak{q} = \mathfrak{q'}+1$.

We first prove the inductive step for $m_{\mathfrak{q'}+1} =0$. The proof of this assertion we divide, in turn, into several
cases: 1) $\ell_{\mathfrak{q'}+1} =1$, 2) $\ell_{\mathfrak{q'}+1} = 2$, \ldots, $\ell_{\mathfrak{q'}+1}$) 
$\ell_{\mathfrak{q'}+1} = \sum \ell_i - \ell_{\mathfrak{q'}+1}$.

Let us consider now the case 1). Let us denote by $\big| \overset{i}{w}\big\rangle$ the vector which arises from
the vector 
\[
|w\rangle = c_{{}_{\ell_1,m_1}}^+c_{{}_{\ell_2,m_2}}^+ \ldots c_{{}_{\ell_\mathfrak{q'},m_\mathfrak{q'}}}^+|u\rangle 
\in \mathcal{H}_{{}_{c_{{}_{\ell,0}}^+|u\rangle}} \subset \mathcal{H}_{{}_{c_{{}_{\ell+1,0}}^+|u\rangle}}, 
\]
by removing in $|w\rangle$ the factor $c_{{}_{\ell_i,m_i}}^+$. The inclusion follows from lemma \ref{Hcalpha'|u>,Hcalpha''|u>}.
By the inductive assumption  
\[
\big| \overset{i}{w}\big\rangle \in \mathcal{H}_{{}_{c_{{}_{\ell-\ell_i,0}}^+|u\rangle}} \subset \mathcal{H}_{{}_{c_{{}_{\ell+1,0}}^+|u\rangle}}.
\]
Let us denote by $\big|\overset{\ell_i-1}{w}\big\rangle$, $\big| \overset{\ell_i+1}{w}\big\rangle$, the vectors 
\begin{align*}
\big|\overset{\ell_i-1}{w}\big\rangle = c_{{}_{\ell_1,m_1}}^+ \ldots c_{{}_{\ell_i-1,m_i}}^+ \ldots c_{{}_{\ell_\mathfrak{q'},m_\mathfrak{q'}}}^+|u\rangle,
\\
\big|\overset{\ell_i+1}{w}\big\rangle = c_{{}_{\ell_1,m_1}}^+ \ldots c_{{}_{\ell_i+1,m_i}}^+ \ldots c_{{}_{\ell_\mathfrak{q'},m_\mathfrak{q'}}}^+|u\rangle.
\end{align*}
By the inductive assumption
\begin{align*}
\big|\overset{\ell_i+1}{w}\big\rangle \in \mathcal{H}_{{}_{c_{{}_{\ell+1,0}}^+|u\rangle}},
\\
\big|\overset{\ell_i-1}{w}\big\rangle \in \mathcal{H}_{{}_{c_{{}_{\ell-1,0}}^+|u\rangle}} \subset \mathcal{H}_{{}_{c_{{}_{\ell+1,0}}^+|u\rangle}},
\end{align*}
with the last inclusion following from lemma \ref{Hcalpha'|u>,Hcalpha''|u>}.
To the vector $|w\rangle$ we apply $U(\lambda)$,
using (\ref{U(lambda)cli10+...clq+|u>}), and arriving with the expansion
\begin{multline*}
e^{{\textstyle\frac{n^2}{8\pi}}||B(\lambda)||^2}U(\lambda)|w\rangle -|w\rangle
\\
= {\textstyle\frac{nB_{{}_{1,0}}(-\lambda)}{4\pi \mathfrak{e}}}c_{{}_{1,0}}^+|w\rangle
+ {\textstyle\frac{nB_{{}_{2,0}}(-\lambda)}{4\pi \mathfrak{e}}}c_{{}_{2,0}}^+|w\rangle + \ldots
\end{multline*}
\begin{multline*}
+ {\textstyle\frac{1}{2!}}\Bigg[
\left({\textstyle\frac{nB_{{}_{1,0}}(-\lambda)}{4\pi \mathfrak{e}}}\right)
\left({\textstyle\frac{nB_{{}_{1,0}}(-\lambda)}{4\pi \mathfrak{e}}}\right)
c_{{}_{1,0}}^+ c_{{}_{1,0}}^+|w\rangle
\\
+
\left({\textstyle\frac{nB_{{}_{1,0}}(-\lambda)}{4\pi \mathfrak{e}}}\right)
\left({\textstyle\frac{nB_{{}_{2,0}}(-\lambda)}{4\pi \mathfrak{e}}}\right)
c_{{}_{1,0}}^+ c_{{}_{2,0}}^+|w\rangle
\\
+
\left({\textstyle\frac{nB_{{}_{2,0}}(-\lambda)}{4\pi \mathfrak{e}}}\right)
\left({\textstyle\frac{nB_{{}_{1,0}}(-\lambda)}{4\pi \mathfrak{e}}}\right)
c_{{}_{2,0}}^+ c_{{}_{1,0}}^+|w\rangle \,\, + \,\, \ldots 
\Bigg]
\end{multline*}
\begin{multline}\label{U|w>-|w>Expansion'}
\,\,\,\,\,\,\,\,\,\,\,\,\,\,\,\,\,\,\,\,\,\,\,\,\,\,\,\,\,\,
+ {\textstyle\frac{1}{3!}}\Bigg[
\left({\textstyle\frac{nB_{{}_{1,0}}(-\lambda)}{4\pi \mathfrak{e}}}\right)
\left({\textstyle\frac{nB_{{}_{1,0}}(-\lambda)}{4\pi \mathfrak{e}}}\right)
\left({\textstyle\frac{nB_{{}_{1,0}}(-\lambda)}{4\pi \mathfrak{e}}}\right)
c_{{}_{1,0}}^+ c_{{}_{1,0}}^+c_{{}_{1,0}}^+|w\rangle
+ \,\,\, \ldots \Bigg] 
\\
+ \sum\limits_{i=1}^{\mathfrak{q'}} \overline{A_{{}_{\ell'_i+1,m_i \,\,\, \ell_i,m_i}}(\lambda)} \, \big|\overset{\ell_i+1}{w}\big\rangle
+ \sum\limits_{i=1}^{\mathfrak{q'}} \overline{A_{{}_{\ell'_i-1,m_i \,\,\, \ell_i,m_i}}(\lambda)} \, \big|\overset{\ell_i-1}{w}\big\rangle
\\
+ \sum\limits_{i=1}^{\mathfrak{q'}} \delta_{{}_{m_i \, 0}} n\mathfrak{e}\overline{B_{{}_{\ell_i,0}}(\lambda)} \, \big| \overset{i}{w}\big\rangle 
\,\,\, + \,\,\, \ldots,
\end{multline}
where dots denote higher-order terms in $\lambda$. 
We have used the fact that $B_{{}_{l,0}}(-\lambda)$ is a quantity of order $l$, and $A_{{}_{l,m \,\,\, l',m}}(\lambda)$ 
is of order $|l-l'|$, at $\lambda=0$. Thus, the vector
\begin{multline*}
{\textstyle\frac{1}{{\textstyle\frac{-n}{4\pi \mathfrak{e}}}B_{{}_{1,0}}(-\lambda)}}
\Bigg(e^{{\textstyle\frac{n^2}{8\pi}}||B(\lambda)||^2}U(\lambda)|w\rangle -|w\rangle 
\\
- \sum\limits_{i=1}^{\mathfrak{q'}} \overline{A_{{}_{\ell'_i+1,m_i \,\,\, \ell_i,m_i}}(\lambda)} \, \big|\overset{\ell_i+1}{w}\big\rangle
- \sum\limits_{i=1}^{\mathfrak{q'}} \overline{A_{{}_{\ell'_i-1,m_i \,\,\, \ell_i,m_i}}(\lambda)} \, \big|\overset{\ell_i-1}{w}\big\rangle
\\
- \sum\limits_{i=1}^{\mathfrak{q'}} \delta_{{}_{m_i \, 0}} n\mathfrak{e}\overline{B_{{}_{\ell_i,0}}(\lambda)} \, \big| \overset{i}{w}\big\rangle 
\Bigg) \in \mathcal{H}_{{}_{c_{{}_{\ell+1,0}}^+|u\rangle}},
\end{multline*}
has the expansion
\begin{multline*}
c_{{}_{1,0}}^+|w\rangle
+ {\textstyle\frac{B_{{}_{2,0}}(-\lambda)}{B_{{}_{1,0}}(-\lambda)}}c_{{}_{2,0}}^+|w\rangle
+ {\textstyle\frac{B_{{}_{3,0}}(-\lambda)}{B_{{}_{1,0}}(-\lambda)}}c_{{}_{3,0}}^+|w\rangle
 + \ldots
\\
+ {\textstyle\frac{1}{2!}}\Bigg[
\left({\textstyle\frac{nB_{{}_{1,0}}(-\lambda)}{4\pi \mathfrak{e}}}\right)
c_{{}_{1,0}}^+ c_{{}_{1,0}}^+|w\rangle
\\
+
\left({\textstyle\frac{nB_{{}_{2,0}}(-\lambda)}{4\pi \mathfrak{e}}}\right)
c_{{}_{1,0}}^+ c_{{}_{2,0}}^+|w\rangle
\\
+
\left({\textstyle\frac{nB_{{}_{2,0}}(-\lambda)}{4\pi \mathfrak{e}}}\right)
c_{{}_{2,0}}^+ c_{{}_{1,0}}^+|w\rangle \,\, + \,\, \ldots 
\Bigg]
\end{multline*}
\begin{multline*}
\,\,\,\,\,\,\,\,\,\,\,\,\,\,\,\,\,\,\,\,\,\,\,\,\,\,\,\,\,\,
+ {\textstyle\frac{1}{3!}}\Bigg[
\left({\textstyle\frac{nB_{{}_{1,0}}(-\lambda)}{4\pi \mathfrak{e}}}\right)
\left({\textstyle\frac{nB_{{}_{1,0}}(-\lambda)}{4\pi \mathfrak{e}}}\right)
c_{{}_{1,0}}^+ c_{{}_{1,0}}^+c_{{}_{1,0}}^+|w\rangle
+ \,\,\, \ldots \Bigg] 
\end{multline*}
converging to 
\[
c_{{}_{1,0}}^+|w\rangle 
\in \mathcal{H}_{{}_{c_{{}_{\ell+1,0}}^+|u\rangle}},
\]
if $\lambda \rightarrow 0$, because $\mathcal{H}_{{}_{c_{{}_{\ell,0}}^+|u\rangle}}$ is closed and dots 
denote terms of first order at least in $\lambda$. 
Thus, we see that 
\[
c_{{}_{1,0}}^+|w\rangle   
= c_{{}_{\ell_1,m_1}}^+ \ldots c_{{}_{\ell_{\mathfrak{q'}},m_\mathfrak{q'}}}^+c_{{}_{1,0}}^+|u\rangle
\in \mathcal{H}_{{}_{c_{{}_{\ell+1,0}}^+|u\rangle}},
\]
which proves the inductive step in case 1).

Now we proceed to the case 2). We again use the inductive assumption and validity of the inductive step
in case 1). Let $\ell_{\mathfrak{q'}+1}=2$. In addition to $|w\rangle$,  $\big|\overset{\ell_i-1}{w}\big\rangle$, 
$\big| \overset{\ell_i+1}{w}\big\rangle$, introduced in case 1), we need to introduce
\begin{align*}
\big|\overset{\ell_i-2}{w}\big\rangle = c_{{}_{\ell_1,m_1}}^+ \ldots c_{{}_{\ell_i-2,m_i}}^+ \ldots c_{{}_{\ell_\mathfrak{q'},m_\mathfrak{q'}}}^+|u\rangle,
\\
\big|\overset{\ell_i+2}{w}\big\rangle = c_{{}_{\ell_1,m_1}}^+ \ldots c_{{}_{\ell_i+2,m_i}}^+ \ldots c_{{}_{\ell_\mathfrak{q'},m_\mathfrak{q'}}}^+|u\rangle,
\end{align*}
and $\big|\overset{ij}{w}\big\rangle$, which arises from $|w\rangle$ by removing two operators: $c_{{}_{\ell_i,m_i}}^+$,$c_{{}_{\ell_j,m_j}}^+$.
By the inductive assumption
\begin{align*}
\big|\overset{\ell_i+2}{w}\big\rangle \in \mathcal{H}_{{}_{c_{{}_{\ell+2,0}}^+|u\rangle}},
\\
\big|\overset{\ell_i-2}{w}\big\rangle \in \mathcal{H}_{{}_{c_{{}_{\ell-2,0}}^+|u\rangle}} \subset \mathcal{H}_{{}_{c_{{}_{\ell+2,0}}^+|u\rangle}},
\\
\big|\overset{ij}{w}\big\rangle \in \mathcal{H}_{{}_{c_{{}_{\ell-\ell_i-\ell_j,0}}^+|u\rangle}} \subset \mathcal{H}_{{}_{c_{{}_{\ell+2,0}}^+|u\rangle}}.
\end{align*}
Because 
\[
c_{{}_{1,0}}^+|w\rangle \in \mathcal{H}_{{}_{c_{{}_{\ell+1,0}}^+|u\rangle}} \subset \mathcal{H}_{{}_{c_{{}_{\ell+2,0}}^+|u\rangle}}, 
\]
then 
\begin{multline*}
{\textstyle\frac{B_{{}_{1,0}}(-\lambda)}{B_{{}_{2,0}}(-\lambda)}}\Bigg({\textstyle\frac{1}{{\textstyle\frac{-n}{4\pi \mathfrak{e}}}B_{{}_{1,0}}(-\lambda)}}
\Bigg(e^{{\textstyle\frac{n^2}{8\pi}}||B(\lambda)||^2}U(\lambda)|w\rangle -|w\rangle 
\\
- \sum\limits_{i=1}^{\mathfrak{q'}} \overline{A_{{}_{\ell'_i+1,m_i \,\,\, \ell_i,m_i}}(\lambda)} \, \big|\overset{\ell_i+1}{w}\big\rangle
- \sum\limits_{i=1}^{\mathfrak{q'}} \overline{A_{{}_{\ell'_i-1,m_i \,\,\, \ell_i,m_i}}(\lambda)} \, \big|\overset{\ell_i-1}{w}\big\rangle
\\
- \sum\limits_{i=1}^{\mathfrak{q'}} \delta_{{}_{m_i \, 0}} n\mathfrak{e}\overline{B_{{}_{\ell_i,0}}(\lambda)} \, \big| \overset{i}{w}\big\rangle
\end{multline*}
\begin{multline*}
- \sum\limits_{i=1}^{\mathfrak{q'}} \overline{A_{{}_{\ell'_i+2,m_i \,\,\, \ell_i,m_i}}(\lambda)} \, \big|\overset{\ell_i+2}{w}\big\rangle
- \sum\limits_{i=1}^{\mathfrak{q'}} \overline{A_{{}_{\ell'_i-2,m_i \,\,\, \ell_i,m_i}}(\lambda)} \, \big|\overset{\ell_i-2}{w}\big\rangle
\\
- \sum\limits_{i,j=1}^{\mathfrak{q'}} \delta_{{}_{m_i \, 0}} n\mathfrak{e}\overline{B_{{}_{\ell_i,0}}(\lambda)}
\delta_{{}_{m_j \, 0}} n\mathfrak{e}\overline{B_{{}_{\ell_j,0}}(\lambda)}
\, \big| \overset{ij}{w}\big\rangle \Bigg) 
- c_{{}_{1,0}}^+|w\rangle\Bigg) 
\in \mathcal{H}_{{}_{c_{{}_{\ell+2,0}}^+|u\rangle}}.
\end{multline*}
From (\ref{U|w>-|w>Expansion'}) 
it follows that this vector has the expansion
\begin{multline*}
c_{{}_{2,0}}^+|w\rangle
+ {\textstyle\frac{B_{{}_{3,0}}(-\lambda)}{B_{{}_{2,0}}(-\lambda)}}c_{{}_{3,0}}^+|w\rangle
+ {\textstyle\frac{B_{{}_{4,0}}(-\lambda)}{B_{{}_{2,0}}(-\lambda)}}c_{{}_{4,0}}^+|w\rangle
 + \ldots
\\
+ {\textstyle\frac{1}{2!}}\Bigg[
{\textstyle\frac{B_{{}_{1,0}}(-\lambda)}{B_{{}_{2,0}}(-\lambda)}}
\left({\textstyle\frac{nB_{{}_{1,0}}(-\lambda)}{4\pi \mathfrak{e}}}\right)
c_{{}_{1,0}}^+ c_{{}_{1,0}}^+|w\rangle
\\
+
{\textstyle\frac{B_{{}_{1,0}}(-\lambda)}{B_{{}_{2,0}}(-\lambda)}}
\left({\textstyle\frac{nB_{{}_{2,0}}(-\lambda)}{4\pi \mathfrak{e}}}\right)
c_{{}_{1,0}}^+ c_{{}_{2,0}}^+|w\rangle
\\
+
{\textstyle\frac{B_{{}_{1,0}}(-\lambda)}{B_{{}_{2,0}}(-\lambda)}}
\left({\textstyle\frac{nB_{{}_{2,0}}(-\lambda)}{4\pi \mathfrak{e}}}\right)
c_{{}_{2,0}}^+ c_{{}_{1,0}}^+|w\rangle \,\, + \,\, \ldots 
\Bigg]
\end{multline*}
\begin{multline*}
\,\,\,\,\,\,\,\,\,\,\,\,\,\,\,\,\,\,\,\,\,\,\,\,\,\,\,\,\,\,
+ {\textstyle\frac{1}{3!}}\Bigg[
{\textstyle\frac{B_{{}_{1,0}}(-\lambda)}{B_{{}_{2,0}}(-\lambda)}}
\left({\textstyle\frac{nB_{{}_{1,0}}(-\lambda)}{4\pi \mathfrak{e}}}\right)
\left({\textstyle\frac{nB_{{}_{1,0}}(-\lambda)}{4\pi \mathfrak{e}}}\right)
c_{{}_{1,0}}^+ c_{{}_{1,0}}^+c_{{}_{1,0}}^+|w\rangle
+ \,\,\, \ldots \Bigg] 
\\
\,\,\, + \,\,\, \ldots
\end{multline*}
which converges to 
\begin{multline*}
c_{{}_{2,0}}^+|w\rangle 
+ 
 {\textstyle\frac{1}{2!}}
\underset{\lambda \rightarrow 0}{\textrm{lim}}
{\textstyle\frac{{\textstyle\frac{n}{4\pi\mathfrak{e}}}B_{{}_{1,0}}(-\lambda){\textstyle\frac{n}{4\pi\mathfrak{e}}}B_{{}_{1,0}}(-\lambda)}{{\textstyle\frac{n}{4\pi\mathfrak{e}}}B_{{}_{2,0}}(-\lambda)}}
c_{{}_{1,0}}^+ c_{{}_{1,0}}^+|w\rangle 
\\
=
c_{{}_{2,0}}^+|w\rangle 
+ {\textstyle\frac{1}{2}} {\textstyle\frac{b_{{}_{1,0}}b_{{}_{1,0}}}{b_{{}_{2,0}}}}c_{{}_{1,0}}^+c_{{}_{1,0}}^+|w\rangle 
\in \mathcal{H}_{{}_{c_{{}_{\ell+2,0}}^+|u\rangle}}, 
\end{multline*}
if $\lambda \rightarrow 0$, because $\mathcal{H}_{{}_{c_{{}_{\ell,0}}^+|u\rangle}}$ is closed. Here $b_{{}_{l,0}}$
are the coefficients of the lowest order in the expansion (\ref{Bexpansion}).
By the case 1), already proved,
\[
c_{{}_{1,0}}^+c_{{}_{1,0}}^+|w\rangle 
\in \mathcal{H}_{{}_{c_{{}_{\ell+2,0}}^+|u\rangle}},
\]
therefore
\[
c_{{}_{2,0}}^+|w\rangle = c_{{}_{\ell_1,m_1}}^+ \ldots c_{{}_{\ell_{\mathfrak{q'}},m_\mathfrak{q'}}}^+c_{{}_{2,0}}^+|u\rangle 
\in \mathcal{H}_{{}_{c_{{}_{\ell+2,0}}^+|u\rangle}}, 
\]
which proves validity of the inductive step in case 2). 

It is not difficult to see, that we can continue in this way, proving the inductive step in (j+1)-case , using the inductive
assumption and all preceding cases, 1), 2), \ldots, j), constructing in this way the corresponding vectors 
\begin{multline*}
c_{{}_{j+1,0}}^+|w\rangle 
+ {\textstyle\frac{1}{2!}} \sum\limits_{{}_{l_1+l_2=j+1}}{\textstyle\frac{b_{{}_{l_1,0}}b_{{}_{l_2,0}}}{b_{{}_{j+1,0}}}}c_{{}_{l_1,0}}^+c_{{}_{l_2,0}}^+|w\rangle +
\ldots
\\
\ldots 
+ {\textstyle\frac{1}{(j+1)!}} \sum\limits_{{}_{l_1+\ldots +l_{j+1}=j+1}}{\textstyle\frac{b_{{}_{l_1,0}} \ldots b_{{}_{l_{j+1},0}}}{b_{{}_{j+1},0}}}c_{{}_{l_1,0}}^+
\ldots c_{{}_{l_{j+1},0}}^+|w\rangle 
\,\,\,\,
\in \mathcal{H}_{{}_{c_{{}_{\ell+j+1,0}}^+|u\rangle}},
\end{multline*}
where all terms, except the first, belong to $\mathcal{H}_{{}_{c_{{}_{\ell+j+1,0}}^+|u\rangle}}$, by the preceding cases
1), \ldots, j), and thus also with the first 
\[
c_{{}_{j+1,0}}^+|w\rangle =  c_{{}_{\ell_1,m_1}}^+ \ldots c_{{}_{\ell_{\mathfrak{q'}},m_\mathfrak{q'}}}^+c_{{}_{j+1,0}}^+|u\rangle 
\,\,\,\,
\in \mathcal{H}_{{}_{c_{{}_{\ell+j+1,0}}^+|u\rangle}},
\] 
for $\ell_{\mathfrak{q'} +1}=j+1$. Therefore, we have proved, that if the assertion of lemma \ref{azimuth=0}
holds for $\mathfrak{q} \leq \mathfrak{q'}$, then it holds for $\mathfrak{q'}+1$, if
$m_{\mathfrak{q'} +1}=0$, \emph{i.e.} with
\begin{equation}\label{l1m1...lqmqlq+10}
c_{{}_{\ell_1,m_1}}^+ \ldots c_{{}_{\ell_{\mathfrak{q'}},m_\mathfrak{q'}}}^+c_{{}_{\ell_{\mathfrak{q'}+1},0}}^+|u\rangle 
\,\,\,\,
\in \mathcal{H}_{{}_{c_{{}_{\Sigma \ell_i,0}}^+|u\rangle}}, 
\,\,\,
- \ell_{i} \leq m_{i} \leq \ell_{i}, \,\, 1 \leq i \leq \mathfrak{q'},
\end{equation}
following from the assertion of lemma \ref{azimuth=0} for $\mathfrak{q} \leq \mathfrak{q'}$. Here
\[
\Sigma \ell_i = \sum\limits_{i=1}^{\mathfrak{q'}+1} \ell_i. 
\]

Now using (\ref{l1m1...lqmqlq+10}) 
we show that also
\begin{equation}\label{l1m1...lqmqlq+1mq+1}
c_{{}_{\ell_1,m_1}}^+ \ldots c_{{}_{\ell_{\mathfrak{q'}},m_\mathfrak{q'}}}^+c_{{}_{\ell_{\mathfrak{q'}+1},m_{\mathfrak{q'}+1}}}^+|u\rangle 
\,\,\,\,
\in \mathcal{H}_{{}_{c_{{}_{\Sigma \ell_i,0}}^+|u\rangle}}, \,\,\, - \ell_{i} \leq m_{i} \leq  \ell_{i}, \,\, 1 \leq i \leq \mathfrak{q'} + 1,
\end{equation}
which will end the proof of the inductive step.

The subgroup $SU(2,\mathbb{C})$ acts on the states
\[
c_{{}_{\ell_1,m_1}}^+ \ldots c_{{}_{\ell_\mathfrak{q},m_\mathfrak{q}}}^+|u\rangle
\]
through the symmetrized tensor product of  the representation given by right multiplication by the matrix 
\[
\overline{A_{{}_{\ell_i,m_i \,\,\, \ell'_i,m'_i}}(a)} = \delta_{{}_{\ell_i \,\, \ell'_i}}\overline{T^{{}^{\ell_i}}_{{}_{m_i \,\, m'_i}}(a)},
\]
with
\begin{multline*}
H_{\pm} 
c_{{}_{\ell_1,m_1}}^+ \ldots c_{{}_{\ell_\mathfrak{q},m_\mathfrak{q}}}^+|u\rangle  
= \sum\limits_{i}\alpha^{{}^{\pm}}_{{}_{\ell_i,m_i}} \,\, c_{{}_{\ell_1,m_1}}^+ \ldots c_{{}_{\ell_i,m_i \pm 1}}^+ \ldots 
c_{{}_{\ell_\mathfrak{q},m_\mathfrak{q}}}^+|u\rangle,
\\
\alpha^{{}^{+}}_{{}_{\ell_i,m_i}} = \sqrt{(\ell_i+m_i+1)(\ell_i-m_i)}, 
\,\,\,
 \alpha^{{}^{-}}_{{}_{\ell_i,m_i}} =  \sqrt{(\ell_i+m_i)(\ell_i-m_i+1)}.
\end{multline*}
It is not difficult to see that all vectors (\ref{l1m1...lqmqlq+1mq+1}) can be expressed as linear combinations
of vectors which arise by applications of powers of $H_{\pm}$ to the vectors 
(\ref{l1m1...lqmqlq+10}). First we show (\ref{l1m1...lqmqlq+1mq+1}) for $m_{\mathfrak{q'}+1} = \pm 1$, and any $-\ell_1 \leq m_i \leq \ell_i$,
$i\leq \mathfrak{q'}$ in (\ref{l1m1...lqmqlq+1mq+1}). To this end we observe, that all vectors in 
\begin{multline}\label{Hpmc+limi...c+lq'+10}
H_{\pm} 
c_{{}_{\ell_1,m_1}}^+ \ldots c_{{}_{\ell_\mathfrak{q'},m_\mathfrak{q'}}}^+c_{{}_{\ell_{\mathfrak{q'}+1},0}}^+|u\rangle
= \alpha^{{}^{\pm}}_{{}_{\ell_1,m_1}} c_{{}_{\ell_1\pm 1,m_1}}^+ \ldots c_{{}_{\ell_\mathfrak{q},m_\mathfrak{q}}}^+c_{{}_{\ell_{\mathfrak{q}+1},0}}^+|u\rangle
\\
+\alpha^{{}^{\pm}}_{{}_{\ell_2,m_2}} c_{{}_{\ell_1,m_1}}^+c_{{}_{\ell_2 \pm 1,m_2}}^+ \ldots c_{{}_{\ell_\mathfrak{q'},m_\mathfrak{q'}}}^+c_{{}_{\ell_{\mathfrak{q'}+1},0}}^+|u\rangle + \ldots
\\
\ldots + \alpha^{{}^{\pm}}_{{}_{\ell_{\mathfrak{q'}+1},0}} c_{{}_{\ell_1,m_1}}^+c_{{}_{\ell_2,m_2}}^+ \ldots c_{{}_{\ell_\mathfrak{q'},m_\mathfrak{q'}}}^+
c_{{}_{\ell_{\mathfrak{q'}+1},\pm 1}}^+|u\rangle,
\end{multline}
except the last one, are of the form (\ref{l1m1...lqmqlq+10}), having at least one $m_i=0$, and by (\ref{l1m1...lqmqlq+10}), 
they all belong to $\mathcal{H}_{{}_{c_{{}_{\Sigma \ell_i,0}}^+|u\rangle}}$. Because the vector (\ref{Hpmc+limi...c+lq'+10})
belongs to $\mathcal{H}_{{}_{c_{{}_{\Sigma \ell_i,0}}^+|u\rangle}}$, then 
\begin{equation}\label{mq'+1=pm1}
c_{{}_{\ell_1,m_1}}^+c_{{}_{\ell_2,m_2}}^+ \ldots c_{{}_{\ell_\mathfrak{q'},m_\mathfrak{q'}}}^+
c_{{}_{\ell_{\mathfrak{q'}+1},\pm 1}}^+|u\rangle \,\, \in \mathcal{H}_{{}_{c_{{}_{\Sigma \ell_i,0}}^+|u\rangle}},
\end{equation}
thus showing that the vector (\ref{l1m1...lqmqlq+1mq+1}) belongs to $\mathcal{H}_{{}_{c_{{}_{\Sigma \ell_i,0}}^+|u\rangle}}$
if any of $m_i$ is equal $\pm 1$.
We continue similarly, applying  $H_\pm$ to (\ref{mq'+1=pm1}),
and showing that the vector (\ref{l1m1...lqmqlq+1mq+1}) belongs to $\mathcal{H}_{{}_{c_{{}_{\Sigma \ell_i,0}}^+|u\rangle}}$
if any of $m_i$ is equal to $\pm 2$, and so on.

\qed

\subsection{Relation of the space $\mathcal{H}_n$ with the cyclic subspaces $\mathcal{H}_x \subset \mathcal{H}_n$. Proof of theorem}\label{HnandCyclicDomains}

Consider the set theoretical sum
\begin{multline*}
\left(\mathcal{H}_{{}_{|u\rangle}} = \mathcal{H}_{{}_{c_{{}_{1,0}}^+|u\rangle}}\right) \cup \left(\mathcal{H}_{{}_{c_{{}_{2,0}}^+|u\rangle}}
\cup \mathcal{H}_{{}_{c_{{}_{1,0}}^+c_{{}_{1,0}}^+|u\rangle}}\right) \cup
\\
\cup \left(\mathcal{H}_{{}_{c_{{}_{3,0}}^+|u\rangle}}
\cup \mathcal{H}_{{}_{c_{{}_{2,0}}^+c_{{}_{1,0}}^+|u\rangle}} \cup \mathcal{H}_{{}_{c_{{}_{1,0}}^+c_{{}_{1,0}}^+c_{{}_{1,0}}^+|u\rangle}}\right)
\cup \ldots
\end{multline*}
of the domains of the cyclic representations with the cyclic vectors $|u\rangle, c_{{}_{\alpha_i}}^+|u\rangle, \ldots$
of te form (\ref{x}) with the azimuthal numbers $m_i=0$ all equal zero in $\alpha_i = (l_i,m_i)$. From lemma \ref{azimuth=0}
it follows that this sum contains linear span of the orthogonal complete system (\ref{x}). Therefore, the closure of this sum is equal $\mathcal{H}_n$.
From lemma \ref{azimuth=0} it follows that the closure of 
\[
 \mathcal{H}_{{}_{c_{{}_{1,0}}^+|u\rangle}} \cup  \mathcal{H}_{{}_{c_{{}_{2,0}}^+|u\rangle}} \cup
\mathcal{H}_{{}_{c_{{}_{3,0}}^+|u\rangle}} \cup \ldots
\]
is equal $\mathcal{H}_n$ and 
\[
\mathcal{H}_{{}_{c_{{}_{l,0}}^+|u\rangle}} \subset \mathcal{H}_{{}_{c_{{}_{l+1,0}}^+|u\rangle}}  
\]
by lemma \ref{Hcalpha'|u>,Hcalpha''|u>}.

Let $(l_0,l_1)$ be the irreducible representations of \cite{Geland-Minlos-Shapiro}, and let, for simplicity of notation, 
the direct sum/integral decomposition of lemma \ref{L3} for $x = c_{{}_{l,0}}^+|u\rangle$, be denoted by
\[
U\Big|_{{}_{\mathcal{H}_{{}_{c_{{}_{l,0}}^+|u\rangle}} }} =
\overset{l}{\underset{l_0=-l}{\bigoplus}} \int\limits_{0}^{+\infty} (l_0,i\rho) \, \nu(\rho,z) d\rho \,  \bigoplus \nu(z) \, (l_0=0, 1-z),
\,\,\,\,\,\, z = {\textstyle\frac{n^2\mathfrak{e}^2}{\pi}},
\]
with the corresponding positive weights $\nu(\rho,z), \nu(z)$, with $\nu(\rho,z)>0$ a.e. for each $z>0$, and $\nu(z)>0$ for $0<z<1$,
and  $\nu(z)=0$ for $z\geq 1$, with the standard Lebesgue measure $d\rho$ on $\mathbb{R}$. 
Using again lemma \ref{Hcalpha'|u>,Hcalpha''|u>}, we arrive with decomposition (\ref{decomposition}). 
Our theorem follows from (\ref{decomposition}).

\section*{Acknowledgments} 

The first-named author wishes to acknowledge 
the support of the Bogoliubov Laboratory of Theoretical Physics,
Joint Institute for Nuclear Research, 141980 Dubna.
We thank the Reviewer for his suggestions.



\end{document}